%% file: entanglement.tex
\documentclass[11pt]{report}
\pdfoutput=1
\input{preamble.tex}

\begin{document}

\title{The Mathematics of Entanglement\\ \vspace{1cm} {\Large Summer School at Universidad de los Andes}}
\author{Fernando G.~S.~L. Brand\~ao (UCL)\\
Matthias Christandl (ETHZ)\\
Aram W.~Harrow (MIT)\\
Michael Walter (ETHZ)}
\date{27--31 May, 2013}
\maketitle

\section*{Foreword}
These notes are from a series of lectures given at the Universidad de Los Andes in Bogot\'a, Colombia on some topics of current interest in quantum information.
While they aim to be self-contained, they are necessarily incomplete and idiosyncratic in their coverage.
For a more thorough introduction to the subject, we recommend one of the textbooks by Nielsen and Chuang or by Wilde, or the lecture notes of Mermin, Preskill or Watrous.
Our notes by contrast are meant to be a relatively rapid introduction into some more contemporary topics in this fast-moving field.
They are meant to be accessible to advanced undergraduates or starting graduate students.

\vfill
\section*{Acknowledgments}
We would like to thank our hosts Alonso Botero, Andres Schlief and
Monika Winklmeier from the Universidad de Los Andes for inviting us
and putting together the summer school.  We would also like to thank
the enthusiastic students who attended.

\eject

\setcounter{tocdepth}{1}
\tableofcontents

\include{27may_1}
\include{27may_2}
\include{27may_3}
\include{27may_ex}
\include{28may_1}
\include{28may_2}
\include{28may_3}
\include{28may_ex}
\include{29may_1}
\include{29may_2}
\include{29may_3}
\include{30may_1}
\include{30may_2}
\include{30may_3}
\include{30may_ex}
\include{31may_1}

\include{31may_2}

\include{31may_3}

\end{document}

%% file: preamble.tex
\usepackage[ascii]{inputenc}
\usepackage[colorlinks,bookmarksnumbered,hypertexnames=false]{hyperref}
\usepackage{bbm,braket,microtype,aliascnt,mathrsfs,amsmath,amssymb,color,amsthm,fullpage,cite,graphicx,cleveref,booktabs,multirow,tikz,mathtools,ifthen}
\usepackage[T1]{fontenc}
\usepackage[USenglish]{babel}

\newcounter{lecture}\crefname{lecture}{lecture}{lectures}
\newtheorem{thm}{Theorem}[lecture]\crefname{thm}{theorem}{theorems}
\crefname{lem}{lemma}{lemmas}
\newtheorem{cor}[thm]{Corollary}\crefname{cor}{corollary}{corollaries}
\newtheorem{prb}[thm]{Problem}\crefname{prb}{problem}{problems}
\crefname{dfn}{definition}{definitions}
\crefname{prop}{proposition}{propositions}
\newtheorem{ex}[thm]{Exercise}\crefname{ex}{exercise}{exercises}
\theoremstyle{remark}
\newtheorem*{exl*}{Example}
\newtheorem*{ex*}{Exercise}

\numberwithin{equation}{lecture}
\numberwithin{section}{lecture}
\DeclareMathOperator{\poly}{poly}
\DeclareMathOperator{\PPT}{PPT}
\DeclareMathOperator{\rank}{rank}
\DeclareMathOperator{\Sep}{Sep}
\DeclareMathOperator{\SEP}{SEP}
\DeclareMathOperator{\sgn}{sgn}
\DeclareMathOperator{\SL}{SL}
\DeclareMathOperator{\Span}{span}
\DeclareMathOperator{\SU}{SU}
\DeclareMathOperator{\Sym}{Sym}
\DeclareMathOperator{\tr}{tr}
\DeclareMathOperator{\Var}{Var}
\DeclareMathOperator{\id}{id}
\newcommand{\E}{\mathbb E}
\newcommand{\Ptime}{\mathsf P}
\newcommand{\NP}{\mathsf{NP}}
\newcommand{\QMA}{\mathsf{QMA}}
\newcommand{\RR}{\mathbb R}
\newcommand{\CC}{\mathbb C}
\newcommand{\proj}[1]{\ket{#1}\!\bra{#1}}
\newcommand{\ot}{\otimes}
\newcommand{\ra}{\rightarrow}
\renewcommand{\Re}{\operatorname{Re}}
\renewcommand{\L}{\left}
\newcommand{\R}{\right}
\newcommand{\eps}{\epsilon}
\newcommand{\Psym}{\Pi_{\text{sym}}}
\def\be#1\ee{\begin{equation}#1\end{equation}}
\def\bea#1\eea{\begin{eqnarray}#1\end{eqnarray}}
\def\beas#1\eeas{\begin{eqnarray*}#1\end{eqnarray*}}
\def\bas#1\eas{\begin{align*}#1\end{align*}}
\def\bsm#1\esm{\begin{smallmatrix}#1\end{smallmatrix}}
\def\bpm#1\epm{\begin{pmatrix}#1\end{pmatrix}}
\def\bsm#1\esm{\left(\begin{smallmatrix}#1\end{smallmatrix}\right)}
\def\benum{\begin{enumerate}}
\def\eenum{\end{enumerate}}
\def\bit{\begin{itemize}}
\def\eit{\end{itemize}}

\newcommand{\handout}[6]{
  \setcounter{lecture}{#1}
  \addtocounter{lecture}{-1}
  \refstepcounter{lecture}
  \setcounter{section}{0}
  \setcounter{equation}{0}
  \setcounter{thm}{0}
  \addtocontents{}{}{}
  \phantomsection
  \addcontentsline{toc}{chapter}{#6}
  \addtocontents{}{}{}
  \noindent
  \begin{center}
    \framebox{\vbox{
    \hbox to 5.78in { {\bf The Mathematics of Entanglement - Summer 2013} \hfill #2 }
    \vspace{4mm}
    \hbox to 5.78in { {\Large \hfill #5  \hfill} }
    \vspace{2mm}
    \hbox to 5.78in { {\it #3 \hfill #4}}}}
  \end{center}
  \vspace*{4mm}
}
\newcommand{\lecture}[4]{\renewcommand{\thelecture}{\arabic{lecture}}\handout{#1}{#2}{Lecturer: #3}{Lecture #1}{#4}{Lecture #1 - #4}}
\newcommand{\problemsession}[3]{\renewcommand{\thelecture}{\Roman{lecture}}\handout{#1}{#2}{Lecturer: #3}{}{Problem Session \Roman{lecture}}{Problem Session \Roman{lecture}}}

\ifthenelse{\isundefined{\nosolutions}}
  {\newenvironment{sol}[1][Solution]{\small\begin{proof}[\rm \bf #1]}{\end{proof}}}
  {\newenvironment{sol}{\comment}{\endcomment}}

%% file: 27may_1.tex
\lecture{1}{27 May, 2013}{Fernando G.S.L. Brand\~ao}{Quantum states}

Entanglement is a quantum-mechanical form of correlation which appears in many areas, such as condensed matter physics, quantum chemistry, and other areas of physics.  This week we will discuss a perspective from quantum information, which means we will abstract away the underlying physics, and make statements about entanglement that apply independent of the underlying physical system.  This will also allow us to discuss information-processing applications, such as quantum cryptography.

\section{Probability theory and tensor products}
Before discussing quantum states, we explain some aspects of probability theory, which turns out to have many similar features.

Suppose we have a system with $d$ possible states, for some integer $d$, which we label by $1,\dots,d$. Thus a deterministic state is simply an element of the set $\{1,\dots,d\}$.  The probabilistic states are probability distributions over this set, i.e.\ vectors in $\RR_+^d$ whose entries sum to 1.  The notation $\RR_+^d$ means that the entries are nonnegative.  Thus, a probability distribution $p = (p(1), \dots, p(d))$ satisfies $\sum_{x=1}^d p(x)=1$ and $p(x)\geq 0$ for each $x$.
Note that we can think of a deterministic state $x \in \{1,\dots,d\}$ as the probability distribution where $p(x) = 1$ and all other probabilities are zero.

\subsection{Composition and tensor products}
If we bring a system with $m$ states together with a system with $n$ states then the composite system has $mn$ states, which we can identify with the pairs $(1,1), (1,2), \dots, (m,n)$.
Thus a state of the composite system is given by a probability distribution $p = (p(x,y))$ in $\RR^{mn}_+$.
The states of the subsystems can be described by the marginal distributions $p(x) = \sum_{y=1}^n p(x,y)$ and $p(y) = \sum_{x=1}^m p(x,y)$.

Conversely, given two probability distributions $p\in \RR_+^m$ and $q\in \RR_+^n$, we can always form a joint distribution of the form
$$p \ot q := \bpm p(1)q(1) \\ p(1)q(2) \\ \vdots \\ p(1)q(n) \\ \vdots \\ p(m)q(n) \epm.$$
That is, the probability of a pair $(x,y)$ is equal to $p(x)q(y)$.
In this case we say that the states of the two systems are \emph{independent}.

Above, we have introduced the notation $\ot$ to denote the {\em tensor product}, which in general maps a pair of vectors with dimensions $m,n$ to a single vector with dimension $mn$.
Later we will also consider the tensor product of matrices.  If $M_n$ denotes the space of $n\times n$ matrices, and we have $A\in M_m, B\in M_n$ then $A\ot B\in M_{mn}$ is the matrix whose entries are all possible products of an entry of $A$ and an entry of $B$.  For example, if $m=2$ and $A = \bsm a_{11} & a_{12} \\ a_{21} & a_{22} \esm$ then $A\ot B$ is the block matrix
$$\bpm a_{11}B & a_{12}B \\ a_{21}B & a_{22}B \epm.$$
One useful fact about tensor products, which simplifies many calculations, is that
$$(A \ot B) (C\ot D) = AC \ot BD.$$

We also define the tensor product of two vector space $V\ot W$ to be the span of all $v\ot w$ for $v\in V$ and $w\in W$.  In particular, observe that $\CC^m \ot \CC^n =\CC^{mn}$.

\section{Quantum mechanics}
We will use Dirac notation in which a ``ket'' $\ket\psi$ denote a column vector in a complex vector space, i.e.
$$\ket\psi = \bpm \psi_1 \\ \psi_2 \\ \vdots \\ \psi_d \epm \in \CC^d.$$
The ``bra'' $\bra\psi$ denotes the conjugate transpose, i.e.
$$\bra\psi = \bpm \psi_1^* & \psi_2^* & \cdots & \psi_d^* \epm.$$
Combining a bra and a ket gives a ``bra[c]ket'', meaning an inner product
$$\braket{\varphi|\psi} = \sum_{i=1}^d \varphi_i^* \psi_i .$$
In this notation the norm is
$$\|\psi\|_2 = \sqrt{\braket{\psi|\psi}} = \sqrt{\sum_{i=1}^d |\psi_i|^2}.$$

Now we can define a quantum state.  The quantum analogue of a system with $d$ states is the $d$-dimensional Hilbert space $\CC^d$.
For example, a quantum system with $d=2$ is called a \emph{qubit}.
Unit vectors $\ket\psi\in\CC^d$, where $\braket{\psi|\psi}=1$, are called {\em pure states}. They are the analogue of deterministic states in classical probability theory.
For example, we might define the following pure states of a qubit:
$$
  \ket 0 = \bpm 1 \\ 0 \epm,
  \ket 1 = \bpm 0 \\ 1 \epm,
  \ket + = \frac{1}{\sqrt{2}} ( \ket 0 + \ket 1),
  \ket - = \frac{1}{\sqrt{2}} ( \ket 0 - \ket 1).
$$
Note that both pairs $\ket 0, \ket 1$ and $\ket +, \ket -$ form orthonormal bases of a qubit.
It is also customary to write $\ket\uparrow = \ket0$ and $\ket\downarrow = \ket1$.

\subsection{Measurements}
A {\em projective measurement} is a collection of projectors $\{P_k\}$ such that $P_k \in M_d$ for each $k$, $P_k^\dag = P_k$, $P_kP_{k'} = \delta_{k,k'}P_k$ and $\sum_k P_k = I$.  For example, we might measure in the {\em computational basis}, which consists of the unit vectors $\ket k$ with a one in the $k^{\text{th}}$ position and zeros elsewhere. Thus define
$$P_k = \proj k = \bpm
0 & & & & & & 0 \\
& \ddots & \\
& & 0\\
& & & 1\\
& & & & 0\\
& & & & & \ddots\\
0 & & & & & & 0
\epm,$$
which is the projector onto the one-dimensional subspace spanned by $\ket k$.

\emph{Born's rule} states that $\Pr[k]$, the probability of measurement outcome $k$, is given by
\begin{equation}
\label{eq:born}
  \Pr[k] = \bra{\psi}P_k \ket{\psi}.
\end{equation}
As an exercise, verify that this is equal to $\tr(P_k \proj\psi)$.
In our example, this is simply $|\psi_k|^2$.

\begin{exl*}
If we perform the measurement $\{\proj 0, \proj 1\}$ on $\ket +$, then $\Pr[0] = \Pr[1] = 1/2$.  If we perform the measurement $\{\proj +, \proj -\}$, then $\Pr[+] = 1$ and $\Pr[-] = 0$.
\end{exl*}

\section{Mixed states}
Mixed states are a common generalization of probability theory and pure quantum mechanics. In general, if we have an ensemble of pure quantum states $\ket{\psi_x}$ with probabilities $p(x)$, then define the {\em density matrix} to be
$$\rho = \sum_x p(x) \proj{\psi_x}.$$
The vectors $\ket{\psi_x}$ do not have to be orthogonal.

Note that $\rho$ is always Hermitian, meaning $\rho = \rho^\dag$.  Here $^\dag$ denotes the conjugate transpose, so that $(A^\dag)_{i,j} = A_{j,i}^*$.
In fact, $\rho$ is positive semi-definite (``PSD''). This is also denoted $\rho \geq 0$.
Two equivalent definitions (assuming that $\rho=\rho^\dag$) are:
\benum
\item For all $\ket\psi$, $\bra\psi\rho\ket\psi \geq 0$.
\item All the eigenvalues of $\rho$ are nonnegative.  That is,
\be \rho = \sum_i \lambda_i \proj{\varphi_i}\label{eq:rho-decomp}\ee
 for an orthonormal basis $\{\ket{\varphi_1},\dots,\ket{\varphi_d}\}$ with each $\lambda_i \geq 0$.
\eenum

\begin{ex*}
	Prove that these definitions are equivalent.
\end{ex*}

A density matrix should also have trace one, since $\tr\rho = \sum_x p(x) \braket{\psi_x|\psi_x} =\sum_x p(x)=1$.

\smallskip

Conversely, any PSD matrix with trace one can be written in the form $\sum_x p(x)\proj{\psi_x}$ for some probability distribution $p$ and some unit vectors $\{\ket{\psi_x}\}$, and hence is a valid density matrix. This is just based on the eigenvalue decomposition: we can always take $p(x)=\lambda_x$ in \cref{eq:rho-decomp}.

Note that this decomposition is not unique in general. For example, consider the \emph{maximally mixed state} $\rho = I / 2 = \bsm 1/2 & 0 \\ 0 & 1/2 \esm$.  This can be decomposed either as $\frac{1}{2} \proj 0 + \frac{1}{2}\proj 1$ or as
$\frac{1}{2} \proj + + \frac{1}{2}\proj -$, or indeed as $\frac{1}{2} \proj u + \frac{1}{2}\proj v$ for any orthonormal basis $\{\ket{u}, \ket v\}$.

\smallskip

For mixed states, if we measure $\{P_k\}$ then the probability of outcome $k$ is given by
\[ \Pr[k] = \tr(\rho P_k). \]
This follows from \emph{Born's rule}~\eqref{eq:born} and linearity.


\section{Composite systems and entanglement}
Here and throughout most of these lectures we will work with distinguishable particles.
A pure state of two quantum systems is given by a unit vector $\ket\eta$ in the tensor product Hilbert space $\CC^n \ot \CC^m \cong \CC^{nm}$.

For example, if particle A is in the pure state $\ket{\psi_A}$ and particle B is in the pure state $\ket{\varphi_B}$ then their joint state is $\ket{\eta_{AB}} = \ket{\psi_A} \ot \ket{\varphi_B}$.
If $\ket{\psi_A}\in\CC^n$ and $\ket{\varphi_B}\in\CC^m$, then we will have $\ket{\eta_{AB}} \in \CC^{mn}$.

This should have the property that if we measure one system, say A, then we should obtain the same result in this new formalism that we would have had if we treated the states separately.  If we perform the projective measurement $\{P_k\}$ on system A then this is equivalent to performing the measurement $\{P_k \ot I\}$ on the joint system.  We can then calculate
\bas \Pr[k] &= \braket{\eta_{AB} | P_k \ot I | \eta_{AB}} = \bra{\psi_A} \bra{\varphi_B} (P_k \ot I) \ket{\psi_A} \ket{\varphi_B}
= \bra{\psi_A} P_k \ket{\psi_A} \braket{\varphi_B | \varphi_B}  = \bra{\psi_A} P_k \ket{\psi_A}. \eas

In probability theory, any deterministic distribution of two random variables can be written as a product $p(x,y) = \delta_{x,x_0} \delta_{y,y_0}$.
In contrast, there are pure quantum states which cannot be written as a tensor product $\ket{\psi_A} \ot \ket{\varphi_B}$ for any choice of $\ket{\psi_A}, \ket{\varphi_B}$.
We say that such pure states are {\em entangled}.
For example, consider the ``EPR pair''
$$\ket{\Phi^+} = \frac{\ket 0 \ot \ket 0 + \ket 1 \ot \ket 1}{\sqrt 2}.$$

Entangled states have many counterintuitive properties.  For example, suppose we measure the state $\ket{\Phi^+}$ using the projectors $\{P_{j,k} = \proj j \ot \proj k\}$.  Then we can calculate
\bas
\Pr[(0,0)] &= \bra{\Phi^+} P_{0,0} \ket{\Phi^+}
= \bra{\Phi^+} \, \proj 0 \ot \proj 0 \, \ket{\Phi^+} = \frac{1}{2}, \\
\Pr[(1,1)] &= \frac{1}{2}, \quad
\Pr[(0,1)] = 0, \quad
\Pr[(1,0)] = 0
\eas
The outcomes are perfectly correlated.

However, observe that if we measure in a different basis, we will also get perfect correlation.
Consider the measurement
$$\{ \proj{++}, \proj{+-}, \proj{-+},\proj{--} \},$$
where we have used the shorthand  $\ket{{++}} := \ket + \ot \ket +$, and similarly for the other three.
Then one can calculate (and doing so is a good exercise) that, given the state $\ket{\Phi^+}$, we have
$$\Pr[(+,+)]  =\Pr[(-,-)] = \frac{1}{2},$$
meaning again there is perfect correlation.

\subsection{Partial trace}
Suppose that $\rho_{AB}$ is a density matrix on $\CC^n \ot \CC^m$.
We would like a quantum analogue of the notion of a marginal distribution in probability theory.
Thus we define the \emph{reduced state} of $A$ to be
$$\rho_A := \tr_B(\rho_{AB}) := \sum_k (I_A \ot \bra{k_B}) \rho_{AB} (I_A \ot {\ket k_B}),$$
where $\{\ket{k_B}\}$ is any orthonormal basis on B. The operation $\tr_B$ is called the \emph{partial trace} over $B$.

We observe that if we perform a measurement $\{P_j\}$ on A, then we have
$$\Pr[j] = \tr ((P_j \ot I_B)\rho_{AB}) = \tr ( P_j \tr_B (\rho_{AB})) = \tr (P_j \rho_A).$$
Thus the reduced state $\rho_A$ perfectly reproduces the statistics of any measurement on the system $A$.

%% file: 27may_2.tex
\lecture{2}{27 May, 2013}{Matthias Christandl}{Quantum operations}
\label{lec:quantum ops}

In this lecture we will talk about dynamics in quantum mechanics. We will start again with measurements, and then go to unitary evolutions and general quantum dynamical processes.

\section{Measurements and POVMs}

\begin{figure}[b]
\centerline{\includegraphics[width=0.8\textwidth]{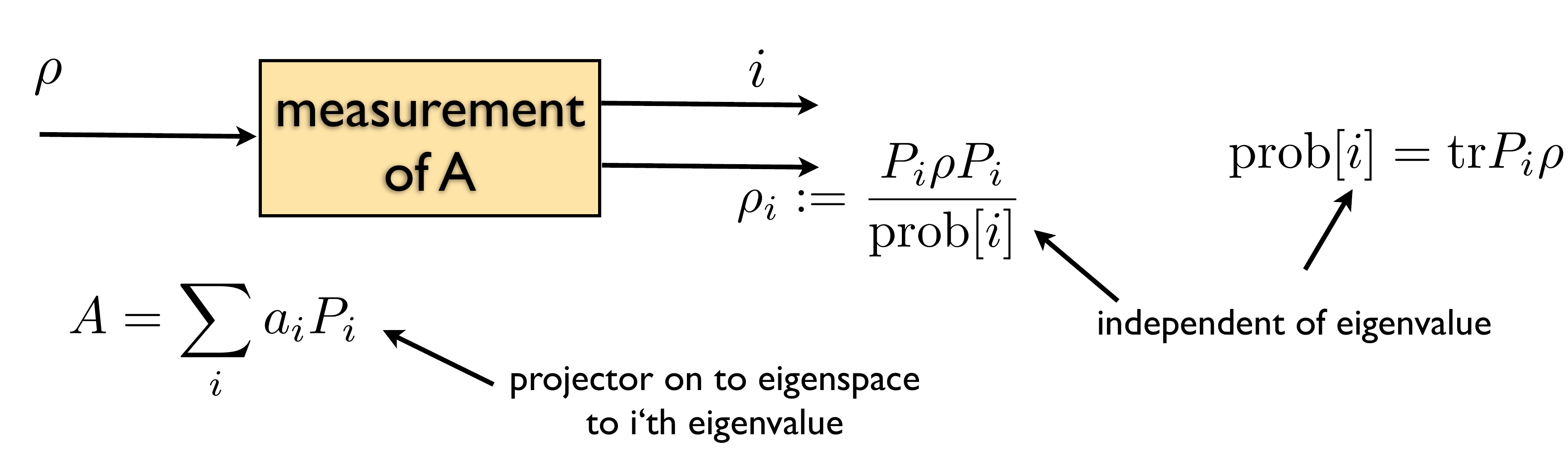}}
\caption{Sketch of a projective measurement $\{P_i\}$ with corresponding observable $A$.\label{fig:proj}}
\end{figure}

Consider a quantum measurement as a box, applied to a mixed quantum state $\rho$, with possible outcomes labelled by $i$.
In the previous lecture, we considered projective measurements given by orthogonal projectors $\{ P_i \}$, with Born's rule $\Pr[i] = \tr(P_i \rho)$ (\cref{fig:proj}).

Another common way to think about these is the following: We can associate to any projective measurement an \emph{observable} $A$ with eigendecomposition $A = \sum_{i} a_i P_i$, where we think of the $a_i$ as the values that the observable attains for each outcome (e.g., the value the measurement device displays, the position of a pointer, \dots). Then the expectation value of $A$ in the state $\rho$ is $\tr(A \rho) = \sum_i a_i \tr(P_i \rho)$.

\begin{figure}
\centerline{\includegraphics[width=0.8\textwidth]{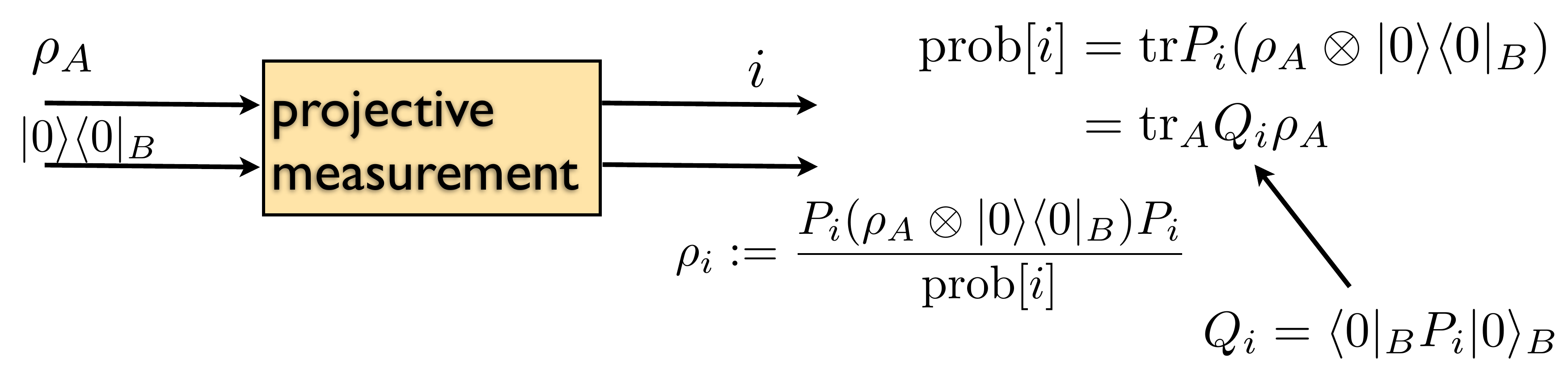}}
\caption{Sketch of a POVM measurement $\{Q_i\}$ built from a projective measurement $\{P_i\}$ on a larger system.\label{fig:povm}}
\end{figure}

But is this the most general measurement allowed in quantum mechanics? It turns out that this is not the case. Suppose we have a quantum state $\rho_A$ on $\CC^d$ and we consider the joint state $\rho_A \otimes \proj 0_B$, with $\ket 0_B \in \CC^d$ the state of an ancillary particle. Let us perform a projective measurement $\{ P_i \}$ on the joint system $\CC^{d} \otimes \CC^{d'} = \CC^{dd'}$ (\cref{fig:povm}).
Then the probability of measuring $i$ is
$$
  \Pr[i] = \tr \left( P_i \, \left( \rho_A \otimes \proj 0_B \right) \right).
$$
Using the partial trace, we can rewrite this as follows:
\begin{align*}
\Pr[i] &= \tr_{A} \left(  \tr_{B} \left( P_i \, \left( \rho_{A} \otimes \proj 0_B \right) \right)   \right) \\
&= \tr_{A} \left( \braket{0_B | P_i | 0_B} \, \rho_A \right) \\
&= \tr(Q_i \rho_A),
\end{align*}
where $Q_i := \braket{0_B | P_i | 0_B}$. Thus the operators $\{Q_i\}$ allow us to describe the measurement statistics without having to consider the state of the ancillary system. What are the properties of $Q_i$? First, it is PSD:
$$
  \braket{\phi_A | Q_i | \phi_A} = \bra{\phi_A} \bra{0_B} P_i \ket{\phi_A} \ket{0_B} \geq 0,
$$
since $P_i \geq 0$. Second, the $Q_i$ sum up to the identity:
$$
  \sum_{i} Q_i = \sum_{i} \braket{0_B | P_i | 0_B} = \braket{0_B | \sum_i P_i | 0_B} = \braket{0_B | I_{AB} | 0_B} = I_A.
$$

The converse of the above is also true, as you will show in \cref{ex:povm}.
Whenever we are given a set of PSD matrices $Q_i \geq 0$ with $\sum_i Q_i = I$, we can always find a projective measurement $\{ P_i \}$ on a larger system $A \otimes B$ such that
\begin{equation*}
  \tr(Q_i \rho_A) = \tr(P_i \, (\rho_A \otimes \proj 0_B)).
\end{equation*}
The generalized quantum measurements we obtain in this way are called \emph{positive operator-valued measure(ment)s}, or POVMs.
Note that since the $Q_i$'s are not necessarily orthogonal projections, there is no upper bound on the number of elements in a POVM.

\begin{exl*}
Consider two projective measurements, e.g.\ $\{ \proj 0, \proj 1 \}$ and $\{ \proj +, \proj - \}$. Then we can define a POVM as a  mixture of these two:
\begin{align*}
  Q_0 &= \frac 1 2 \proj 0, \\
  Q_1 &= \frac 1 2 \proj 1, \\
  Q_2 &= \frac 1 2 \proj +, \\
  Q_3 &= \frac 1 2 \proj -.
\end{align*}
It is clear that $\sum_k Q_k = I$. One way of thinking about this POVM is that with probability 1/2 we measure in the computational basis, and with probability 1/2 in the $\ket{\pm}$ basis.
\end{exl*}

\begin{figure}
\centerline{\includegraphics[width=0.8\textwidth]{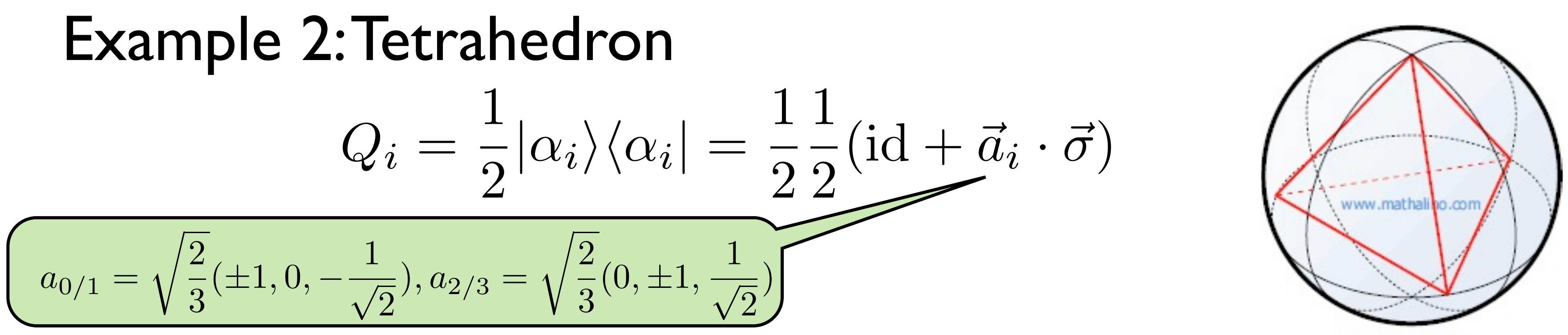}}
\caption{Sketch of a POVM measurement $\{Q_i\}$ constructed from four pure states that form a tetrahedron on the Bloch sphere.\label{fig:tetra}}
\end{figure}

\begin{exl*}
The quantum state $\rho$ of a qubit can always be written in the form
$$
\rho = \rho(\vec r) = \frac 1 2 \left( I + r_x \sigma_x + r_y \sigma_y + r_z \sigma_z \right),
$$
with the Pauli matrices
$$
  \sigma_x = \bpm 0 & 1 \\ 1 & 0 \epm,
  \sigma_z = \bpm 1 & 0 \\ 0 & -1 \epm,
  \sigma_y = \bpm  0 & -i \\ i & 0 \epm.
$$
Since the Pauli matrices are traceless, $\rho$ has indeed trace one. We can the describe the state by a 3-dimensional vector $\vec{r} = (r_x, r_y, r_z) \in \mathbb{R}^3$. It turns out that $\rho$ is PSD if, and only if, $\lVert r \rVert \leq 1$. Therefore, any quantum state of a qubit corresponds to a point in a 3-dimensional ball, called the \emph{Bloch ball}. A state $\rho$ is pure if, and only if, $\lVert r \rVert = 1$, i.e.\ if it is an element of the \emph{Bloch sphere}. The maximally mixed state $I/2$ corresponds to the origin $\vec{r} = (0,0,0)$.

Now consider a collection of four pure states
$\{ \proj{a_i} \}_{i=1,\dots,4}$
that form a tetrahedron on the Bloch sphere (\cref{fig:tetra}).
Then, by symmetry of the tetrahedron, $\sum_i \proj{a_i} = I$, so they form indeed a POVM.
\end{exl*}

\section{Unitary dynamics}

Let $\ket{\psi}$ be a quantum state and consider its time evolution according to the Schr\"odinger equation for a time-independent Hamiltonian $H$. Then the state after some time $t$ is given by
$$
  \ket{\psi_t} = e^{-i H t} \ket{\psi},
$$
where we have set $\hbar = 1$. The matrix $U = e^{-i H t}$ describing the evolution of the system is a unitary matrix, i.e.\ $U U^{\dagger} = U^{\dagger} U = I$.

\begin{exl*}
$U_t = e^{it \vec{e} \cdot \vec{\sigma}/2}$ with $\vec{e} \in \mathbb{R}^3$ a unit vector and $\vec{\sigma} = (\sigma_x, \sigma_y, \sigma_z)$ the vector of Pauli matrices. We have
$$
 U_t \, \rho(\vec r) \, U_{t}^\dagger = \rho(R_t \, \vec{r}),
$$
where $R_t$ denotes the matrix describing a rotation by an angle $t$ around the axis $\vec{e}$.
\end{exl*}

\begin{exl*}
The Hadamard unitary is given by $H = \frac{1}{\sqrt{2}} \bsm 1 & 1 \\ 1 & -1 \esm$.
Its action on the computational basis vectors is
\begin{align*}
  H \ket{0} = \ket{+}, \quad
  H \ket{1} = \ket{-}.
\end{align*}
\end{exl*}

\section{General time evolutions}

\begin{figure}
\centerline{\includegraphics[width=0.8\textwidth]{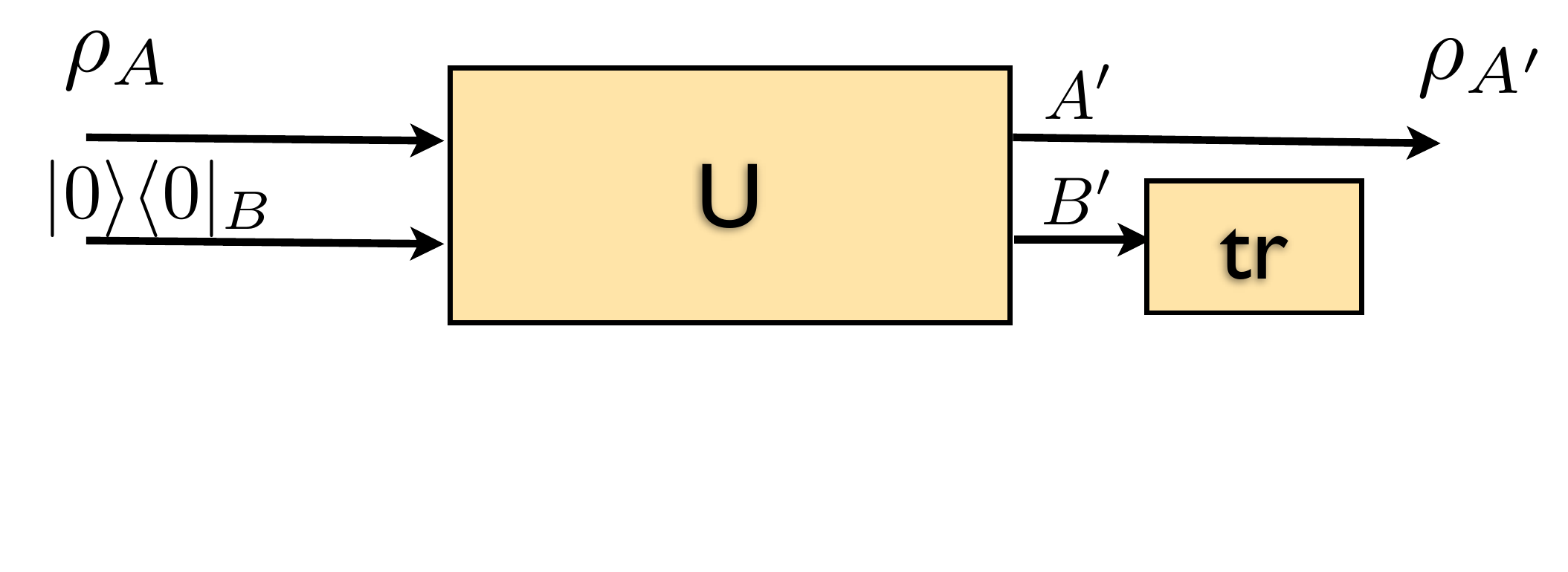}}
\caption{Sketch of a quantum operation built from a unitary time evolution on a larger system.\label{fig:cptp}}
\end{figure}

There are more general possible dynamics in quantum mechanics than unitary evolution. One possibility is that we add an acilla state $\proj 0_B$ to $\rho_A$ and consider a unitary dynamics $U_{AB\to A'B'}$ on the joint state.
Thus the resulting state of the $A'B'$ system is
$$U_{AB\to A'B'} ( \rho_A \otimes {\proj 0}_B ) U_{AB\to A'B'}^\dagger.$$
Suppose now that we are only interested in the final state of the subsystem $A'$. Then
$$\rho_{A'} = \tr_{B'} \left( U_{AB\to A'B'} ( \rho_A \otimes {\proj 0}_B ) U_{AB\to A'B'}^\dagger \right),$$
where we traced out over subsystem $B'$. We can associate a map $\Lambda$ to this evolution by
$$\Lambda(\rho_A) = \rho_A' = \tr_{B'} \left( U_{AB\to A'B'} ( \rho_A \otimes {\proj 0}_B ) U_{AB\to A'B'}^\dagger \right),$$
see \cref{fig:cptp}.

What are the properties of $\Lambda$? First it maps PSD matrices to PSD matrices. We call this property \emph{positivity}. Second, it preserves the trace---we say the map is \emph{trace-preserving}. In fact, even the map $\Lambda \otimes \id$, where $\id$ is the identity map on an auxiliary space of arbitrary dimension, is positive. We call this property \emph{completely positivity}.

An important theorem, \emph{Stinespring's dilation theorem}, is that the converse also holds:
Any map $\Lambda$ which is completely positive and trace-preserving can be written as
\begin{equation}
\Lambda(\rho_A) = \tr_{B'} \left( U_{AB\to A'B'} ( \rho_A \otimes {\proj 0}_B ) U_{AB\to A'B'}^\dagger \right)
\end{equation}
for some suitable unitary $U_{AB\to A'B'}$.  Therefore, any general quantum dynamics can be realized by a completely positive, trace-preserving map, also called a \emph{quantum operation} or \emph{quantum channel}.

\begin{exl*}
A basic example of a quantum operation is the so-called \emph{depolarizing channel},
$$\Lambda(\rho) = (1 - p) \rho + p \frac I d.$$
With probability $1-p$, the state is preserved; with probability $p$ the state is ``destroyed'' and replaced by the maximally mixed one, modeling a simple type of noise.
\end{exl*}

%% file: 27may_3.tex
\lecture{3}{27 May, 2013}{Aram Harrow}{Quantum entropy}

\section{Shannon entropy}

In this part, we want to understand quantum information in a quantitative way. One of the important concepts is entropy. But let us first look at classical entropy.

Given is a probability distribution $p \in \RR^d_+$, $\sum_x p(x)=1$. The \emph{Shannon entropy} of $p$ is defined to be $$H(p)=-\sum_x p(x) \log p(x).$$
Here, and in the following, the logarithm is always taken to base two, corresponding to the unit ``bit''. Moreover, we set $0 \log 0 := \lim_{s\rightarrow 0} s \log s = 0$.

Entropy quantifies uncertainty. We have maximal certainty for a deterministic distribution $p(x) = \delta_{x,x_0}$, where $H(p)=0$. The distribution with maximal uncertainty is the uniform distribution $p(x) \equiv \frac 1 d$, for which $H(p)=\log d$.

In the following we want to give Shannon entropy an operational meaning by considering the problem of \emph{data compression}.
For this, imagine you have a binary alphabet ($d=2$) and you sample $n$ times independently from the distribution $p=(\pi, 1-\pi)$. We say that the corresponding random variables $X_1, \dots, X_n$ are independent and identically distributed (\emph{i.i.d.}).

Typically, the number of 0's in the string $X_1 \cdots X_n$ is $n\pi \pm O(\sqrt n)$ and the number of 1's is $n(1-\pi) \pm O(\sqrt{n})$. To see why this is the case, consider the sum $S=X_1 + \dots + X_n$ (i.e., the number of 1's in the string).
The expectation value of this random variable is
$$\E[S]=\E[X_1]+\dots +\E[X_n]=n(1-\pi),$$
where we have used the linearity of the expectation value. Furthermore, the variance of $S$ is
$$\Var[S] = \Var[X_1]+ \dots + \Var[X_n]=n \Var[X_1]= n \pi (1-\pi)\leq \frac n 4.$$
Here, we have used the independence of the random variables $X_i$ in the first equality and $Var[X_1]=\E[X_1^2]-\E[X_1]^2=(1-\pi)-(1-\pi)^2=\pi(1-\pi)$ in the third.
Thus the standard deviation of $S$ is smaller than $\frac{\sqrt{n}}{2}$.

What does this have to do with compression? The total number of strings of $n$ bits is $\lvert\{0,1\}^n\rvert=2^n$.
In contrast, the number of strings with $n \pi$ 0's is
$${\binom{n}{\pi n}}=\frac{n!}{(\pi n)! \, ((1-\pi) n)!} \approx \frac{(n/e)^n}{(\pi n/e)^{\pi n} \, ((1-\pi) n/e)^{(1-\pi) n}}= \pi^{-n\pi } (1-\pi)^{-n(1-\pi)},$$
where we have used Stirling's approximation.
We can rewrite this as
$$\exp(-n \pi \log \pi - n (1-\pi) \log (1-\pi)) = \exp (nH(p)).$$
Hence we only need to store around $\exp (nH(p))$ possible strings, which we can do in a memory having around $n H(p)$ bits.
(Note that so far we have ignored the fluctuations; if we took them into account, we would need an additional $O(\sqrt{n})$ bits.) This analysis easily generalises to arbitrary alphabets (not only binary).

\section{Typical sets}

I now want to give you a different way of looking at this problem, a way that is both more rigorous and will more easily generalise to the quantum case. This we will do with help of typical sets.

Again let $X_1, \dots, X_n$ be i.i.d distributed with distribution $p$ in some alphabet $\Sigma$. The probability of a string is then given by
$$\Pr[X_1 = x_1, \dots, X_n = x_n] = p(x_1) \cdots p(x_n) = p^{\otimes n}(x^n),$$
where we have introduced the notation $p^{\otimes n} = p \otimes \dots \otimes p$ and $x^n = (x_1, \dots, x_n) \in \Sigma^n$.
Note that
$$\log p^{\otimes n}(x^n)=\sum_{i=1}^n\log p(x_i)\approx n \E[\log p(x_i)]\pm \sqrt{n} \sqrt{Var[\log p(x_i)]}= -nH(p) \pm O(\sqrt{n})$$
where we have used that
$$\E[\log p(x_i)]=\sum_i p(x_i) \log p(x_i) =-H(p)$$.

Let us now define the \emph{typical set} as the set of strings
$$T_{p, n, \delta}=\{x^n \in \Sigma : |-\log p^{\otimes n }(x^n) - n H(p) | \leq n \delta\}.$$
Then, for all $\delta > 0$, we have that
\begin{equation*}
  \lim_{n \rightarrow \infty} \Pr[X^n \in T_{p,n,\delta}] =
  \lim_{n \rightarrow \infty} \sum_{x^n \in T_{p,n,\delta}} p^{\otimes n}(x^n) = 1.
\end{equation*}
Our compression algorithm simply keeps all the strings that are in the typical set and throws away all others.
Hence, all we need to know the size of the typical set. For this, note that
$$\exp (-n H(p)-n \delta) \leq p^{\otimes n}(x^n) \leq \exp (-n H(p)+n \delta)$$
for all typical strings $x^n \in T_{p, n, \delta}$.
Therefore,
$$1 \geq \Pr[X^n \in T_{p,n,\delta}] \geq |T_{p, n, \delta }| \ \min_{x^n \in T_{p,n,\delta}} p^{\otimes n}(x^n) \geq |T_{p,n,\delta}| \ \exp (-n H(p)-n \delta),$$
which implies that
$$\log  |T_{p, n, \delta }| \leq n (H(p) + \delta).$$

In \cref{ex:source compression}, you will make the above arguments more precise and show that this rate is optimal.
That is, we cannot compress to $nR$ bits for $R < H(p)$ unless the error does not go to zero as $n$ goes to infinity.

\section{Quantum compression}

When compressing quantum information, probability distributions are replaced by density matrices $\rho^{\otimes n} = \rho \otimes \cdots \otimes \rho$. If $\rho$ is a state of a qubit then this state acts on a $2^n$-dimensional Hilbert space.
The goal of quantum data compression is to represent this state on a lower-dimensional subspace.
In analogy to the case of bits, we now measure the size of this subspace in terms of the number of qubits that are needed to represent vectors in that subspace, i.e\ by the log of the dimension.

It turns out that it is possible (and optimal) to use $n(S(\rho) + \delta)$ qubits. Here, $S(\rho)$ is the \emph{von Neumann entropy} of the quantum state $\rho$, defined by
$$S(\rho) = -\tr \rho \log \rho = -\sum \lambda_i \log \lambda_i = H(\lambda),$$
where the $\lambda_i$ denote the eigenvalues of $\rho$.

%% file: 27may_ex.tex
\problemsession{1}{27 May, 2013}{Michael Walter}

\begin{ex}[POVM measurements]
\label{ex:povm}

Given a POVM $\{Q_i\}$, show that we can always find a projective measurement $\{ P_i \}$ on a larger system $A \otimes B$ such that
\begin{equation*}
  \tr(Q_i \rho_A) = \tr \left(P_i \, (\rho_A \otimes \proj 0_B) \right).
\end{equation*}

\begin{sol}
  Let $B$ denote an ancilla system of dimension $n$ and consider the map
  \[
    \ket{\phi_A} \ket{0_B} \mapsto \sum_{i=1}^n \sqrt{Q_i} \ket{\phi_A} \otimes \ket{i_B}.
  \]
  This map is an isometry on the subspace $A \otimes \ket{0_B}$, since
  \begin{align*}
    \bigl( \sum_{i=1}^n \bra{\phi_A} \sqrt{Q_i} \otimes \bra{i_B} \bigr) \bigl( \sum_{j=1}^n \sqrt{Q_j} \ket{\phi_A} \otimes \ket{j_B} \bigr)
    = \sum_{i,j} \braket{\phi_A | \sqrt{Q_i} \sqrt{Q_j} | \phi_A} \braket{i_B | j_B}
    = \sum_i \braket{\phi_A | Q_i | \phi_A}
    = \braket{\phi_A | \phi_A}.
  \end{align*}
  It can thus be extended to a unitary $U_{AB}$.
  We can thus define a projective measurement $(P_i)$ by $P_i = U_{AB}^\dagger (I_A \otimes \proj j_B) U_{AB}$.
  Then:
  \begin{align*}
    &\quad \tr \left( P_i \, (\rho_A \otimes \proj 0_B) \right)
    = \tr \left( (I_A \otimes \proj i_B) U_{AB} (\rho_A \otimes \proj 0_B) U_{AB}^\dagger \right) \\
    &= \tr \left( \braket{i_B | U_{AB} (\rho_A \otimes \proj 0_B) U_{AB}^\dagger | i_B} \right)
    = \tr \left( \sqrt{Q_i} \rho_A \sqrt{Q_i} \right)
    = \tr(Q_i \rho_A). \qedhere
  \end{align*}
\end{sol}

\end{ex}

\bigskip

\begin{ex}[Source compression]
\label{ex:source compression}
Let $\Sigma = \{1,\dots,\lvert\Sigma\rvert\}$ be an alphabet, and $p(x)$ a probability distribution on $\Sigma$.
Let $X_1,X_2,\dots$ be i.i.d.\ random variables with distribution $p(x)$ each.
In the lecture, typical sets were defined by
\begin{equation*}
  T_{p,n,\delta} = \{ (x_1,\dots,x_n) \in \Sigma^n : \lvert - \frac 1 n \log p^{\otimes n}(x_1, \dots, x_n) - H(p) \rvert \leq \delta \}.
\end{equation*}

\benum
\item Show that $\Pr[X^n \in T_{p,n,\delta}] \rightarrow 1$ as $n \rightarrow \infty$.

\emph{Hint: Use Chebyshev's inequality.}

\begin{sol}
  \begin{align*}
    \Pr[X^n \in T_{p,n,\delta}]
    = \Pr[\lvert - \frac 1 n \log p^{\otimes n}(X^n) - H(p) \rvert \leq \delta]
    = \Pr[\lvert \underbrace{- \frac 1 n \sum_{i=1}^n \log p(X_i)}_{=: Z} - H(p) \rvert \leq \delta].
  \end{align*}
  The expectation of the random variable $Z$ is equal to the entropy of the distribution $p(x)$,
  \begin{align*}
    \E[Z] = \E[-\log p(X_i)] = H(p),
  \end{align*}
  because the $X_i$ are i.i.d.\ according to $p(x)$.
  Moreove, since the $X_i$ are independent, its variance is given by
  \begin{equation*}
    \Var[Z] = \frac 1 {n^2} \Var[\sum_{i=1}^n \log p(X_i)] = \frac 1 n \Var[\log p(X_1)].
  \end{equation*}
  Using Chebyshev's inequality, we find that
  \begin{equation*}
    \Pr[T_{p,n,\delta}]
    = 1 - \Pr[\lvert Z - H(p) \rvert > \delta]
    \geq 1 - \frac {\Var[Z]} {\delta^2}
    = 1 - \frac 1 n \frac {\Var[\log p(X_1)] } {\delta^2}
    = 1 - O(1/n)
  \end{equation*}
  as $n \rightarrow \infty$ (for fixed $p$ and $\delta$).  (One can
  further show, although it is not necessary, that $\Var[\log
  p(X_1)] \leq \log^2(d)$.)
\end{sol}

\item Show that the entropy of the source is the optimal compression rate. That is, show that we cannot compress to $nR$ bits for $R < H(p)$ unless the error does not go to zero as $n \rightarrow \infty$.

\emph{Hint: Pretend first that all strings are typical.} 

\begin{sol}
  Suppose that we have a (deterministic) compression scheme that uses $n R$ bits, where $R < H(X)$. (For simplicity, we assume that $nR$ is an integer.)
  Denote by $\mathcal E_n \colon \Sigma^n \rightarrow \{1,\dots,2^{nR}\}$ the compressor, by $\mathcal D_n \colon \{1,\dots,2^{nR}\} \rightarrow \Sigma^n$ the decompressor, and by $A_n = \{ x^n : x^n = \mathcal D_n(\mathcal E_n(x^n)) \}$ the set of strings that can be compressed correctly. Note that $A_n$ has no more than $2^{nR}$ elements.
  The probability of success of the compression scheme is given by
  \begin{equation*}
    p_{\text{success}}
    = \Pr[X^n = \mathcal D_n(\mathcal E_n(X^n))]
    = \Pr[X^n \in A_n].
  \end{equation*}
   Now,
  \begin{align}
  \nonumber
    \Pr[X^n \in A_n]
    &= \Pr[X^n \in A_n \cap T_{p,n,\delta}] + \Pr[X^n \in A_n \cap T_{p,n,\delta}^c] \\
  \label{eq:two pieces}
    &\leq \Pr[X^n \in A_n \cap T_{p,n,\delta}] + \Pr[X^n \in T_{p,n,\delta}^c]
  \end{align}
  For any fixed choice of $\delta$, the right-hand side probability converges in~\eqref{eq:two pieces} to zero as $n \rightarrow \infty$ (by the previous exercise).
  On the other hand, the set $A_n \cap T_{p,n,\delta}$ has at most $2^{nR}$ elements, since this is even true for $A_n$.
  Moreover, since all its elements are typical, we have that $p^{\otimes n}(x^n) \leq 2^{n (-H(X) + \delta)}$.
  It follows that the left-hand side probability in~\eqref{eq:two pieces} can be bounded from above by
  \begin{equation*}
    \Pr[X^n \in A_n \cap T_{p,n,\delta}] \leq 2^{n (R - H(X) + \delta)}.
  \end{equation*}
  If we fix a $\delta$ such that $R < H(X) - \delta$ then this probability likewise converges to zero.
  It follows that the probability of success of the compression scheme, $p_{\text{success}}$, in fact goes to zero as $n \rightarrow \infty$.
\end{sol}

\eenum

\end{ex}

%% file: 28may_1.tex
\lecture{4}{28 May, 2013}{Fernando G.S.L. Brand\~ao}{Teleportation and entanglement transformations}
\label{lec:entanglement trafos}

\paragraph{Prologue: Post-measurement states.}
One loose thread from the previous lecture is to explain what happens to a quantum state after the measurement.
Consider a projective measurement $\{P_k\}$.  (We saw in \cref{ex:povm} in yesterday's problem session that in fact these can simulate even generalized measurements.)
Recall that outcome $k$ occurs with probability $\Pr[k] = \tr(P_k\rho)$.
Then if this measurement outcome occurs, quantum mechanics postulates that we are left with the state
\be \frac{P_k \rho P_k}{\tr(P_k\rho)}. \label{eq:post-measurement} \ee
Observe that this has the property that repeated measurements always produce the same answer (although the same is not necessarily true of generalized measurements).

For a pure state $\ket\psi$, the post-measurement state is
\be \frac{P_k \ket{\psi}}{\|P_k \ket\psi \|}. \label{eq:pure-measured} \ee
Equivalently, we can write
$P_k \ket\psi = \sqrt{p} \ket{\varphi},$
where $\ket{\varphi}$ is the unit vector~\eqref{eq:pure-measured} representing the post-measurement state, and $p$ is the probability of that outcome.

\section{Teleportation}
Suppose that Alice has a qubit $\ket\psi_{A'} = c_0 \ket 0 + c_1 \ket 1$ that she would like to transmit to Bob.  If they have access to a quantum channel, such as an optical fiber, she can of course simply give Bob the physical system $A'$ whose state is $\ket\psi_{A'}$.  This approach is referred to as {\em quantum communication.}
However, if they have access to shared entanglement, then this communication can be replaced with {\em classical communication} (while using up the entanglement).  This is called {\em teleportation.}

The procedure is as follows.  Suppose Alice and Bob share the state
$$\ket{\Phi^+}_{AB} = \frac{\ket{00} + \ket{11}}{\sqrt 2},$$
and Alice wants to transmit $\ket{\psi}_{A'}$ to Bob.   Then Alice first measures systems $AA'$ in the basis
$\{\ket{\Phi^+}, \ket{\Phi^-}, \ket{\Psi^+}, \ket{\Psi^-}\}$, defined as
\bas
\ket{\Phi^\pm} &= \frac{\ket{00} \pm \ket{11}}{\sqrt 2}, \\
\ket{\Psi^\pm} &= \frac{\ket{01} \pm \ket{10}}{\sqrt 2}.
\eas
For ease of notation, define $\{\ket{\eta_0}, \ket{\eta_1}, \ket{\eta_2}, \ket{\eta_3}\} :=
\{\ket{\Phi^+}, \ket{\Phi^-}, \ket{\Psi^+}, \ket{\Psi^-}\}$.

For example, outcome 0 corresponds to the unnormalized state
$$\L( \proj{\Phi^+}_{A'A} \ot I_B\R) (\ket{\psi}_{A'} \ot \ket{\Phi^+}_{AB} )=
\frac{1}{2} \ket{\Phi^+}_{A'A} \ot \ket{\psi}_B,$$
meaning the outcome occurs with probability $1/4$ and when it does, Bob gets $\ket\psi$ (cf.\ the discussion in the prologue).

One can show (and you will calculate in the exercises) that outcome $i$ (for $i\in\{0,1,2,3\}$) corresponds to
$$\L( \proj{\eta_i} \ot I_B\R) \ket{\psi}_{A'} \ot \ket{\Phi^+}_{AB} =
\frac{1}{2} \ket{\eta_i}_{A'A} \ot \sigma_i\ket{\psi}_B,$$
where $\{\sigma_0,\sigma_1, \sigma_2, \sigma_3\}$ denote the four Pauli matrices $\{I, \sigma_x, \sigma_y, \sigma_z\}$.  The 1/2 means that each outcome occurs with probability $1/4$.
Thus, transmitting the outcome $i$ to Bob allows him to apply the correction $\sigma_i^\dagger = \sigma_i$ and recover the state $\ket\psi$.

This protocol has achieved the following transformation of resources:
\begin{center}
  \emph{1 ``bit'' entanglement $+$ 2 bits classical communication $\geq$ 1 qubit quantum communication}
\end{center}

As a sanity check, we should verify that entanglement alone cannot be used to communicate.  To check this, the joint state after Alice's measurement is
$$\rho_{A'AB} = \frac{1}{4} \sum_{i=0}^3 \proj{\eta_i}_{A'A} \ot
\sigma_i\proj\psi \sigma_i^\dagger.$$
Bob's state specifically is
$$\rho_B = \tr_{A'A}(\rho_{A'AB}) = \frac{1}{4} \sum_{i=0}^3
\sigma_i\proj\psi \sigma_i^\dagger = \frac{I_B}{2}.$$

\paragraph{Teleporting entanglement.}  This protocol also works if applied to qubits that are entangled with other states.  For example, Alice might locally prepare an entangled state $\ket{\psi}_{RA'}$ and then teleport qubit $A'$ to Bob's system $B$.  Then the state $\ket\psi$ will be shared between Alice's system $R$ and Bob's system $B$. Thus, teleportation can be used to create shared entanglement.  Of course, it consumes entanglement at the same rate, so we are not getting anything for free here.

\section{LOCC entanglement manipulation}
Suppose that Alice and Bob can freely communicate classically and can manipulate quantum systems under their control, but are limited in their ability to communicate quantumly.   This class of operations is called \emph{LOCC}, meaning ``local operations and classical communication''.  It often makes sense to study entanglement in this setting, since LOCC can modify entanglement from one type to another, but cannot create it where it didn't exist before.   What types of entanglement manipulations are possible with LOCC?

One example is to map a pure state $\ket\psi_{AB}$ to $(U_A \ot V_B)\ket\psi_{AB}$, for some choice of unitaries $U_A, V_B$.

A more complicated example is that Alice might measure her state with a projective measurement $\{P_k\}$ and transmit the outcome $k$ to Bob, who then performs a unitary $U_k$ depending on the outcome.  This is essentially the structure of teleportation.  The resulting map is
$$\rho_{AB} \mapsto \sum_k (P_k \ot U_k)\rho(P_k \ot U_k^\dag).$$

One task for which we might like to use LOCC is to extract pure entangled states from a noisy state.  For example, we might want to map a given state $\rho_{AB}$ to the maximally entangled state $\proj{\Phi^+}$.  This problem is in general called {\em entanglement distillation}, since we are distilling pure entanglement out of noisy entanglement.  However, we typically consider it with a few variations.  First, as with many information-theoretic problems, we will consider asymptotic transformations in which we map $\rho_{AB}^{\ot n}$ to $\proj{\Phi_+}^{\ot m}$, and seek to maximize the ratio $m/n$ as $n\ra \infty$.  Additionally, we will allow a small error that goes to zero as $n\ra \infty$.  Semi-formally, the {\em distillable entanglement} of $\rho$ is thus defined as
$$E_D(\rho_{AB}) = \lim_{n\ra \infty} \max \L\{\frac{m}{n} :
\rho^{\ot n} \xrightarrow{\text{LOCC}} \sigma_m \approx \proj{\Phi_+}^{\ot m}\R\}.$$
In order to make this definition precise, we need to formalize the notion of closeness (``$\approx$'').

\section{Distinguishing quantum states}

One operationally meaningful way to define a distance between two quantum states $\rho$, $\sigma$ is in terms of the maximum distinguishing bias that any POVM measurement can achieve,
$$D(\rho,\sigma) = \max_{0 \leq M \leq I} |\tr (M(\rho-\sigma))|.$$
It turns out that $$D(\rho,\sigma) = \frac 1 2 \|\rho-\sigma\|_1,$$ where $\|X\|_1$ is the {\em trace norm}, defined as $\|X\|_1 = \tr(\sqrt{X^\dag X})$.  For this reason, the distance $D(\rho,\sigma)$ is also called the {\em trace distance}.

Using this language, we can define the distillable entanglement $E_D$ properly as
$$E_D(\rho_{AB}) = \lim_{\eps\ra 0} \lim_{n\ra \infty} \max \L\{\frac{m}{n} :
\rho^{\ot n} \xrightarrow{\text{LOCC}} \sigma_m, \ \|\sigma_m - \proj{\Phi_+}^{\ot m}\|_1
\leq \eps\R\}.$$

\section{Entanglement dilution}
Suppose now that we wish to create a general entangled state $\rho_{AB}$ out of pure EPR pairs.  As with distillation, we will aim to maximize the asymptotic ratio achievable while the error goes to zero.  Define the {\em entanglement cost}
$$E_c(\rho_{AB}) = \lim_{\eps\ra 0} \lim_{n\ra \infty}
\min \L\{\frac{m}{n} :
\proj{\Phi_+}^{\ot m} \xrightarrow{\text{LOCC}} \sigma_n, \ \|\sigma_n - \rho_{AB}^{\ot n}\|_1 \leq \eps \R\}.$$
In general, $E_c$ and $E_D$ are both hard to compute.
However, if $\rho_{AB}$ is pure then there is a simple beautiful formula, which you will discuss in \cref{ex:E_c E_D}.

\begin{thm}
For any pure state $\ket\psi_{AB}$,
$$E_c(\proj\psi_{AB}) = E_D(\proj\psi_{AB}) = S(\rho_A) = S(\rho_B).$$
\end{thm}

%% file: 28may_2.tex
\lecture{5}{28 May, 2013}{Matthias Christandl}{Introduction to the quantum marginal problem}

\begin{figure}[h]
\centerline{\includegraphics[width=0.3\textwidth]{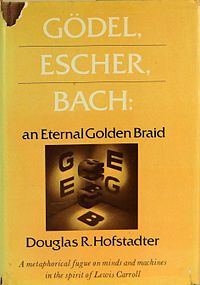}}
\caption{Cover of the book \emph{G\"odel, Escher, Bach} by Douglas Hofstadter taken from the Wikipedia page.\label{fig:geb}}
\end{figure}

In \cref{fig:geb} is the cover of the book \emph{G\"odel, Escher and Bach}. You see that the projection of the wooden object is either B, G or E---depending on the direction of the light shining through. Is it possible to project any triple of letters in this way? It turns out that the answer is no. For example, by geometric considerations there is no way of projecting ``A'' everywhere.\footnote{I am grateful to Graeme Mitchison who introduced me to the idea of illustrating the classical marginal problem in this way.}

The goal of this lecture is to introduce a quantum version of this problem!

\section{The quantum marginal problem or quantum representability problem}

\begin{figure}
\centerline{\includegraphics[width=0.6\textwidth]{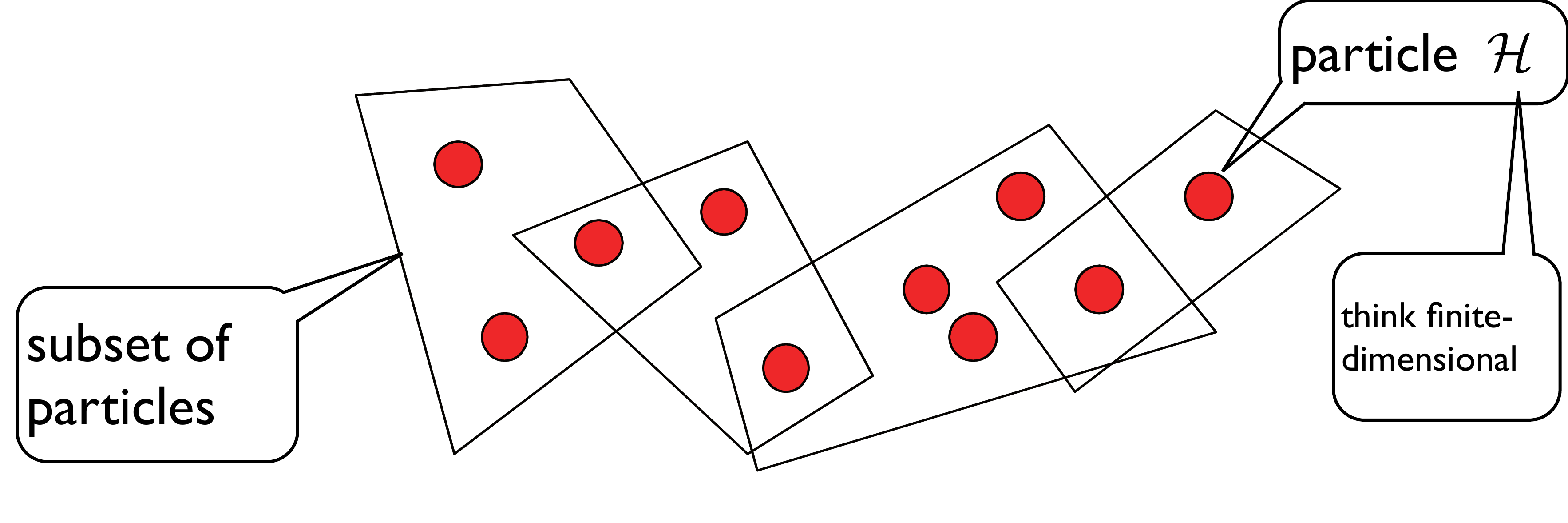}}
\caption{General quantum marginal problem.\label{fig:qmp}}
\end{figure}

Consider a set of $n$ particles with $d$ dimensions each. The state lives in $(\mathbb{C}^{d})^{\otimes n}$. We consider different subsets of the particles $S_i \subseteq \{ 1, \dots, N \}$ and suppose we are given quantum states $\rho_{S_i}$ for each of this sets. The question we want to address is whether these ``marginals'' are compatible, i.e.\ does there exist a quantum state $\rho_{\{1, \dots, n \}}$ that has the $\rho_{S_i}$ as its reduced density matrices, i.e.,
\begin{equation*}
\tr_{S_{i}^c} \left( \rho   \right) = \rho_{S_i},
\end{equation*}
with $S_{i}^c$ the complement of $S_i$ in $\{1, \dots, n \}$.
This is called the \emph{quantum marginal problem}, or \emph{quantum representability problem}.

\subsection{Physical motivation}

\begin{figure}
\centerline{\includegraphics[width=0.6\textwidth]{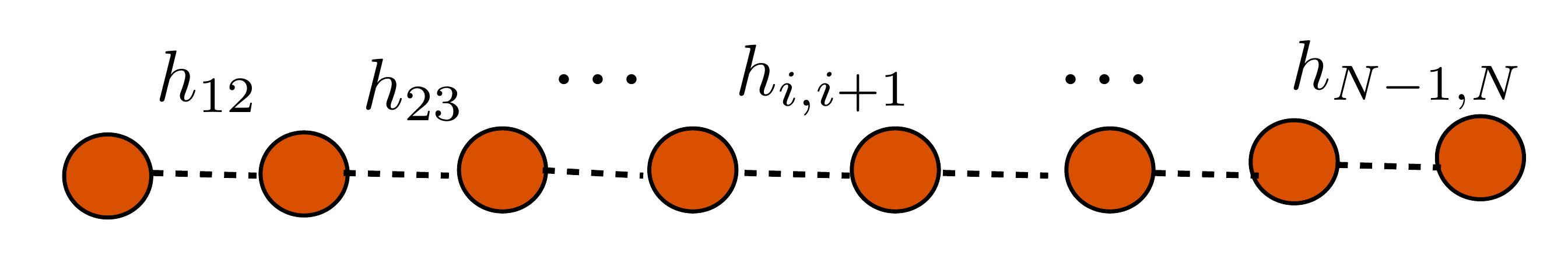}}
\centerline{\includegraphics[width=0.6\textwidth]{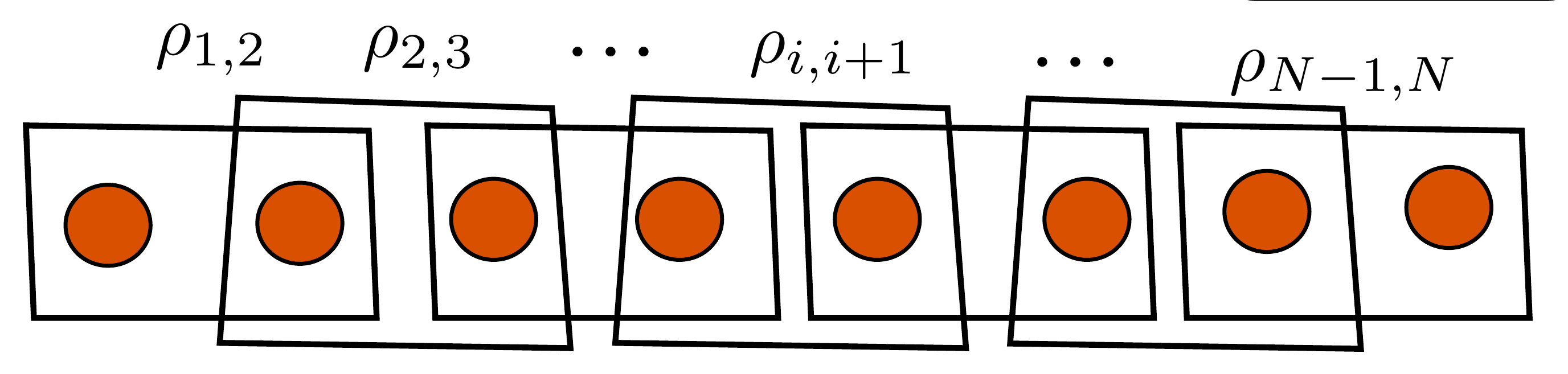}}
\caption{Nearest-neighbor Hamiltonian and the corresponding quantum marginal problem.\label{fig:line}}
\end{figure}

This is a interesting problem from a mathematical point of view, but it is also a prominent problem in the context of condensed matter physics and quantum chemistry. Consider a one-dimensional system with nearest-neighbour Hamiltonian
$$H = \sum_{i} H_{i, i+1},$$
where $H_{i, i+1} := h_{i, i+1} \otimes I_{\{1, \dots, n\} \setminus {i, i+1}}$ only acts on qubits $i$ and $i+1$.
A quantity of interest is the \emph{ground state energy} of the model, given by the minimum eigenvalue of $H$. We can write it variationally as
\begin{equation*}
E_{g} = \min_{\ket{\psi}} \bra{\psi} H \ket{\psi} = \min_{\rho_{1, \dots, n}} \tr(\rho_{1, \dots, n} H)
\end{equation*}
since the set of quantum states is convex and its extremal points are the pure states (e.g., the Bloch sphere). Considering the specific form of the Hamiltonian, we find
\begin{align*}
  E_g = \min_{\rho_{1, \dots, n}} \tr(\rho_{1, \dots, n} H)
  = \min_{\rho_{1, \dots, n}} \sum_i \tr(\rho_{1, \dots, n} H_{i,i+1})
  = \min_{\rho_{1, \dots, n}} \sum_i \tr(\rho_{i,i+1} h_{i,i+1})
\end{align*}
and therefore
\begin{equation}
  \label{compat}
  E_g = \min_{\{\rho_{i,i+1}\} \text{ compatible}} \sum_i \tr(\rho_{i,i+1} h_{i,i+1}),
\end{equation}
where the minimization is over sets of two-body density matrices $\{\rho_{i,i+1}\}$ which are compatible with the existence of a global state $\rho_{1,\dots,n}$ (\cref{fig:line}).

Observe that the initial maximization is over $\ket{\psi} \in \left(\mathbb{C}^d \right)^{\otimes n}$, i.e.\ over a $d^n$-dimensional space.
In contrast, the minimization in \cref{compat} is over $O(n d^2)$ variables.
Therefore, if we could solve the compatibility problem, then we could solve the problem of computing the ground state energy in a much more efficient way.
Unfortunately this is not a good strategy and in fact one can show that the compatibility problem is computationally hard ($\NP$-hard and even $\QMA$-hard).

\smallskip
There is an interesting connection between the representability problem and quantum entropies.
For example, an important relation satisfied by the von Neumann entropy of quantum states of tripartite systems ABC is its \emph{strong subadditivity},
\begin{equation*}
S(AB) + S(BC) \geq S(B) + S(ABC).
\end{equation*}
Clearly this inequality puts restrictions on compatible states.
More interestingly, one can also use results from the quantum marginal problem to give a proof of this inequality.

\section{Pure-state quantum marginal problem}
\label{sec:pure state qmp}

A particular case of the quantum marginal problem is the following: given three quantum states $\rho_A$, $\rho_B$ and $\rho_C$, are they compatible? In this case it is that the answer is yes, just consider $\rho_{ABC} = \rho_A \otimes \rho_B \otimes \rho_C$.
But what if we require that the global state $\rho_{ABC}$ is pure? That is, we would like to find a \emph{pure} state $\ket{\psi}_{ABC}$ such that
\begin{equation*}
  \tr_{AB} \left( \proj\psi_{ABC} \right) = \rho_C, \quad
  \tr_{AC} \left( \proj\psi_{ABC} \right) = \rho_B, \quad
  \tr_{BC} \left( \proj\psi_{ABC} \right) = \rho_A.
\end{equation*}
Taking the tensor product of the reduced states is then not an option any more, since it will in general lead to a mixed state.

\begin{exl*}
$\rho_A = \rho_B = \rho_C = I/2$ are compatible with the \emph{GHZ state}
$$\ket{\mathrm{GHZ}}_{ABC} = \frac 1 {\sqrt 2} \left( \ket{000} + \ket{111}) \right).$$
\end{exl*}

\begin{exl*}
Suppose $\rho_A$, $\rho_B$ and $\rho_C$ are compatible. Are $\rho'_A = U_A \rho_A U_{A}^{\dagger}$, $\rho'_B = U_B \rho_B U_{B}^{\dagger}$, and $\rho'_C = U_C \rho_C U_{C}^{\dagger}$ compatible too? The answer is yes. Indeed if $\ket{\psi}_{ABC}$ was an extension of $\rho_A$, $\rho_B$ and $\rho_C$, then $(U_{A} \otimes U_{B} \otimes U_C) \ket{\psi}_{ABC}$ is an extension of $\rho'_A$, $\rho'_B$ and $\rho'_C$.
\end{exl*}

We conclude from the latter example that the property of $\rho_A, \rho_B, \rho_C$ being compatible only depends on the spectra $\lambda_A$, $\lambda_B$ and $\lambda_C$ of $\rho_A$, $\rho_B$ and $\rho_C$. Recall that the \emph{spectrum} of a matrix $\rho_A$ is its collection of eigenvalues $\lambda_A := (\lambda_{A, 1}, \dots, \lambda_{A, d})$, where by convention $\lambda_{A,1} \geq \dots \geq \lambda_{A,d}$.

\subsection{Warm-up: Two parties}

Given $\rho_A$ and $\rho_B$, are they compatible with a pure state $\ket{\psi_{AB}}$?

A useful way of writing a bipartite pure state $\ket{\psi_{AB}}$ is in terms of its \emph{Schmidt decomposition},
\begin{equation}
\label{schmidt}
\ket{\psi_{AB}} = \sum_{i} s_i \ket{e_i} \otimes \ket{f_i},
\end{equation}
for orthogonal bases $\{ \ket{e_i} \}$ and $\{ \ket{f_i} \}$ of $A$ and $B$, respectively.
The numbers $\{ s_i \}$, which can always be chosen to be real and nonnegative, are called \emph{Schmidt coefficients} of $\ket{\psi}_{AB}$.
The reduced density matrices of $\ket{\psi}_{AB}$ are
\begin{equation*}
  \rho_A = \sum_i \lvert s_i\rvert^2 \proj{e_i}
\end{equation*}
and
\begin{equation*}
  \rho_B = \sum_i \lvert s_i\rvert^2 \proj{f_i}
\end{equation*}
Therefore we conclude that the eigenvalues of $\rho_A$ and $\rho_B$ are equal (including multiplicity) and given by $\{ \lvert s_i\rvert^2 \}$ (here, we have used that the dimensions of $A$ and $B$ are equal -- otherwise, the multiplicity of the eigenvalue 0 can be different).

Conversely, given $\rho_A$ and $\rho_B$ which have the same spectrum is clear that we can always find an extension $\ket{\psi_{AB}}$ by using \cref{schmidt}.
Thus $\rho_A$ and $\rho_B$ are compatible if, and only if, they have the same spectrum.

\subsection{Outlook: Three qubits}

Consider $\rho_A$, $\rho_B$ and $\rho_C$ each acting on $\mathbb{C}^2$. Then since $\lambda_A = (\lambda_{\max}^A, 1- \lambda_{\max}^A)$, the compatible region can be considered as a subset $\{ (\lambda_{\max}^A, \lambda_{\max}^B, \lambda_{\max}^C) \} \subseteq \mathbb{R}^{3}$.
We will see in the next lecture that this set has a simple algebraic characterization.
Apart from the ``trivial'' constraints $1/2 \leq \lambda_{\max}^A, \lambda_{\max}^B, \lambda_{\max}^C \leq 1$, a triple of spectra is compatible if and only if
\begin{align*}
  \lambda_{\max}^A + \lambda_{\max}^B &\leq 1 + \lambda_{\max}^C
\end{align*}
and its permutations hold.

%% file: 28may_3.tex
\lecture{6}{28 May, 2013}{Aram Harrow}{Monogamy of entanglement}
\label{lec:monogamy}

Today, I will discuss a property of entanglement known as \emph{monogamy}.
Consider a Hamiltonian that has two-body interactions,
$$H= \sum_{\langle i,j \rangle} H_{ij},$$
where the sum is over all edges $\langle i,j \rangle$ of the interaction graph.
We will consider the rather crude approximation that every particle interacts with any other particle in the same way.
This approximation is known as the \emph{mean field approximation},
$$H \approx \frac{1}{n} \sum_{1 \leq i<j \leq n} H_{ij}.$$
It is then folklore that the ground state has the form $\approx \rho^{\otimes n}$.

\begin{exl*}
Suppose that all $H_{ij}=F_{ij}$, where $F$ is the \emph{swap operator} defined by
$$F \ket{\alpha}\ket{\beta}=\ket{\beta}\ket{\alpha}.$$
The $(+1)$-eigenspace of $F$ is spanned by the \emph{triplet} basis
$\ket{\uparrow\uparrow}$, $\ket{\downarrow\downarrow}$ and $(\ket{\uparrow\downarrow}+\ket{\downarrow\uparrow})/\sqrt 2$.
The $(-1)$-eigenspace is one-dimensional and spanned by the \emph{singlet} $(\ket{\uparrow\downarrow}-\ket{\downarrow\uparrow})/\sqrt 2$.

Thus to find the ground state energy of the Hamiltonian, every two-particle reduced density matrix should be in the singlet state.
However, if a global state $\ket{\psi_{ABC}}$ has the singlet as its reduced density matrix $\rho_{AB}$ then it is necessarily of the form
$$\ket\psi_{ABC} = \frac 1 {\sqrt 2} \left( \ket{\uparrow\downarrow}_{AB} - \ket{\downarrow\uparrow}_{AB} \right) \ot \ket{\phi}_C.$$
Thus we immediately see that the other pairs of particles cannot be entangled!
(Note that the same conclusion is true if $\rho_{AB}$ is an arbitrary pure state.)
\end{exl*}

This turns out to be a general feature of such systems.

\begin{thm}[Quantum de Finetti]\label{thm:deF}
	Let $\ket{\psi}$ be a permutation-symmetric state on $(\mathbb C^D)^{\otimes k+n}$ (i.e., $\ket\psi$ is left unchanged by the permutation action defined in \cref{eq:permutation action} below).
	Then,
$$\tr_n \proj{\psi} \approx \int d\mu(\sigma) \sigma^{\otimes k},$$
where $\mu$ is a probability distribution over density matrices on $\mathbb C^D$.
\end{thm}

This is a quantum version of de Finetti's theorem from statistics. The important consequence of this theorem is that the remaining $k$ particles are not entangled. Since the ground states of mean field systems are permutation invariant this means that these ground states are not entangled, and hence in some sense classical.

We will now introduce some mathematical tools needed to prove this theorem, the first of which is the symmetric subspace.

We remark that the quantum de Finetti theorem can be extended to
\emph{permutation-invariant} mixed states, i.e., density matrices $\rho$ that merely commute with the permutation action.
We will discuss how this can be done in \cref{subsec:perm inv}.

\section{Symmetric subspace}

Let $S_n$ be the group of permutations of $n$ objects. Note that it contains $n!$ elements. Now fix $D$ and a permutation $\pi \in S_n$. Let us define an action of the permutation $\pi$ on $(\mathbb C^D)^{\otimes n}$ by
\begin{equation}
\label{eq:permutation action}
	P_\pi \, \ket{i_1}\otimes \cdots \otimes \ket{i_n}=\ket{i_{\pi^{-1}(1)}}\otimes \cdots \otimes \ket{i_{\pi^{-1}(n)}}.
\end{equation}
The \emph{symmetric subspace} is defined as the set of vectors that are invariant under the action of the symmetric group,
$$\Sym^n(\mathbb C^D)=\{\ket{\Psi} \in (\mathbb C^D)^{\otimes n}: P_\pi\ket{\Psi}= \ket{\Psi} \quad \forall \pi \in S_n \}.$$

\medskip

\begin{exl*}[$D=2, n=2$]
$$\Sym^2(\mathbb C^2)= \Span\ \{ \ket{00},  \ket{11},  \ket{01} + \ket{10}\} $$
\end{exl*}

\begin{exl*}[$D=2, n=3$]
$$\Sym^3(\mathbb C^2)= \Span\ \{ \ket{000},  \ket{111},  \ket{001} + \ket{010} + \ket{100}, \ket{101} + \ket{011} + \ket{110} \} $$
\end{exl*}

The general construction is as follows.
Define the \emph{type} of a string $x^n=(x_1, \dots, x_n) \in \{1,\dots,D\}^n$ as
$$\mathrm{type}(x^n) = \sum_i e_{x_i},$$
where $e_j$ is the basis vector with a one in the $j$'th position.
Note that $t=(t_1, \dots, t_D)$ is a type if and only if $t_1 + t_2 + \dots + t_D = n$ and the $t_i$ are non-negative integers.
For every type $t$, the unit vector
$$\ket{\gamma_t}={\binom n t}^{-1/2} \sum_{\mathrm{type}(x^n) = t} \ket{x^n}$$
is permutation-symmetric, and $\Sym^n(\mathbb C^D)= \Span \{\ket{\gamma_t}\}$.

We can now compute the dimension of the symmetric subspace.
Note that we can interpret this number as the number of ways in which you can arrange $n$ balls into $D$ buckets. There are ${\binom {n+D-1} n}$ ways of doing this, which is therefore the dimension.

A useful way for calculations involving the symmetric subspace are the following two characterisations of the projector onto the symmetric subspace:

\benum
\item $\Psym^{D,n} = \frac{1}{n!}\sum_{\pi \in S_n} P_\pi$.
\item $\frac{\Psym^{D,n}}{\tr \Psym^{D,n}} = \int d\phi \, \proj{\phi}^{\otimes n}$,
where we integrate over the unit vectors in $C^D$ with respect to the uniform probability measure $d\phi$.
Note that $\tr \Psym^{D,n} =\dim \Sym^n(C^D) = {\binom {n+D-1} n}$
\eenum

\begin{exl*}[$n=1$]
$$\Psym^{D,1} = I = D \int d\phi \, \proj{\phi}$$
\end{exl*}

\begin{exl*}[$n=2$]
$$\Psym^{D,2} = \frac {I + F} 2 = \frac 2 {D (D+1)} \int d\phi \, \proj{\phi}^{\otimes 2}$$
\end{exl*}

We can verify 1.\ directly by checking that the following three conditions are satisfied:
\benum
\item[a)] $\Psym^{D,n} \ket\psi \in \Sym^n(\mathbb C^D)$ for all $\ket\psi \in (\mathbb C^D)^{\otimes n}$.
\item[b)] $\Psym^{D,n} \ket\psi = \ket\psi$ for all $\ket\psi \in \Sym^n(\mathbb C^D)$.
\item[c)] $\Psym^{D,n} = (\Psym^{D,n})^\dagger$.
\eenum
To prove 2., either use representation theory (using Schur's lemma) or rewrite the integral over unit vectors as an integral over Gaussian vectors and then using Wick's theorem to solve the integral.

\section{Application to estimation}
Given $n$ copies of a pure state, $\ket{\psi}^{\otimes n}$, we want to output a (possibly random) estimate $\ket{\hat \psi}$ that approximates $\ket{\psi}$. We could now use different notions of approximation. Here, we want to maximise the average overlap
$\E[\lvert \langle \hat\psi | \psi\rangle \rvert^{2k}]$
for some fixed $k$.

In order to do this, we will use the continuous POVM $\{Q_{\hat{\psi}}\}$, where
$$Q_{\hat \psi} = {\binom {n+D-1} n} \, \proj{\hat{\psi}}^{\otimes n}.$$
Note that $\int d\hat{\psi} \, Q_{\hat{\psi}} = \Psym^{D,n}$.
The average overlap of this estimation scheme is given by
$$\E[|\langle \hat\psi | \psi\rangle|^{2k}] = \int d\hat\psi \, p(\hat{\psi} | \psi) \, |\langle \hat\psi | \psi\rangle|^{2k}$$
where $p(\hat{\psi} | \psi) = \tr (Q_{\hat\psi} \, \proj{\psi}^{\otimes n})$ is the probability density of the estimate $\ket{\hat\psi}$ given state $\ket\psi^{\otimes n}$.
This in turn equals
\begin{align*}
&\quad {\binom {n+D-1} n} \int d\hat\psi \, |\langle \hat\psi | \psi\rangle|^{2(k+n)} \\
&= {\binom {n+D-1} n} \int d\hat\psi \, \tr \left( \proj{\hat\psi}^{\otimes k+n} \proj{\psi}^{\otimes k+n} \right) \\
&= {\binom {n+D-1} n} \tr \left( \proj{\psi}^{\otimes k+n} \int d\hat\psi \, \proj{\hat\psi}^{\otimes k+n} \right) \\
&= {\binom {n+D-1} n} \tr \left( \proj{\psi}^{\otimes k+n} \frac {\Psym^{D,k+n}} {\binom {n+k+D-1} {n+k}} \right) \\
&= \frac {\binom {n+D-1} n} {\binom {n+k+D-1} {n+k}}
=  \frac {(n+D-1) \cdots (n+1)} {(n+k+D-1) \cdots (n+k+1)} \\
&\geq \left( \frac {n+1} {n+k+1} \right)^{D-1}
=  \left( 1 - \frac {k} {n+k+1} \right)^{D-1} \\
&\geq 1-Dk/n.
\end{align*}

%% file: 28may_ex.tex
\problemsession{2}{28 May, 2013}{Michael Walter}

\begin{ex}[Typical subspaces]
\label{ex:typical subspace}

Let $\rho = \sum_x \lambda_x \proj x$ be a density operator. Define projectors
\begin{equation*}
  P_{\rho,n,\delta}
  = \sum_{(x_1,\dots,x_n) \in T_{\lambda,n,\delta}} \proj {x_1} \otimes \dots \otimes \proj {x_n}
  = \sum_{x^n \in T_{\lambda,n,\delta}} \proj {x^n}.
\end{equation*}
The range of $P_{\rho,n,\delta}$ is called a \emph{typical subspace}.

\benum
\item Show that the rank of $P_{\rho,n,\delta}$ (i.e., the dimension of a typical subspace) is at most $2^{n (S(\rho) + \delta)}$.

\begin{sol}
  The size of the typical set $T_{\lambda,n,\delta}$ is at most $2^{n (H(\lambda) + \delta)}$, and $H(\lambda) = S(\rho)$.
\end{sol}

\item Show that $\tr \rho^{\otimes n} P_{\rho,n,\delta} \rightarrow 1$ as $n \rightarrow \infty$.

\begin{sol}
  \begin{align*}
    \tr \rho^{\otimes n} P_{\rho,n,\delta}
    = \sum_{(x_1,\dots,x_n) \in T_{\lambda,n,\delta}} \lambda_{x_1} \dots \lambda_{x_n}
    = \sum_{(x_1,\dots,x_n) \in T_{\lambda,n,\delta}} \lambda^{\otimes n}(x^n)
    = \Pr[X^n \in T_{\lambda,n,\delta}],
  \end{align*}
  where $X_1, \dots, X_n$ are i.i.d. random variables each distributed according to $\lambda$.
  This probability converges to one as $n \rightarrow \infty$, as we saw in \cref{ex:source compression}.
\end{sol}

\eenum

\end{ex}

\bigskip

\begin{ex}[Entanglement cost and distillable entanglement]
\label{ex:E_c E_D}

In this exercise, we will show that for a bipartite pure state $\ket{\psi_{AB}}$, both the entanglement cost $E_c$ and the distillable entanglement $E_D$ are equal to the von Neumann entropy of the reduced density matrices:
$$E_c(\ket\psi_{AB}) = E_D(\ket\psi_{AB}) = S(\rho_A) = S(\rho_B).$$
We first show that $E_c(\ket\psi_{AB}) \leq S(\rho_A)$. For this, we fix $\delta > 0$ and consider the state
$\ket{\widetilde\psi_{A^nB^n}} \propto \left( P_{\rho_A,n,\delta} \otimes I_B \right) \ket{\psi_{AB}}^{\otimes n}$.

\benum
\item Show that $\lVert \ket{\widetilde\psi_{A^nB^n}} - \ket{\psi_{AB}}^{\otimes n} \rVert_1 \rightarrow 0$.

\begin{sol}
  Note that
  \begin{equation*}
    \bra{\psi_{AB}}^{\otimes n} \left( P_{\rho_A,n,\delta} \otimes I_B \right) \ket{\psi_{AB}}^{\otimes n}
    = \tr \rho_A^{\otimes n} P_{\rho_A,n,\delta} \rightarrow 1
  \end{equation*}
  by the second part of \cref{ex:typical subspace}. That this implies that the trace distance between $\ket\psi_{AB}^{\otimes n}$ and the post-measurement state $\ket{\widetilde\psi}_{A^nB^n}$ converges to 0 is a special case of the so-called \emph{gentle measurement lemma}:

  Let $\ket\psi$ be a pure state, $P$ a projection and $\bra\psi P \ket\psi \geq 1 - \eps$. The post-measurement state is $\ket{\widetilde\psi} = \tfrac {P\ket\psi} {\lVert P\ket\psi\rVert}$, and the overlap (\emph{fidelity}) between it and the original state can be lower-bounded by
  \begin{equation*}
    \lvert \langle \psi | \widetilde\psi \rangle \rvert^2 =
    \frac {\lvert \langle \psi | P | \psi \rangle \rvert^2} {\lVert P \ket\psi \rVert^2} =
    \lvert \langle \psi | P | \psi \rangle \rvert \geq 1 - \eps.
  \end{equation*}
  Now use that
  \begin{equation*}
    \frac 1 4 \lVert \proj\psi - \proj{\widetilde\psi} \rVert_1^2 = 1 - \lvert\langle\psi | \widetilde\psi\rangle\rvert,
  \end{equation*}
  which we leave as an exercise (but see \cref{eq:trace distance vs fidelity pure states} in \cref{lec:de finetti proof}).
\end{sol}

\item Show that the rank of $\widetilde\rho_{A^n}$ is at most $2^{n (S(\rho_A) + \delta)}$.

\begin{sol}
  Since $\widetilde\rho_{A^n} \propto P_{\rho_A,n,\delta} \rho_A^{\otimes n} P_{\rho_A,n,\delta}$,
  this follows directly from the first part of \cref{ex:typical subspace}.
\end{sol}

\item Show that $\ket{\widetilde\psi_{AB}}$ can be produced by LOCC from $n (S(\rho_A) + \delta)$ EPR pairs.
Conclude that $E_c(\ket\psi_{AB}) \leq S(\rho_A) + \delta$.

\emph{Hint: Use quantum teleportation.}

\begin{sol}
  Consider the following protocol: Alice first prepares the bipartite state $\ket{\widetilde\psi_{A^nB^n}}$ on her side, and
  then teleports the $B$-part to Bob. To do so, she needs approx.\ $\log_2 \rank \rho_B = \log_2 \rank \rho_A = \log_2 2^{n (S(\rho_A) + \delta)} = n (S(\rho_A) + \delta)$ EPR pairs.
\end{sol}

\eenum

We now show that $E_D(\ket{\psi_{AB}}) \geq S(\rho_A)$.
For this, consider the spectral decomposition $\rho_A = \sum_k \lambda_k \proj k$. For each type $t = (t_1,\dots,t_d)$, define the ``type projector''
\begin{equation*}
  P_{n,t} = \sum_{(k_1,\dots,k_n) \text{ of type $t$}} \proj {k_1} \otimes \dots \otimes \proj {k_n}.
\end{equation*}
Note that $\sum_t P_{n,t} = I_{A^n}$, so that the $(P_{n,t})$ constitute a projective measurement.

\benum
\item[4.] Suppose that Alice measures $(P_{n,t})$ and receives the output $t$.
Show that all non-zero eigenvalues of her post-measurement state $P_{n,t} \rho_A P_{n,t}$ are equal.
How many EPR pairs can Alice and Bob produce from the global post-measurement state?

\begin{sol}
  The vectors $\ket{x^n}$ are the eigenvectors of $\rho^{\otimes n}$. Note that the corresponding eigenvalue, $\lambda_{x_1} \dots \lambda_{x_n}$, only depends on the type of the string $x^n$. Thus the non-zero eigenvalues of the post-measurement state $\widetilde\rho_{A^n}$ on Alice's side are all equal, and the rank of $\widetilde\rho_{A^n}$ is equal to the number of strings with type $t$ (and hence given by a binomial coefficient, see Aram's lecture). In view of the Schmidt decomposition, the global post-measurement state is equivalent to approx.\ $\log_2 \rank \widetilde\rho_{A^n}$ EPR pairs.
\end{sol}

\item[5.] For any fixed $\delta > 0$, conclude that this scheme allows Alice and Bob to produce at least $n (S(\rho_A) - \delta)$ EPR pairs with probability going to one as $n \rightarrow \infty$. Conclude that $E_D(\ket{\psi_{AB}}) \geq S(\rho_A) - \delta$.

\begin{sol}
  With high probability, the measured type $t$ is typical.
\end{sol}

\eenum

\end{ex}

\bigskip

\begin{ex}[Pauli principle]

Consider a system of $N$ fermions with single-particle Hilbert space $\CC^d$ $(d \geq N)$.
The quantum state of such a system is described by a density matrix $\rho$ on the antisymmetric subspace $\bigwedge^N \CC^d = \{ \ket\psi \in (\CC^d)^{\otimes N} : P_\pi \ket \psi = \det P_{\pi} \, \ket \psi \}$.

\benum
\item Since $\bigwedge^N \CC^d \subseteq (\CC^d)^{\otimes N}$, we know how to compute the reduced state of any of the fermions.
Show that all single-particle reduced density matrices $\rho_1, \dots, \rho_N$ are equal.

\begin{sol}
  Since $\rho$ is supported on the anti-symmetric subspace, we have
  \begin{equation*}
    P_\pi \rho P_\pi^\dagger = (\det P_\pi) \rho (\det P_\pi)^* = \rho.
  \end{equation*}
  By choosing $\pi = (k \; l)$, the permutation that exchanges $k$ and $l$, it follows that
  \begin{align*}
    \tr \rho_k A
    &= \tr \rho (I^{\otimes k-1} \otimes A \otimes I^{\otimes N-k})
    = \tr P_\pi \rho P_\pi^\dagger (I^{\otimes k-1} \otimes A \otimes I^{\otimes N-k}) \\
    &= \tr \rho P_\pi^\dagger (I^{\otimes k-1} \otimes A \otimes I^{\otimes N-k}) P_\pi
    = \tr \rho (I^{\otimes l-1} \otimes A \otimes I^{\otimes N-l})
    = \tr \rho_l A. \qedhere
  \end{align*}
\end{sol}

\item The original \emph{Pauli principle} asserts that occuption numbers of fermionic quantum states are no larger than one, i.e.
\begin{equation*}
  \tr a_i^\dagger a_i \rho \leq 1.
\end{equation*}
Show that this is equivalent to a constraint on the single-particle reduced density matrices of $\rho$.

\begin{sol}
  The matrix elements of the single-particle reduced density matrix of a fermionic state are given by
  \begin{equation}
  \label{second quantization}
    \braket{i | \rho_1 | j} =
    \frac 1 N \tr a_j^\dagger a_i
  \end{equation}
  You check this e.g.\ by considering the occupation number basis of the antisymmetric subspace (this basis is also useful for proving the Pauli principle itself; note that $a_i^\dagger a_i = n_i$ is a number operator). By using \cref{second quantization}, the Pauli principle can be restated as the following constraint on the diagonal elements of $\rho_1$ with respect to an arbitrary basis $\ket i$:
  \begin{equation*}
    \braket{i | \rho_1 | i} \leq \frac 1 N
  \end{equation*}
  Since this holds for an arbitrary basis, this is in turn equivalent to demanding that the largest eigenvalue of $\rho_1$ be no larger than $1/N$.
\end{sol}

\eenum

\end{ex}

%% file: 29may_1.tex
\lecture{7}{29 May, 2013}{Fernando G.S.L. Brand\~ao}{Separable states, PPT and Bell inequalities}

Recall from yesterday the following theorem.
\begin{thm}\label{thm:BBPS}
For any pure state $\ket\psi_{AB}$,
$$E_c(\proj\psi_{AB}) = E_D(\proj\psi_{AB}) = S(\rho_A) = S(\rho_B),$$
where $S(\rho) = -\tr\rho\log\rho$.
\end{thm}
As a result, many copies of a pure entangled state can be (approximately) reversibly transformed into EPR pairs and back again.  Up to a small approximation error and inefficiency, we have $\ket\psi_{AB}^{\ot n} \underleftrightarrow{\text{\scriptsize ~~LOCC~~}} \ket{\Phi^+}^{\ot n S(\rho_A)}$.

\section{Mixed-state entanglement}
For pure states, an entangled state is one that is not a product state.  This is easy to check, and we can even quantify the amount of entanglement (using \cref{thm:BBPS}) by looking at the entropy of one of the reduced density matrices.

But what about for mixed states?  Here the situation is more complicated.
We define the set of {\em separable states} $\Sep$ to be the set of all $\rho_{AB}$ that can be written as a convex combination
\be \sum_i p_i \, \proj{\psi_i}_A \ot \proj{\varphi_i}_B.
\label{eq:sep-form} \ee
A state is called {\em entangled} if it is not separable.

We should check that this notion of entanglement makes sense in terms of LOCC.  And indeed, separable states can be created using LOCC: Alice samples $i$ according to $p$, creates $\ket{\psi_i}$ and sends $i$ to Bob, who uses it to create $\ket{\varphi_i}$.  On the other hand, entangled states cannot be created from a separable state by using LOCC.  That is, the set Sep is closed under LOCC.

\section{The PPT test}
It is in general hard to test whether a given state $\rho_{AB}$ is separable. Naively we would have to check for all possible decompositions of the form~\eqref{eq:sep-form}.  So it is desirable to find efficient tests that work at least some of the time.

One such test is the {\em positive partial transpose test}, or PPT test.
If
$$X_{AB} = \sum_{i,j,k,l} c_{i,j,k,l} \, \ket i\!\bra j_A \ot \ket k\!\bra l_B$$
then the \emph{partial transpose} of $X_{AB}$ is
\bas X_{AB}^{T_A} &= \sum_{i,j,k,l} c_{i,j,k,l} \, \ket i\!\bra j_A^T \ot \ket k\!\bra l_B
= \sum_{i,j,k,l} c_{i,j,k,l} \, \ket j\!\bra i_A \ot \ket k\!\bra l_B
\eas
More abstractly, the partial transpose can be thought of as $(\mathcal T \ot \id)$, where $\mathcal T$ is the transpose map with respect to the computational basis.

The PPT test asks whether $\rho^{T_A}$ is positive semidefinite.  If so, we say that $\rho$ is PPT.

Observe that all separable states are PPT.
This is because if $\rho=\sum_i p_i \, \proj{\psi_i}_A \ot \proj{\varphi_i}_B$ is a separable state, then
$$\rho^{T_A} = \sum_i p_i \, \proj{\psi^*_i}_A \ot \proj{\varphi_i},$$
where the $\ket{\psi^*_i}$ are pure states whose coefficients in the computational basis are the complex conjugates of those of $\ket{\psi_i}$.
This is still a valid density matrix and in particular is positive semidefinite (indeed, it is also in $\Sep$).

Thus, $\rho\in \Sep$ implies $\rho\in\PPT$.   The contrapositive is that $\rho\not\in\PPT$ implies $\rho\not\in\Sep$.  This gives us an efficient test that will detect entanglement in some cases.

Are there in fact any states that are {\em not} in PPT?  Otherwise this would not be a very interesting test.

\begin{exl*}
$\ket{\Phi^+}_{AB} = \frac{\ket{00} + \ket{11}}{\sqrt 2}$.  Then
\begin{align*}
  \proj{\Phi^+}_{AB}
  &= \frac 1 2 \left( \proj{00} + \ket{00}\!\bra{11} + \ket{11}\!\bra{00} + \proj{11} \right)
  = \frac 1 2 \, {\bpm 1 & 0 & 0 & 1 \\ 0 & 0 & 0 & 0 \\ 0 & 0 & 0 & 0 \\ 1 & 0 & 0 & 1 \epm}, \\
  \proj{\Phi^+}_{AB}^{T_A}
  &= \frac 1 2 \left( \proj{00} + \ket{10}\!\bra{01} + \ket{01}\!\bra{10} + \proj{11} \right)
  = \frac 1 2 \, {\bpm 1 & 0 & 0 & 0 \\ 0 & 0 & 1 & 0 \\ 0 & 1 & 0 & 0 \\ 0 & 0 & 0 & 1 \epm}
  = \frac F 2,
\end{align*}
where $F$ is the swap operator.
Thus the partial transpose has eigenvalues $(1/2, 1/2, 1/2, -1/2)$, meaning that $\proj{\Phi^+}_{AB}\not\in \PPT$.
Of course, we already knew that $\ket{\Phi^+}$ was entangled.
\end{exl*}

\begin{exl*}
Let's try an example where we do not already know the answer, e.g.\ a noisy version of $\ket{\Phi^+}$.
Let
$$\rho = p \proj{\Phi^+} + (1-p) \frac{I}{4}.$$
Then one can calculate $\lambda_{\min}(\rho^{T_A}) = -\frac p 2 + \frac{1-p}{4}$ which is $<0$ if and only if $p>1/3$.
\end{exl*}

Maybe $\PPT = \Sep$?  Unfortunately not.  For $\CC^2 \ot \CC^3$, all PPT states are separable. But for larger systems, e.g.\ $\CC^3 \ot \CC^3$ or $\CC^2 \ot \CC^4$, there exist PPT states that are not separable.

\subsection{Bound entanglement}

While there are PPT states that are entangled, no EPR pairs can be distilled from such states by using LOCC:

\begin{thm}\label{thm:bound}
If $\rho_{AB} \in \PPT$ then $E_D(\rho)= 0$.
\end{thm}

To prove this we will establish two properties of the set PPT:
\benum
\item {\em PPT is closed under LOCC:}
Consider a general LOCC protocol.  This can be thought of as Alice and Bob alternating general measurements and sending each other the outcomes.  When Alice makes a measurement, this transformation is
$$\rho_{AB} \mapsto
\frac{(M_A \ot I_B) \rho_{AB} (M_A^\dag \ot I_B)}{\tr ((M_A^\dag M_A \ot I_B)\rho_{AB})}.$$
After Bob makes a measurement as well, depending on the outcome, the state is proportional to
$$(M_A \ot N_B) \rho_{AB} (M_A^\dag \ot N_B^\dag),$$
and so on.  The class SLOCC (stochastic LOCC) consists of outcomes that can be obtained with some positive probability, and we will see later that this can be characterized in terms of $(M_A \ot N_B) \rho_{AB} (M_A^\dag \ot N_B^\dag)$.

We claim that if $\rho_{AB}\in \PPT$ then $(M_A \ot N_B) \rho_{AB} (M_A^\dag \ot N_B^\dag)\in\PPT$.
Indeed
\[
  \bigl( (M_A \ot N_B) \rho_{AB} (M_A^\dag \ot N_B^\dag) \bigr)^{T_A}
  = (M_A^* \ot N_B) \rho_{AB}^{T_A} (M_A^* \ot N_B)^\dag
  \geq 0,
\]
since $\rho_{AB}^{T_A} \geq 0$ and $X Y X^\dagger\geq 0$ whenever $Y\geq 0$.

\item {\em PPT is closed under tensor product:}
If $\rho_{AB}, \sigma_{A'B'}\in \PPT$, then $(\rho_{AB} \ot \sigma_{A'B'})\in \PPT$ with respect to $AA':BB'$.
Why?  Because
$$(\rho_{AB} \ot \sigma_{A'B'})^{T_{AA'}} =
\rho_{AB}^{T_A} \ot \sigma_{A'B'}^{T_{A'}} \geq 0.$$
\eenum

\begin{proof}[Proof of \cref{thm:bound}]
Assume towards a contradiction that $\rho\in\PPT$ and  $E_D(\rho)>0$.  Then for any $\eps>0$ there exists $n$ such that $\rho_{AB}^{\ot n}$ can be transformed to $\ket{\Phi^+}$ using LOCC up to error $\eps$.    Since $\rho\in\PPT$, $\rho^{\ot n}$ is also PPT and so is the output of the LOCC protocol, which we call $\sigma$.  Then $\sigma^{T_A} \geq 0$ and $\|\sigma - \proj{\Phi^+}\|_1 \leq \eps$.  If we had $\eps=0$, then this would be a contradiction, because $\sigma$ is in PPT  and $\proj{\Phi^+}$ is not.  We can use an argument based on continuity (of the partial transpose and the lowest eigenvalue) to show that a contradiction must appear even for some sufficiently small $\eps>0$.
\end{proof}

If $\rho$ is entangled but $E_D(\rho)=0$, then we say that $\rho$ has \emph{bound entanglement} meaning that it is entangled, but no pure entanglement can be extracted from it. By \cref{thm:bound}, we know that any state in PPT but not Sep must be bound entangled.

A major open question (the ``NPT bound entanglement'' question)  is whether there exist bound entangled states that have a non-positive partial transpose.

\section{Entanglement witnesses}

The set of separable states $\Sep$ is convex, meaning that if $\rho,\sigma\in\Sep$ and $0\leq p\leq 1$ then
$p \rho + (1-p)\sigma\in\Sep$.  Thus the separating hyperplane theorem implies that for any $\rho\not\in\Sep$, there exists a Hermitian matrix $W$ such that
\benum
\item $\tr(W\sigma) \geq 0$ for all $\sigma \in \Sep$.
\item $\tr (W\rho) < 0$.
\eenum

\begin{exl*}
  Consider the state $\rho = \proj{\Phi^+}$.  Let $W = I - 2 \proj{\Phi_+}$.  As an exercise, show that $\tr (W\sigma) \geq 0$ for all $\sigma\in\Sep$.  We can also check that $\tr(W\rho) = -1$.
\end{exl*}

Observe that an entanglement witness $W$ needs to be chosen with a specific $\rho$ in mind.  As an exercise, show that no $W$ can be a witness for {\em all} entangled states of a particular dimension.

\section{CHSH game}
\label{sec:chsh}

One very famous type of entanglement witness is called a {\em Bell inequality}.  In fact, these bounds rule out not only separable states but even classically correlated distributions over states that could be from a theory more general than quantum mechanics.  Historically, Bell inequalities have been important in showing that entanglement is an inescapable, and experimentally testable, part of quantum mechanics.

The game is played by two players, Alice and Bob, together with a Referee.  The Referee choose bits $r,s$ at random and sends $r$ to Alice and $s$ to Bob.  Alice then sends a bit $a$ back to the Referee and Bob sends the bit $b$ to the Referee.

\begin{center}
\begin{tikzpicture}
\node (a) [circle, draw] at (-2,0) {A};
\node (b) [circle, draw] at (2,0) {B};
\node (r) [circle, draw] at (0,2) {R};
\draw [->] (a) edge [bend right] node[below] {a} (r);
\draw [->] (r) edge [bend right] node[above] {r} (a);
\draw [->] (b) edge [bend left] node[below] {b} (r);
\draw [->] (r) edge [bend left] node[above] {s} (b);
\end{tikzpicture}\end{center}

Alice and Bob win if $a\oplus b = r\cdot s$, i.e.\ they want $a\oplus b$ to be chosen according to this table:
\begin{center}
\begin{tabular}{c|c|c}
$r$ & $s$ & desired $a\oplus b$\\
\hline
0 & 0 & 0 \\
0 & 1 & 0 \\
1 & 0 & 0 \\
1 & 1 & 1 \\
\end{tabular}
\end{center}

In the next lecture we will show that if Alice and Bob use a deterministic or randomized classical strategy, their success probability will be $\leq 3/4$.  In contrast, using entanglement they can achieve a success probability of $\cos^2(\pi/8) \approx 0.854\dots > 3/4$.  This strategy, together with the ``payoff'' function (+1 if they win, -1 if they lose), yields an entanglement witness, and one that can be implemented only with local measurements.

%% file: 29may_2.tex
\lecture{8}{29 May, 2013}{Matthias Christandl}{Exact entanglement transformations}
\label{lec:exact}

\section{Three qubits, part two}

Last lecture we considered pure quantum states of three qubits, $\ket{\psi_{ABC}} \in \mathbb{C}^2 \otimes \mathbb{C}^2 \otimes \mathbb{C}^2$. We had claimed that if $\rho_A$, $\rho_B$, and $\rho_C$ are the reductions of $\ket{\psi_{ABC}}$ then
\begin{equation}
\label{cond}
\lambda^{A}_{\max} + \lambda^B_{\max} \leq 1 + \lambda^C_{\max}.
\end{equation}
Let us prove it. We have
\begin{align*}
  &\quad \lambda_{\max}^A + \lambda_{\max}^B
  = \max_{\phi_A} \braket{\phi_A | \rho_A | \phi_A}
   + \max_{\phi_B} \braket{\phi_B | \rho_B | \phi_B}
  = \max_{\phi_A} \tr \rho_A \proj{\phi_A}
   + \max_{\phi_B} \tr \rho_B \proj{\phi_B} \\
  &= \max_{\phi_A, \phi_B} \tr \rho_{AB} \left( \proj{\phi_A} \otimes I_B + I_A \otimes \proj{\phi_B} \right)
  \leq \max_{\phi_A, \phi_B} \tr \rho_{AB} \left( I_{AB} + \proj{\phi_A} \otimes \proj{\phi_B} \right) \\
  &= 1 + \max_{\phi_A, \phi_B} \tr \rho_{AB} \proj{\phi_A} \otimes \proj{\phi_B}
  \leq 1 + \max_{\phi_{AB}} \tr \rho_{AB} \proj{\phi_{AB}}
  = \lambda_{\max}^{AB}
  = \lambda_{\max}^{C},
\end{align*}
where in the last equality we have used that $\ket{\psi_{ABC}}$ is pure.

\begin{figure}
\centerline{\includegraphics[width=0.6\textwidth]{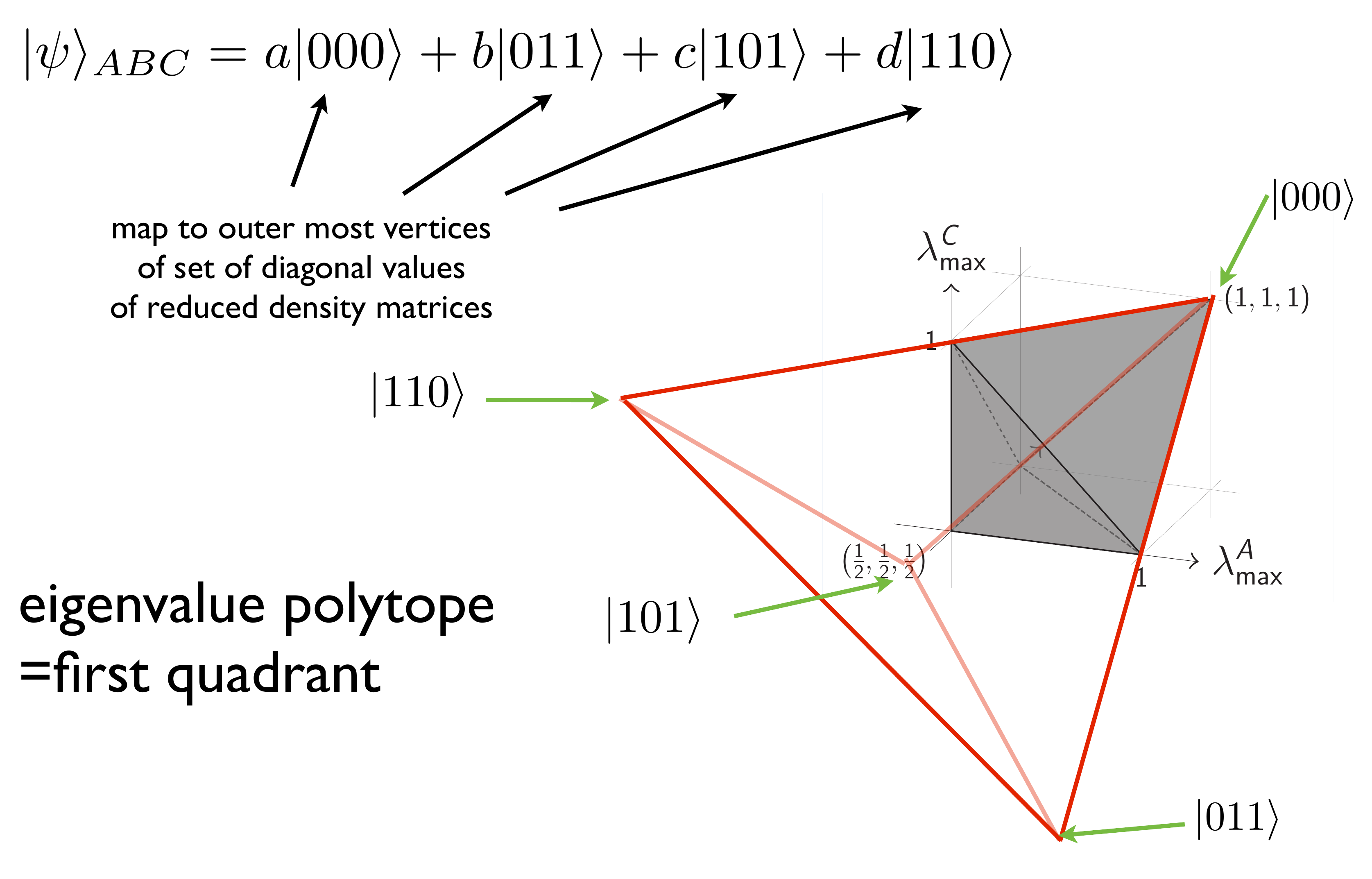}}
\caption{Sufficiency of the inequalities~\eqref{cond}.\label{fig:suff}}
\end{figure}

To show that inequality~\eqref{cond} together with its two permutations are also sufficient for a triple of eigenvalues to be compatible, let us consider the following ansatz:
\begin{equation*}
\ket{\psi_{ABC}} = a\ket{000} + b\ket{011} + c\ket{101} + d \ket{110}
\end{equation*}
with real parameters $a$, $b$, $c$, $d$ whose squares sum to one. The one-body reduced density matrices are
\begin{align*}
\rho_A &= (a^2 + b^2) \, \proj 0 + (c^2 + d^2) \, \proj 1, \\
\rho_B &= (a^2 + c^2) \, \proj 0 + (b^2 + d^2) \, \proj 1, \\
\rho_C &= (a^2 + d^2) \, \proj 0 + (b^2 + c^2) \, \proj 1.
\end{align*}
This leads to a system of equations
\begin{align*}
  a^2 + b^2 = \lambda_1, \quad
  a^2 + c^2 = \lambda_2, \quad
  a^2 + d^2 = \lambda_3,
\end{align*}
which can be solved if~\eqref{cond} and its permutations are satisfied (details omitted).
See \cref{fig:suff} for a graphical version of this proof.

In the next lecture we will see how algebraic geometry and representation theory are useful for studying the quantum marginal problem in higher dimensions.

\section{Exact entanglement transformation}

In \cref{lec:entanglement trafos}, Fernando considered asymptotic and approximate entanglement transformations. Here we will consider a different regime, namely single-copy and exact transformations.

Given two multipartite states, $\ket{\phi}$ and $\ket{\psi}$, which one is more entangled? One way to put an order on the set of quantum states is to say that $\ket{\phi}$ is at least as entangled as $\ket{\psi}$ if we can transform $\ket{\phi}$ into $\ket{\psi}$ by LOCC.

\begin{figure}
\centerline{\includegraphics[width=0.4\textwidth]{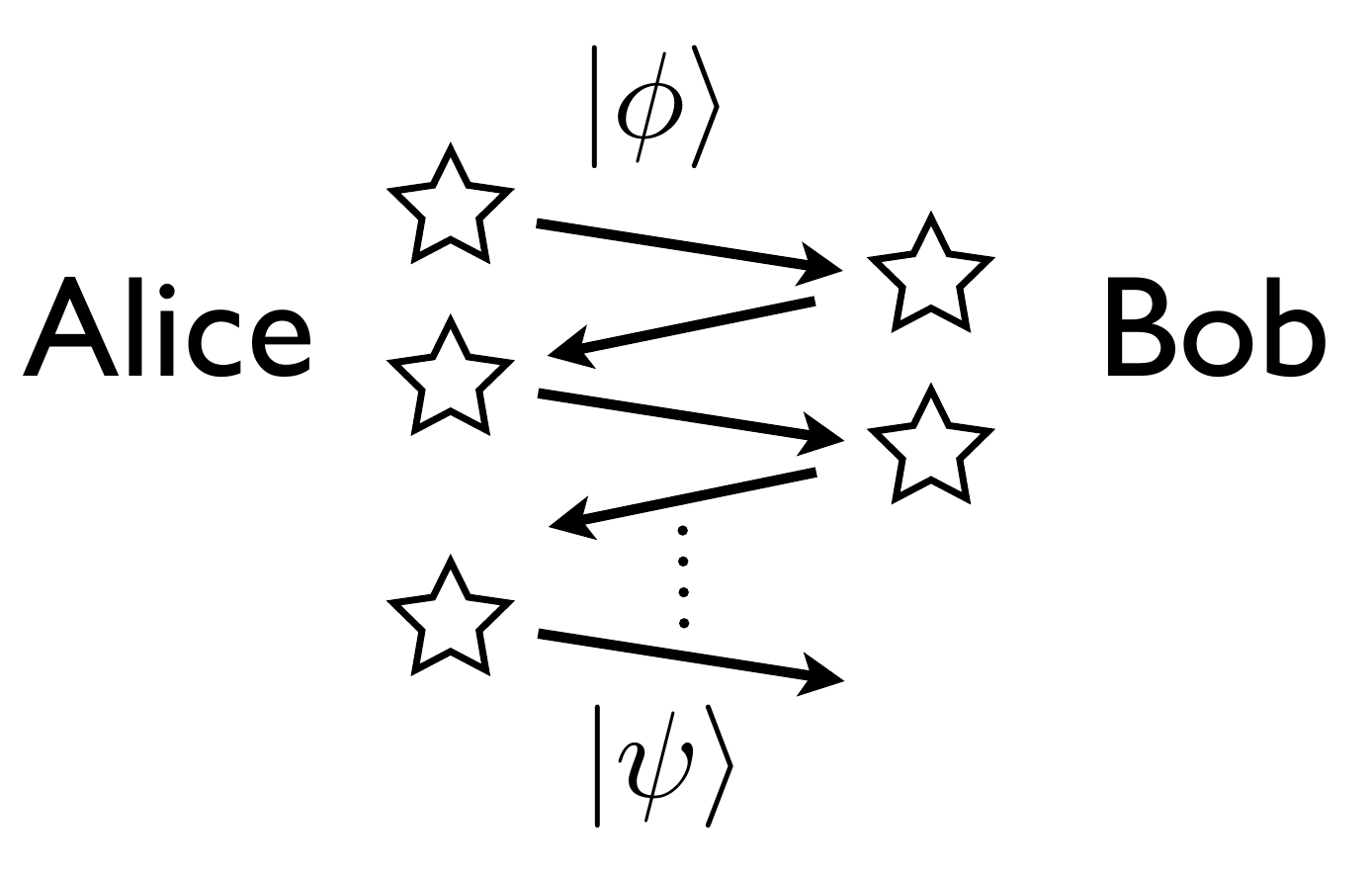}}
\caption{Sketch of an LOCC protocol transforming $\ket\phi$ into $\ket\psi$.\label{fig:trafo}}
\end{figure}

An LOCC protocol is given by a sequence of measurements by one of the parties and classical communication of the outcome obtained to the other (\cref{fig:trafo}).

\subsection{Quantum instrument}

Consider a quantum operation
\begin{equation*}
\Lambda(\rho_A) = \tr_{B'} (U (\rho_A \otimes \proj 0_B)U^{\dagger}) = \sum_{i} \braket{i_{B'} | U | 0_B} \rho_A \braket{0_B | U^{\dagger} | i_{B'}} = \sum_i E_i \rho_A E_i^{\dagger},
\end{equation*}
with $E_i := \braket{i_{B'} | U | 0_B}$. The $E_i$'s are called the \emph{Kraus operators} of $\Lambda$.

Note that the partial trace is the same as performing a projective measurement on $B'$ and forgetting the outcome obtained. Suppose now that we would record the outcome instead. Then, conditioned on outcome $i$, the (unnormalized) state is $E_i \rho_A E_i^{\dagger}$. We can associate the following operation to it:
\begin{equation*}
  \Gamma(\rho_A) = \sum_i E_i \rho_A E_i^{\dagger} \otimes \proj i.
\end{equation*}
The operation $\Gamma$ is also called a \emph{quantum instrument}.

\subsection{LOCC as quantum operations}

Going back to the LOCC protocol, Alice first measurement can be modelled by a set of Kraus operators $\{ A_{i_1} \}$. Then Bob's measurement, which can depends on Alice's outcome, will be given by $\{ B_{i_1, i_2} \}$, and so on. In terms of a quantum operation, a general $n$-round LOCC protocol can be written as
\begin{align} \label{loccmap}
\Lambda(\rho) &= \sum_{i_1, \dots, i_n} (A_{i_1, \dots, i_n} \cdots A_{i_1} \otimes B_{i_1, \dots, i_n} \cdots B_{i_1i_2}) \rho ( A_{i_1, \dots, i_n} \cdots A_{i_1} \otimes  B_{i_1, \dots, i_n} \cdots B_{i_1i_2})^{\dagger}  \nonumber \\
&\otimes \ket{i_1, \dots, i_n} \bra{i_1, \dots, i_n}_{A'}  \otimes  \ket{i_1, \dots, i_n} \bra{i_1, \dots, i_n}_{B'}
\end{align}

\subsection{SLOCC: Stochastic LOCC}
\label{subsec:slocc matthias}

The general form~\eqref{loccmap} of an LOCC operation is daunting. It turns out that the whole picture simplifies if we restrict our attention to the transformation of pure states to pure states and consider transformations to be successful if they transform the state with nonzero probability $p>0$ (as opposed to unit probability). We write
$$\ket\psi \xrightarrow{\text{SLOCC}} \ket\phi$$ and
call such an operation \emph{stochastic LOCC}, or SLOCC for short. Let us now derive a mathematical characterisation of SLOCC. First note that any LOCC operation can be written as a separable map, that is, as a map with Kraus operators $A_i \otimes B_i \otimes C_i$. That it only succeeds with nonzero probability means that we loosen the normalisation constraint
$$\sum_i A_i^\dagger A_i \otimes B_i^\dagger B_i \otimes C_i^\dagger C_i=\id$$
to
$$\sum_i A_i^\dagger A_i \otimes B_i^\dagger B_i \otimes C_i^\dagger C_i\leq \id.$$
If such an operation is to transform $\ket\psi$ into $\ket\phi$, then
$$p \proj\phi=\sum_i A_i \otimes B_i \otimes C_i \proj\psi (A_i \otimes B_i \otimes C_i)^\dagger.$$
Since the LHS is a pure state, all terms on the RHS must be proportional to each other and since we are only interested in the transformation to succeed with non-zero probability, we thus see that
$$\ket\phi=A \otimes B \otimes C \ket\psi$$
for some $A, B, C$ (which are proportional to $A_i, B_i, C_i$ for some $i$). Conversely, if
$$\ket\phi=A \otimes B \otimes C \ket\psi$$
then it is possible to implement this transformation with local transformation with nonzero probability: First find strictly positive constants $a, b, c$, s.th.
$$\tilde A= a A, \tilde B= b B, \tilde C= c C,$$
satisfy
$$\tilde A^\dagger A \leq \id, \tilde B^\dagger B \leq \id , \tilde C^\dagger C \leq \id.  $$
Then implement the local operation corresponding to the application of the local CPTP maps with Kraus operators $\{\tilde A, \sqrt{\id -\tilde A^\dagger \tilde A}\} $. In summary, we find that
$$\ket\psi \xrightarrow{\text{SLOCC}} \ket\phi$$
iff
$$\ket\phi=A \otimes B \otimes C \ket\psi$$
for some matrices $A, B, C$.

We say that $\ket\psi$ and $\ket\phi$ have the ``same type of entanglement'' if both
$\ket\psi \xrightarrow{\text{SLOCC}} \ket\phi$ and $\ket\phi \xrightarrow{\text{SLOCC}} \ket\psi$.
It is then easy to see that this is the case if, and only if, there exist invertible matrices $A$, $B$ and $C$ such that
\begin{equation}
\label{eq:slocc equivalence matthias}
  \ket\phi = (A \otimes B \otimes C) \ket\psi.
\end{equation}
Since we do not care about normalization, we can w.l.o.g.\ take $A, B, C$ to be matrices in $\SL(d)$, the group of $d\times d$ matrices of unit determinant.
Therefore we see that the problem of characterizing different entanglement classes is equivalent to the problem of classifying the orbit classes of $\SL(d) \times \SL(d) \times \SL(d)$.

In general the number of orbits is huge. Indeed the dimension of the Hilbert space scales as $d^3$, but the group only has approximately $3 d^2$ parameters.

But the case of three qubits turns out to be simple and we only have 6 different classes.
One is the class of (fully) separable states, with representative state
\begin{equation*}
\ket{000}_{ABC}.
\end{equation*}
Then there are three states where only two parties are entangled, with representative states
\begin{equation*}
  \ket{\Phi^+}_{AB} \otimes \ket{0}_C, \quad
  \ket{\Phi^+}_{AC} \otimes \ket{0}_B, \quad
  \ket{\Phi^+}_{BC} \otimes \ket{0}_A.
\end{equation*}
The fifth class is the so-called \emph{GHZ-class}, represented by the GHZ state
\begin{equation*}
  \ket{\text{GHZ}}_{ABC} = \frac 1 {\sqrt 2} \left( \ket{000} + \ket{111} \right).
\end{equation*}
The last class is the so-called \emph{W-class}, with representative state
\begin{equation*}
  \ket{\text{W}}_{ABC} = \frac{1}{\sqrt{3}} \left( \ket{100} + \ket{010} + \ket{001} \right).
\end{equation*}
In \cref{lec:entanglement polytopes}, Michael will show you that the local spectra of the quantum states in a fixed class of entanglement form a subpolytope of the polytope of spectra that we have discussed in the context of the pure-state quantum marginal problem.

%% file: 29may_3.tex
\lecture{9}{29 May, 2013}{Aram Harrow}{Quantum de Finetti theorem}
\label{lec:de finetti proof}

\section{Proof of the quantum de Finetti theorem}

Let us remind ourselves that the quantum de Finetti theorem (\cref{thm:deF}) states that $\tr_n \proj{\psi} \approx \int d\mu(\sigma) \sigma^{\otimes k}$ for all $\ket{\psi} \in \Sym^{n+k}(\CC^D)$ and $n$ large.

The intuition here is that measuring the last $n$ systems and finding that they are each in state $\sigma$ implies that the remaining $k$ systems are also in state $\sigma$.

Let us now do the math. Recall that
$$\int d\phi \, \proj{\phi}^{\otimes m} = \frac{\Psym^{D,m}}{\binom{D+m-1}{D-1}}.$$
Therefore, if $\ket{\psi} \in \Sym^{n+k}(\CC^D)$ then
\begin{align*}
  &\quad \tr_n \proj{\psi}
  = \tr_n \left(\left( I^{\otimes k} \otimes \Psym^{D,n} \right) \proj\psi \right) \\
  &= \int d\phi \, {\binom {D+n-1} {D-1}} \tr_n \left( \left( I^{\otimes k} \otimes \proj\phi^{\otimes n} \right) \proj\psi \right) \\
  &= \int d\phi \, \proj{\widetilde v_\phi},
\end{align*}
where we defined $\ket{\widetilde{v}_\phi} := \sqrt{\binom {D+n-1} {D-1}}(I^{\otimes k} \otimes \bra{\phi}^{\otimes n})\ket{\psi}$.
Let us write $\ket{\widetilde{v}_\phi} = \sqrt{p_\phi} \ket{v_\phi}$, with $\ket{v_\phi}$ a unit vector.
We claim that $\ket{v_\phi} \approx \ket{\phi}^{\otimes k}$ on average:
\begin{align}
\nonumber
&\quad \int d\phi \, p_\phi \lvert \bra{v_\phi}\ket{\phi}^{\otimes k} \rvert^2
= \int d\phi \, \lvert \bra{\widetilde{v}_\phi}\ket{\phi}^{\otimes k} \rvert^2\\
\nonumber
& =  {\binom {D+n-1} {D-1}} \int d\phi \, \lvert \bra{\psi} \ket{\phi}^{\otimes (n+k)} \rvert^2
= {\binom {D+n-1} {D-1}} \tr ( \proj{\psi} \int d\phi \, \proj{\phi}^{\otimes n+k} )\\
\label{eq:de finetti lower bound}
&= \frac {\binom {D+n-1} {D-1}} {\binom {D+n+k-1} {D-1}}  \tr ( \proj{\psi} \, \Psym^{D,n+k} )
= \frac {\binom {D+n-1} {D-1}} {\binom {D+n+k-1} {D-1}} \geq 1 - kD/n
\end{align}
(The lower bound was proved at the end of \cref{lec:monogamy}.)

Note that this bound is polynomial in $n$. This is tight. There exists, however, an improvement to an exponential dependence in $n$ at the cost of replacing product states by almost-product states.

In order to conclude the proof of the quantum de Finetti theorem, we need to relate the trace distance to the average we computed.
For this, we consider the fidelity $\lvert \braket{\alpha|\beta} \rvert^2$ between states $\ket{\alpha}$ and $\ket{\beta}$.
If now $\lvert\braket{\alpha | \beta}\rvert = 1 - \eps$ and we expand $\ket{\beta}= \sqrt{1-\eps}\ket{\alpha} +\sqrt{\eps}\ket{\alpha}$, then
\begin{equation}
\label{eq:trace distance vs fidelity pure states}
\lVert \proj{\alpha}-\proj{\beta} \rVert_1= \L\| \bpm 1 & 0 \\ 0 & 0 \epm -
\bpm 1-\eps & \sqrt{\eps(1-\eps)} \\ \sqrt{\eps(1-\eps)} & \eps\epm
\R\|_1= 2\sqrt{\eps}.
\end{equation}

\subsection{Permutation-invariant mixed states}
\label{subsec:perm inv}
Suppose that $\rho_{Q_1 \dots Q_{n}}$ (with each $\dim Q_i = D$)  is permutation-invariant, meaning that $P_\pi \rho P_\pi^\dag = \rho$ for all $\pi \in S_{n}$.  (We use $n$ instead of $n+k$ here to simplify notation.) This is a weaker condition than having support in $\Sym^{n}(\CC^D)$.  Sometimes being permutation-invariant is called being ``symmetric'' and having support in $\Sym^{n}(\CC^D)$ is called being ``Bose-symmetric.''

If $\rho$ is merely permutation-invariant, then we cannot directly apply the above theorem.  However we will show that $\rho$ has a purification $\ket{\psi_{Q_1\dots Q_nR_1\dots R_n}}$ (with $\dim R_i= D$) that lies in $\Sym^n(\CC^{D^2})$, so that we can apply the de Finetti theorem proved above.   This was also proved in Lemma 4.2.2 of [Renner; \href{http://arxiv.org/abs/quant-ph/0512258}{arXiv:quant-ph/0512258}], but we give an alternate proof here.  Our proof is in a sense equivalent but uses a calculating style that is more widely used in quantum information theory.

Diagonalize $\rho$ as $\rho = \sum_\lambda \lambda \Pi_\lambda$ where each $\lambda$ in the sum is distinct and the $\Pi_\lambda$ are projectors.
Since $[\rho, P_\pi]=0$ for all $\pi$ it follows that each $P_\pi$ commutes with each $\Pi_\lambda$; i.e.
\bas  P_\pi \Pi_\lambda = \Pi_\lambda P_\pi, \eas
for all $\pi, \lambda$.
Define $M := \sum_\lambda \sqrt{\lambda} \, \Pi_\lambda$.  Then we also have
\be P_\pi M  = M P_\pi \label{eq:M-symmetric}\ee
for all $\pi$.

Also define $\ket{\Phi}_{Q^n R^n} := \sum_{x=1}^{d^n} \ket x_{Q^n} \ot \ket x_{R^n}$, where we have abbreviated $Q^n := Q_1\dots Q_n$, $R^n := R_1,\dots, R_n$ and where $\ket{x}$ is the usual product basis. (This definition of $\ket\Phi$ is somewhat unconventional in that $\ket\Phi$ is an unnormalized state.)  One useful feature of $\ket\Phi$ is that for any $d^n\times d^n$ matrix $A$,
\be (I \ot A)\ket\Phi = (A^T \ot I)\ket\Phi, \label{eq:trans-id}\ee
as can be verified by expanding the product in the basis of $\ket\Phi$.  Observe also that $\tr_{R^n}\proj{\Phi} = I_{Q^n}$.

At last, define  $\ket\psi_{Q^n R^n} := (M \ot I)\ket\Phi$.  First we check that $\ket\psi$ is a purification of $\rho$.  Indeed
\be \tr_{R^n} \proj\psi = M \tr_{R^n} (\proj\Phi) M^\dag = MM^\dag = \rho\ee.
Next we show that $\ket\psi \in \Sym^n(\CC^{D^2})$.  If $\pi \in S_n$ permutes the $(Q_1,R_1),\dots,(Q_n,R_n)$ systems and we order them as $Q_1,\dots,Q_n,R_1,\dots, R_n$, then its action can be written as $P_\pi\ot P_\pi$.  Thus we need to check whether $\ket\psi$ is invariant under each $P_\pi\ot P_\pi$.
Indeed,
\bas
&\quad (P_\pi  \ot P_\pi)\ket\psi \\
& =  (P_\pi  \ot P_\pi)(M \ot I)\ket\Phi \\
& =  (P_\pi M  \ot I)(I \ot P_\pi)\ket\Phi \\
& =  (P_\pi M P_\pi^T  \ot I)\ket\Phi & \text{using \cref{eq:trans-id}} \\
& =  (MP_\pi  P_\pi^T  \ot I)\ket\Phi & \text{using \cref{eq:M-symmetric}} \\
&= (M \ot I)\ket\Phi = \ket\psi & \text{since $P_\pi$ are real.}
\eas

\section{Quantum key distribution}

A surprising application of entanglement is quantum key distribution. Suppose Alice and Bob share an EPR pair
$\ket{\Phi^+}=\frac{1}{\sqrt{2}}( \ket{00} + \ket{11} )$. Then the joint state $\ket{\psi}_{ABE}$ of Alice, Bob and a potential eavesdropper Eve  is such that $\tr_E \proj{\psi}_{ABE}= \proj{\Phi^+}_{AB}$, and hence necessarily of the form $\ket{\psi}_{ABE}= \ket{\Phi^+}_{AB} \otimes \ket{\gamma}_E$

By measuring in their standard basis, Alice and Bob thus obtain a secret random bit $r$. They can use this bit to send a bit securely with help of the Vernam one-time pad cipher:
Let's call Alice's message $m$. Alice sends the cipher $c= m \oplus r$ to Bob. Bob then recovers the message by adding $r$: $c\oplus r= m \oplus r \oplus r = m$.

How can we establish shared entanglement between Alice and Bob? Alice could for instance create the state locally and send it to Bob using a quantum channel (i.e.\ a glass fibre).

But how can we now verify that the joint state that Alice and Bob have after the transmission is an EPR state?

Here is a simple protocol:
\benum
\item Alice sends halves of $n$ EPR pairs to Bob.
\item They choose randomly half of them and perform CHSH tests (see sections~\ref{sec:chsh} and~\ref{sec:chsh more}).
\item They obtain a secret key from the remaining halves.
\eenum
There are many technical details that I am glossing over here. One is, how can you be confident that the other halves are in this state? By the de Finetti theorem---the choice was permutation invariant!

Unfortunately, the version that we discussed above requires the number of key bits  $k$ to scale as $n^c$ for some $c<1$; otherwise the lower bound in \cref{eq:de finetti lower bound} will not approach one.
Ideally we would have $k/n$ approach a constant, which can be achieved by using the stronger bounds from the exponential de Finetti theorem (Renner) or the post-selection technique (Christandl, K\"onig, Renner).

Another issue is that there might be noise on the line. It is indeed possible to do quantum key distribution even in this case, but here one needs some other tools mainly relating to classical information theory (information reconciliation or privacy amplification).

%% file: 30may_1.tex
\lecture{10}{30 May, 2013}{Fernando G.S.L. Brand\~ao}{Computational complexity of entanglement}
\label{lec:complex sep}

\section{More on the CHSH game}
\label{sec:chsh more}

We continue our discussion of the CHSH game.
\begin{center}
\begin{tikzpicture}
\node (a) [circle, draw] at (-2,0) {A};
\node (b) [circle, draw] at (2,0) {B};
\node (r) [circle, draw] at (0,2) {R};
\draw [->] (a) edge [bend right] node[below] {a} (r);
\draw [->] (r) edge [bend right] node[above] {r} (a);
\draw [->] (b) edge [bend left] node[below] {b} (r);
\draw [->] (r) edge [bend left] node[above] {s} (b);
\end{tikzpicture}\end{center}

\noindent Alice and Bob win if $a\oplus b = r\cdot s$, i.e.\ they want $a\oplus b$ to be chosen according to this table:
\begin{center}
\begin{tabular}{c|c|c}
$r$ & $s$ & desired $a\oplus b$\\
\hline
0 & 0 & 0 \\
0 & 1 & 0 \\
1 & 0 & 0 \\
1 & 1 & 1 \\
\end{tabular}
\end{center}

\paragraph{Deterministic strategies.}
Consider a deterministic strategy.   This means that if Alice receives
$r=0$, she outputs the bit $a_0$ and if she receives $r=1$, she
outputs the bit $a_1$.  Similarly, Bob outputs $b_0$ if he receives
$s=0$ and $b_1$ if he receives $s=1$.

There are four possible inputs.  If they set $a_0=a_1=b_0=b_1=0$, then
they will succeed with probability 3/4.  Can they do better?  For a
deterministic strategy this can only mean winning with probability 1.
But this implies that
\bas
a_0 \oplus b_0 & = 0 \\
a_0 \oplus b_1 & = 0 \\
a_1 \oplus b_0 & = 0 \\
a_1 \oplus b_1 & = 1
\eas
Adding this up (and using $x\oplus x = 0$) we find $0=1$, a
contradiction.

\paragraph{Randomized strategies.}  What if Alice and Bob share some
correlated random variable and choose a deterministic strategy based
on this?  Then the payoff is the average of the payoffs of each of the
deterministic strategies.  Thus, there must always be at least one
deterministic strategy that does at least as well as the average.  So
we can assume that an optimal strategy does not need to make use of
randomness.

\begin{ex*}
What if they use uncorrelated randomness?  Can this help?
\end{ex*}

\paragraph{Quantum strategies.}
Now suppose they share an EPR pair $\ket{\Phi^+}$.
Define
\bas
\ket{\phi_0(\theta)} & = \hphantom{-} \cos(\theta) \ket 0 + \sin(\theta) \ket 1 \\
\ket{\phi_1(\theta)} & = -\sin(\theta) \ket 0 + \cos(\theta) \ket 1
\eas
Observe that $\{\ket{\phi_0(\theta)}, \ket{\phi_1(\theta)}\}$ is an
orthonormal basis for any choice of $\theta$.

The strategy is as follows.   Alice and Bob will each measure
their half of the entangled state in the basis
$\{\ket{\phi_0(\theta)}, \ket{\phi_1(\theta)}\}$ for some choice of
$\theta$ that depends on their inputs.  They will output 0 or 1,
depending on their measurement outcome.   The choices of $\theta$ are

\begin{center}
\begin{tabular}{ccc}
\toprule
\multirow{2}{*}{Alice} &
$r=0$ & $\theta=0$ \\
& $r=1$ & $\theta=\pi/4$ \\
\midrule
\multirow{2}{*}{Bob} &
$s=0$ & $\theta=\pi/8$ \\
& $s=1$ & $\theta=-\pi/8$ \\
\bottomrule
\end{tabular}
\end{center}

\begin{ex*}
Show that $\Pr[\text{win}] = \cos^2(\pi/8) = \frac{1}{2} + \frac{1}{2\sqrt 2} > 3/4$.
\end{ex*}

Another way to look at the quantum strategy is in terms of local, $\pm1$-valued
observables.  Alice and Bob's strategy can be described in terms of
the matrices
\bas
A_0 &= {\bpm 1 & 0 \\ 0 & -1 \epm} &
B_0 &= \frac{1}{\sqrt 2} {\bpm 1 & 1 \\ 1 & -1 \epm} \\
A_1 &= {\bpm 0 & 1 \\ 1 & 0 \epm} &
B_1 &= \frac{1}{\sqrt 2} {\bpm 1 & -1 \\ -1 & -1 \epm} \\
\eas
Given a state $\ket\psi$, the value of the game can be expressed in terms of the ``bias''
$$\frac{1}{4}
\bra\psi (A_0 \ot B_0 + A_0 \ot B_1 + A_1 \ot B_0 - A_1 \ot B_1)\ket \psi
= \Pr[\text{win}] -\Pr[\text{lose}]
= 2\Pr[\text{win}] - 1$$
(see \cref{ex:tsirel} for details).
We can define a Hermitian matrix $W'$ by
$$W' = \frac{1}{4}(A_0 \ot B_0 + A_0 \ot B_1 + A_1 \ot B_0 - A_1 \ot
B_1).$$
Then, for any $\sigma \in \Sep$,
$$\frac{1}{4} \tr(W'\sigma) \leq 2 \max_{\sigma \in \Sep} \Pr[\text{win}] - 1 = 2 \frac 34 - 1 
= \frac 1 2.$$
Thus if we define $W = \frac I 2 - \frac 1 4 W'$ then for all $\sigma\in\Sep$,
$\tr(W\sigma) \geq 0$, while $\tr(W\proj{\Phi^+}) = -\frac{1}{\sqrt 2}
< 0$.

In this way, Bell inequalities define entanglement witnesses; moreover, ones
that distinguish an entangled state even from separable states over
unbounded dimension that are measured with possibly different
measurement operators!

There has been some exciting recent work on the CHSH game.  One recent
line of work has been on the {\em rigidity} property, which states
that any quantum strategy that comes within $\eps$ of the optimal
value $\frac{1}{2} + \frac{1}{2\sqrt 2}$ must be within $\eps'$ of the
ideal strategy (up to some trivial changes).
This is relevant to the field of {\em device-independent} quantum
information processing, which attempts to draw conclusions about an
untrusted quantum device based only on local measurement outcomes.
(For more references see [McKague, Yang, Scarani; \href{http://arxiv.org/abs/1203.2976}{arXiv:1203.2976}] and [Scarani; \href{http://arxiv.org/abs/1303.3081}{arXiv:1303.3081}].)

\section{Computational complexity}

\begin{prb}[Weak membership for $\Sep$]
\label{prb:weak mem sep}
Given a quantum state $\rho_{AB}$ on $\CC^n \ot \CC^m$, $\eps>0$, and the promise that
either
\benum \item $\rho_{AB} \in \Sep$, or
\item $D(\rho,\Sep) = \min_{\sigma\in\Sep}D(\rho,\sigma) \geq \eps$,
\eenum
decide which is the case.
\end{prb}

This problem is called the ``weak'' membership problem because of the $\eps>0$ parameter,
which means we don't have to worry too much about numerical precision.

There are many choices of distance measure $D(\cdot,\cdot)$.  We could
take $D(\rho,\sigma) = \frac 1 2 \| \rho - \sigma \|_1$, as we did
earlier.  Or we could use $\|\rho-\sigma\|_2$, where $\|X\|_2 :=
\sqrt{\tr(X^\dag X)}$.

Another important problem related to $\Sep$ is called the {\em support
  function}.  Like weak membership, it can be defined for any set, but
we will focus on the case of $\Sep$.

\begin{prb}[Support function of $\Sep$]
\label{prb:supp sep}
Given a Hermitian matrix $M$ on $\CC^n \ot \CC^m$ and $\eps>0$, compute $h_{\Sep}(M)
\pm \eps$, where
$$h_{\Sep}(M) := \max_{\sigma \in\Sep} \tr (M\sigma).$$
\end{prb}

There is a sense in which \cref{prb:weak mem sep} $\cong$ \cref{prb:supp sep}, meaning that an
efficient solution for one can be turned into an efficient solution to
the other.  We omit the proof of this fact, which is a classic result
in convex optimization [M. Gr\"otschel, L. Lov\'asz, A. Schrijver.
{\em Geometric Algorithms and Combinatorial Optimization}, 1988].

\paragraph{Efficiency.}  What does it mean for a problem to be
``efficiently'' solvable?   If we parametrize a problem by the size of
the input, then we say a problem is efficient if inputs of size $n$ can be
solved in time polynomial in $n$, i.e.\ in time $\leq c_1 n^{c_2}$ for
some constants $c_1,c_2$.  This class of problems is called $\Ptime$, which
stands for \emph{Polynomial time}.  Examples include multiplication, finding
eigenvalues, solving linear systems of equations, etc.

Another important class of problems are those where the solution can
be efficiently {\em checked}.  This is called $\NP$, which stands for
\emph{Nondeterministic Polynomial time}.  (The term ``nondeterministic'' is
somewhat archaic, and refers to an imaginary computer that randomly
checks a possible solution and needs only to succeed with some
positive, possibly infinitesimal, probability.)

One example of a problem in $\NP$ is called 3-SAT.  A 3-SAT instance is a
formula over variables $x_1,\dots,x_n\in\{0,1\}$ consisting of an AND
of $m$ clauses, where each clause is an OR of three variables or their
negations.  Denoting OR with $\vee$, AND with $\wedge$, and NOT $x_i$
with $\bar x_i$, an example of a formula would be
$$\phi(x_1,\dots,x_n) = (x_1 \vee \bar x_4 \vee x_17) \wedge (\bar x_2 \vee \bar x_7 \vee
x_10)
\wedge \dots.$$
Given a formula $\phi$, it is not {\em a priori} obvious how we can
figure out if it is satisfiable.  One option is to check all possible
values of $x_1,\dots,x_n$.  But there are $2^n$ assignments to check,
so this approach requires exponential time.  Better algorithms are
known, but none has been proven to run in time better than $c^n$ for
various constants $c>1$. However, 3-SAT is in $\NP$ because if $\phi$
is satisfiable, then there exists a short ``witness'' proving this
fact that we can quickly verify.  This witness is simply a satisfying
assignment $x_1,\dots,x_n$.  Given $\phi$ and $x_1,\dots,x_n$
together, it is easy to verify whether indeed $\phi(x_1,\dots,x_n)=1$.

\paragraph{NP-hardness.}  It is generally very difficult to prove that a
problem {\em cannot} be solved efficiently.  For example, it is
strongly believed that 3-SAT is not in $\Ptime$, but there is no proof of
this conjecture.  Instead, to establish hardness we need to settle for finding evidence that
falls short of a proof.

Some of the strongest evidence we are able to obtain for this is to
show that a problem is  $\NP$-hard, which means that any problem in
$\NP$ be efficiently reduced to it.  For example, 3-SAT is $\NP$-hard.  This means
that if we could solve 3-SAT instances of length $n$ in time $T(n)$,
then any other problem in $\NP$ could be solved in time $\leq
\poly(T(\poly(n)))$.    In particular, if 3-SAT were in $\Ptime$ then it
would follow that $\Ptime = \NP$.

It is conjectured that $\Ptime \neq \NP$, because it seems harder to
{\em find} a solution in general than to {\em recognize} a solution.
This is one of the biggest open problems in mathematics, and all
partial results in this direction are much much weaker.  However, if
we assume for now that $\Ptime\neq \NP$, then showing a problem is
$\NP$-hard implies that it is not in $\Ptime$.   And since thousands of
problems are known to be $\NP$-hard\footnote{See this list:
\url{http://en.wikipedia.org/wiki/List_of_NP-complete_problems}.
The terminology $\NP$-complete refers to problems that are both
$\NP$-hard and in $\NP$.} it suffices to show a reduction from {\em
any} $\NP$-hard problem in order to show that a new problem is also
$\NP$-hard.  Thus, this can be an effective method of showing that a
problem is likely to be hard.

\begin{thm}
Problems 1 and 2 are $\NP$-hard for $\eps = 1/\poly(n,m)$.
\end{thm}

We will give only a sketch of the proof.

\begin{figure}
\centerline{
\includegraphics[width=0.4\textwidth]{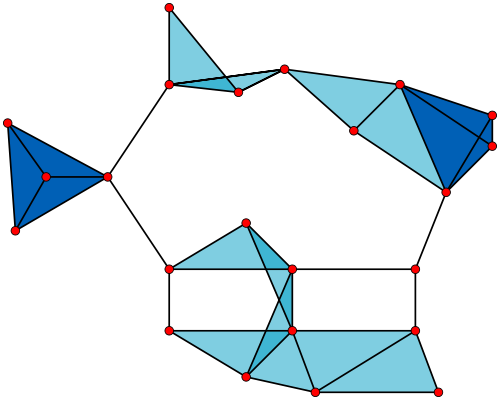}
}
\caption{ This figure is taken from the wikipedia article
\url{http://en.wikipedia.org/wiki/Clique_(graph_theory)}. The
42 2-cliques are the edges, the 19 3-cliques are the triangles colored light
blue and the 2 4-cliques are colored dark blue. There are no 5-cliques.
\label{fig:clique}}
\end{figure}

\benum
\item Argue that MAX-CLIQUE is $\NP$-hard.  This is a classical result that
  we will not reproduce here.  Given a graph $G=(V,E)$ with vertices
  $V$ and edges $E$, a clique is a subset $S\subseteq V$ such that
  $(i,j)\in E$ for each $i,j\in S$, $i\neq j$.  An example is given in
  \cref{fig:clique}.  The MAX-CLIQUE problem asks for the size of the largest
clique in a given graph.
\item MAX-CLIQUE can be related to a bilinear optimization problem
  over probability distributions by the following theorem.
\begin{thm}[Motzkin-Straus]
Let $G=(V,E)$ be a graph, with maximum clique of size $W$.  Then
\be 1 - \frac{1}{W} = 2 \max \sum_{(i,j)\in E} p_i p_j,\label{eq:clique-form}\ee
where the $\max$ is taken over all probability distributions $p$.
\end{thm}
\item
Given a graph, define
$$M = \sum_{(i,j)\in E} \proj{i,j}.$$
Then
$$\max_{\ket \phi}
\braket{\phi,\phi | M | \phi, \phi} =
\max_{\|\phi\|_2=1} \sum_{(i,j)\in E} |\phi_i|^2 |\phi_j|^2.$$
Defining $p_i = |\phi_i|^2$, we recover the RHS of \cref{eq:clique-form}.
\item
We argue that
$$h_{\Sep}(M) = \max_{\|\phi\|_2=\|\psi\|_2=1} \braket{\phi,\psi | M | \phi,\psi}.$$
This is because $\Sep$ is a convex set, its extreme points are of the
form $\proj{\phi,\psi}$, and the maximum of any linear function over a
convex set can be achieved by an extreme point.
\item
Finally, we argue that maximizing over $\ket{\phi,\psi}$ is equivalent in
difficulty to maximizing over $\ket{\phi,\phi}$.
\eenum

What accuracy do we need here?  If we want to distinguish a clique of
size $n$ (where there are $n$ vertices) from size $n-1$, then we need
accuracy $(1-\frac{1}{n-1}) - (1 - \frac{1}{n}) \approx 1/n^2$.  Thus,
we have shown that \cref{prb:supp sep} is $\NP$-hard for $\eps= 1/n^2$.

%% file: 30may_2.tex
\lecture{11}{30 May, 2013}{Michael Walter}{Quantum marginal problem and entanglement}
\label{lec:entanglement polytopes}

\section{Entanglement classes as group orbits}

In \cref{lec:exact}, Matthias introduced SLOCC (stochastic LOCC), where we can post-select on particular outcomes.
We now consider an \emph{entanglement class} of pure quantum states that can be converted into each other by SLOCC,
$$C_{\phi} = \big\{ \ket{\psi_{ABC}} \,:\, \ket{\psi_{ABC}} \stackrel{\text{SLOCC}}\longleftrightarrow \ket{\phi_{ABC}} \big\},$$
where $\ket{\phi_{ABC}}$ is an arbitrary state in the class.
Matthias explained to us that any such class can equivalently be characterized in the following form:
$$C_{\phi} := \big\{ \ket{\psi_{ABC}} \,:\, \ket{\psi_{ABC}} \propto (A \otimes B \otimes C) \ket{\phi_{ABC}} \text{ for some } A, B, C \in \SL(d) \big\}$$
(Here, $\SL(d)$ is the ``special linear group'' of invertible operators of unit determinant, which leads to the proportionality sign rather than the equality that we previously saw in \cref{eq:slocc equivalence matthias}.)

For three qubits there is a simple classification of all such classes of entanglement, which we will discuss in \cref{ex:entanglement classes}.
Apart from product states and states with only bipartite entanglement, there are two classes of ``genuinely'' tripartite entangled states, with the following representative states:
\begin{align*}
  \ket{GHZ} &= \frac{1}{2} \left( \ket{000} + \ket{111} \right) \\
  \ket{W} &= \frac{1}{2} \left( \ket{100} + \ket{010} + \ket{001} \right)
\end{align*}

Let us now introduce the group
$$G = \big\{ A \otimes B \otimes C : A, B, C \in \SL(d) \big\}.$$
Then we can rephrase the above characterization in somewhat more abstract language:
An SLOCC entanglement class $C_\phi$ is simply the orbit $G \cdot \ket{\phi_{ABC}}$ of a representative quantum state $\ket{\phi_{ABC}}$ under the group of SLOCC operations $G$, up to normalization. (That is, it is really an orbit in the projective space of pure states).

It turns out that $G$ is a Lie group just like $\SL(d)$. Indeed, an easy-to-check fact is that $\SL(d) = \{ e^X : \tr(X) = 0 \}$, where $e^X$ denotes the exponential of a $d \times d$ matrix $X$. Therefore,
$$G = \{ e^X \otimes e^Y \otimes e^Z = e^{X \otimes I \otimes I + I \otimes Y \otimes I + I \otimes I \otimes Z} : \tr X = \tr Y = \tr Z = 0 \},$$
and we hence the Lie algebra of $G$ is spanned by the traceless local Hamiltonians.

\section{The quantum marginal problem for an entanglement class}

What are the possible $\rho_A$, $\rho_B$, $\rho_C$ that are compatible with a pure state \emph{in a given entanglement class}?
Note that this only depends on the spectra $\lambda_A$, $\lambda_B$ and $\lambda_C$ of the reduced density matrices, as one can always apply local unitaries and change the basis without leaving the SLOCC class.

\begin{figure}
  \begin{center}
  \includegraphics[width=6cm]{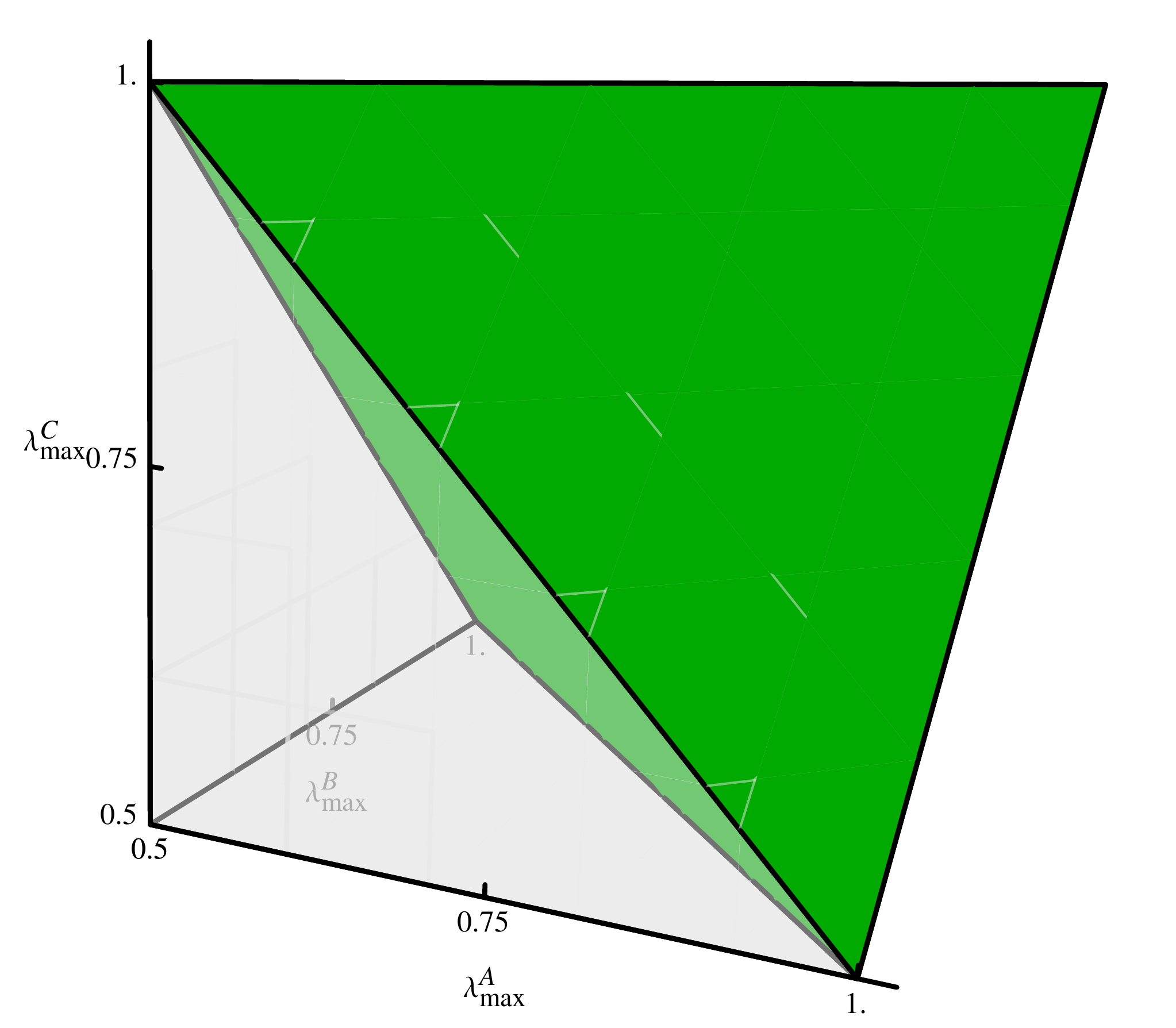}
  \end{center}
  \caption{The entanglement polytope of the W class (green) is the region of all local eigenvalues that are compatible with a state from the W class or its closure.}
  \label{fig:w polytope}
\end{figure}

Are there any new constraints? Yes! For example, the reduced density matrices of the class of product states are always pure, hence its local eigenvalues satisfy $\lambda^A_{\max} = \lambda^B_{\max} = \lambda^C_{\max} = 1$.
A more interesting example is the W class. Here, the set of compatible spectra is given by the equation
\[ \lambda_{\max}^A + \lambda_{\max}^B + \lambda_{\max}^C \geq 2, \]
as we will discuss in \cref{ex:W polytope} in the last problem session (\cref{fig:w polytope}).

\section{Locally maximally mixed states}

Let us start with the following special case of the problem:
Given an entanglement class $G \cdot \ket{\phi_{ABC}}$, does it contain a state $\rho = \proj\psi_{ABC}$ with $\rho_A = \rho_B = \rho_C \propto I/d$?
Such a state is also called \emph{locally maximally mixed}; it corresponds to the ``origin'' in the coordinate system of \cref{fig:w polytope}.
This is equivalent to
$$\tr(\rho_A X) = \tr(\rho_B Y) = \tr(\rho_C Z) = 0$$
for all traceless Hermitian matrices $X, Y, Z$.

Geometrically speaking, this means that the norm square of the state $\ket{\psi_{ABC}}$ should not change (to first order) when we apply an arbitrary infinitesimal SLOCC operation \emph{without afterwards renormalizing the state}.
Indeed:
\begin{align*}
  &\quad \left.\frac{\partial}{\partial t}\right|_{t=0} \Vert e^{Xt} \otimes e^{Yt} \otimes e^{Zt} \ket{\psi_{ABC}}  \Vert^2 \\
  &= \left.\frac{\partial}{\partial t}\right|_{t=0} \bra{\psi_{ABC}} e^{2Xt} \otimes e^{2Yt} \otimes e^{2Zt} \ket{\psi_{ABC}} \nonumber \\
  &= 2 \, \braket{\psi_{ABC} | X \otimes I \otimes I + I \otimes Y \otimes I + I \otimes I \otimes Z | \psi_{ABC}} \\
  &= \tr(\rho_A X) + \tr(\rho_B Y) + \tr(\rho_C Z) = 0
\end{align*}
For example, if $\ket{\psi_{ABC}}$ is a vector of minimal norm in the orbit $G \cdot \ket{\phi_{ABC}}$ then $\rho_A = \rho_B = \rho_C \propto I/d$.

What happens when there is no state in the class with $\rho_A = \rho_B = \rho_C \propto I/d$?
That might seem strange, as it implies by the above that there is no vector of minimal norm in the orbit.
But such situations can indeed occur since the group $G$ is not compact. For example,
\begin{equation}
\bpm \eps & \\ & \frac 1 \eps \epm
\otimes
\bpm \eps & \\ & \frac 1 \eps \epm
\otimes
\bpm \eps & \\ & \frac 1 \eps \epm
\ket W
=
\eps \ket W,
\end{equation}
and when $\eps$ goes to zero, we approaches the zero vector in the Hilbert space.
However, $0$ is \emph{not} an element of the orbit $G \cdot \ket{W}$ (in fact, $\{0\}$ is an orbit on its own).

Although so far we have only proved the converse, this observation is in fact enough to conclude that there exists no quantum state in the W class which is locally maximally mixed.
More generally, we have the following fundamental result in geometric invariant theory:

\begin{thm}[Kempf-Ness]
  The following are equivalent:
  \begin{itemize}
  \item There exists a vector of minimal norm in $G \cdot \ket{\phi_{ABC}}$.
  \item There exists a quantum state in the class $G \cdot \ket{\phi_{ABC}}$ with $\rho_A = \rho_B = \rho_C \propto I/d$.
  \item $G \cdot \ket{\phi_{ABC}}$ is closed.
  \end{itemize}
\end{thm}


How about if we look at the closure of the W class? States in the closure of class are those which can be approximated arbitrarily well by states from the class. Thus they can in practice be used for the same tasks as the class itself, as long as the task is ``continuous''.

\smallskip

It is a fact that the closure of any orbit $G \cdot \ket{\phi}_{ABC}$ is a disjoint union of orbits, among which there is a \emph{unique} closed orbit. There are two options: Either this orbit $\{0\}$, or it is the orbit through some proper (unnormalized) quantum state. Therefore:

\begin{cor}
  There exists a quantum state in the closure of the entanglement class $G \cdot \ket{\phi_{ABC}}$ that is locally maximally mixed if, and only if, $0 \notin \overline{G \cdot \ket{\phi_{ABC}}}$.
\end{cor}

We saw before that 0 is in the closure of the W class. Therefore, the corollary shows that we cannot even approximate a locally maximally mixed state by states from the W class. This agrees with \cref{fig:w polytope}, which shows that the set of eigenvalues that are compatible with the closure of the W class does not contain the locally maximally mixed point (the ``origin'' in the figure).

\subsection{Invariant polynomials}

If we have two closed sets -- such as $\{0\}$ and an orbit closure $\overline{G \cdot \ket{\phi}}$ which not contain the origin -- then we can always find a continuous function which separates these sets. Since both sets are $G$-invariant and we are working in the realm of algebraic geometry, we can in fact choose this function to be a $G$-invariant homogeneous polynomial $P$, such that $P(0) = 0$ and $P(\ket\psi) \neq 0$.\footnote{There is a slight subtlety in that there are two interesting topologies that we may consider when we speak of the ``closure'': the standard topology, induced by the any norm on our Hilbert space, and the Zariski topology, for which the separation result is true. In general, the Zariski closure is larger than the norm closure. But in the case of $G$-orbit closures there is no difference.}
The converse is obviously also true, and so we find that:

\begin{thm}
  There exists a quantum state in the closure of the entanglement class $G \cdot \ket{\phi_{ABC}}$ that is locally maximally mixed if, and only if, there exists a non-constant $G$-invariant homogeneous polynomial such that $P(\ket{\phi_{ABC}}) \neq 0$.
\end{thm}

At first sight, this new characterization does not look particularly useful, since we have to check all $G$-invariant homogeneous polynomials. However, these invariant polynomials form a finitely generated algebra, and so we only have to check a finite number of polynomials.
For three qubits, e.g., there is only a single generator: Every $G$-invariant polynomial is a linear combination of powers of \emph{Cayley's hyperdeterminant}
\begin{align*}
  P(\ket\psi)
  &= \psi_{000}^2\psi_{111}^2
  + \psi_{100}^2\psi_{011}^2
  + \psi_{010}^2\psi_{101}^2
  + \psi_{001}^2\psi_{110}^2 \\
  &- 2 \psi_{000}\psi_{111}\psi_{100}\psi_{011}
  - 2 \psi_{000}\psi_{111}\psi_{010}\psi_{101}
  - 2 \psi_{000}\psi_{111}\psi_{001}\psi_{110} \\
  &- 2 \psi_{100}\psi_{011}\psi_{010}\psi_{101}
  - 2 \psi_{100}\psi_{011}\psi_{001}\psi_{110}
  - 2 \psi_{010}\psi_{101}\psi_{001}\psi_{110} \\
  &+ 4 \psi_{000}\psi_{110}\psi_{101}\psi_{011}
  + 4 \psi_{111}\psi_{001}\psi_{010}\psi_{100}.
\end{align*}
It is non-zero precisely on the quantum states of GHZ class, which can be verified by plugging in representative states of all six classes.

\medskip

We conclude this lecture with some remarks.
The characterization in terms of invariant polynomials brings us into the realm of \emph{representation theory}. Indeed, the space of polynomials on the Hilbert space is a $G$-representation, and the invariant polynomials are precisely the trivial representations contained in it.

It is natural to ask about the meaning of the other irreducible representations. It turns out that, in the same way that the trivial representations correspond to locally maximally mixed states (i.e., local eigenvalues $1/d,\dots,1/d$), the other irreducible representations correspond to the other spectra $(\lambda_A,\lambda_B,\lambda_C)$ that are compatible with the class.
Although we do not have the time to discuss this, this can also be proved using the techniques we have discussed in this lecture.
%
As a direct corollary, one can show that the solution to the quantum marginal problem for the closure of an entanglement class is always convex. It is in fact a convex polytope, which we might call the \emph{entanglement polytope} of the class. Thus, the green polytope in \cref{fig:w polytope} is nothing but the entanglement polytope of the W class. The study of these polytopes as entanglement witnesses was proposed in [Walter, Doran, Gross, Christandl; \href{http://arxiv.org/abs/1208.0365}{arXiv:1208.0365}].

Each entanglement polytope is a subset of the polytope of spectra that we have discussed in the context of the pure-state quantum marginal problem in \cref{sec:pure state qmp}.
However, we can always choose to ignore the entanglement class in the above discussion!
If we do so then we obtain a representation-theoretic characterization of the latter polytopes, i.e.\ of the solution of the pure-state quantum marginal problem.
In \cref{lec:kron}, Matthias will discuss an alternative way of arriving at this characterization that starts directly with representation theory rather than geometry.

%% file: 30may_3.tex
\lecture{12}{30 May, 2013}{Aram Harrow}{High dimensional entanglement}

Today, I will tell you about bizarre things that can happen with entanglement of high dimensional quantum states. Recall from \cref{lec:complex sep} that
$$h_{\Sep}(M)=\max_{\sigma \in \SEP} \tr M \sigma$$
where $\{M, I - M\}$ are the yes/no outcomes of a POVM. He also showed that it is $\NP$-hard to compute this quantity exactly in general. So, here we want to consider approximations to this quantity that we can compute easier.

For this we introduce approximations to the set of separable states based on the concept of $n$-extendibility.
Let $\rho_{AB}$ be a density matrix on $\CC^{d_A} \otimes \CC^{d_B}$.
We say that $\rho_{AB}$ is \emph{(symmetrically) $n$-extendible} if there exists a state $\widetilde{\rho}_{AB_1 \cdots B_n}$ on $\CC^{d_A}\otimes \Sym^n (\CC^{d_B})$ such that $$\rho_{AB}= \tr_{B_2\cdots B_n} ( \widetilde{\rho}_{AB_1 \cdots B_n} ).$$

It turns out that the set of $n$-extendible states is a good outer approximation to the set of separable states that gets better and better as $n$ increases. But let us first check that the set of separable states is contained in it, that is, that every separable $\rho_{AB}$ is $n$-extendible. This can be seen by writing the separable state $\rho_{AB}$ in the form $\sum_i p_i \, \proj{\alpha_i} \otimes \proj{\beta_i}$. A symmetric extension is then given by $ \widetilde{\rho}_{AB_1 \cdots B_n} = \sum_i p_i \, \proj{\alpha_i} \otimes \proj{\beta_i}^{\otimes n}$. The following theorem shows that the $n$-extendible states are indeed an approximation of $\Sep$:

\begin{thm}
\label{thm:ext}
If $\rho_{AB}$ is $n$-extendible, then there is a separable state $\sigma$ with $\frac{1}{2}||\rho-\sigma||_1\leq \frac{d}{n}$.
\end{thm}

The proof of this theorem is very similar to the proof of the quantum de Finetti theorem which we did yesterday (in fact, you could adapt the proof as an exercise if you wish).

As a corollary it now follows that we can approximate $h_{SEP}(M)$ by
$$h_{\mathrm{n-ext}}(M):= \max_{\rho \text{ n-ext}} \tr M \rho.$$

\begin{cor} For all $0 \leq M \leq I$,
$$h_{\Sep}(M) \leq h_{\mathrm{n-ext}}(M)\leq h_{\Sep}(M) + \frac{d}{n}.$$
\end{cor}

The lower bound follows directly from the fact that the set of separable states is contained in the set of $n$-extendible states (it even holds for all Hermitian $M$ without the restriction $0 \leq M \leq id$.
For the upper bound, we use the observation that
$$ \max_{0\leq M \leq I} \tr M (\rho-\sigma)= \frac{1}{2} ||\rho-\sigma||_1$$
and obtain
$$h_{\mathrm{n-ext}}(M)= \max_{\rho \text{ n-ext}} \tr M \rho \leq \max_{\sigma \in \Sep} \tr M \sigma + \frac{d}{n} .$$

 We now want to see how difficult it is to compute $h_{\mathrm{n-ext}}(M)$. We rewrite $ h_{\mathrm{n-ext}}(M)$ in the form
 \begin{align*}
 &\quad h_{\mathrm{n-ext}}(M)
 = \max_{\ket{\psi} \in \CC^{d_A}\otimes \Sym^n(\CC^{d_B})} \bra{\psi} M \otimes I^{\otimes (n-1)}\ket{\psi}\\
 &=\lambda_{\max}[(I_A \otimes \Psym^{d_B,n}) (M \otimes I^{\otimes (n-1)}) (I_A \otimes \Psym^{d_B,n})].
 \end{align*}

Hence, the effort to compute $ h_{\text{n-ext}}(M)$ is polynomial in $d^{n+1}$. In order to obtain an $\eps$ approximation to  $h_{sep}(M) $ we have to choose $\eps= d/n$ according to the corollary. Hence the effort to approximate up to accuracy $\eps$ then the effort scales as $d^{n/\eps}$.

Actually this is optimal for general $M$. In order to see why, we are going to employ a family of quantum states known as the antisymmetric states (it is also known as the universal counterexample to any conjecture in entanglement theory which you may have). The antisymmetric state comes in a pair with the symmetric state:

The \emph{symmetric state} is
$$\rho_{\text{sym}}= \frac{\Psym^{d, 2}}{d(d+1)/2} = \frac{I + F}{d(d+1)}.$$
It is separable, because $\frac{\Psym^{d, 2}}{d(d+1)/2}= \int d\phi \, \proj{\phi}^{\otimes 2}$, as we saw in \cref{lec:monogamy}.

The \emph{antisymmetric state} is
$$\rho_{\text{anti}}= \frac{I - \Psym^{d, 2}}{d(d-1)/2} = \frac{I - F}{d(d-1)}$$
This antisymmetric state it funny because
\benum
\item it is very far from separable: for all separable $\sigma$: $\frac{1}{2}\lVert\rho_{\text{anti}}-\sigma\rVert_1 \geq \frac{1}{2}$, but
\item it is also very extendible: more precisely, two copies $\rho_{\text{anti}}\otimes \rho_{\text{anti}}$ are $(d-1)$-extendible.
\eenum

Let us first see why 1.\ holds. For this, let $M=\Psym^{d, 2}$. Then, $\tr M \rho_{\text{anti}} = 0$, since the symmetric and the antisymmetric subspace are orthogonal. On the other hand,
$$\tr M \sigma = \tr (\sigma/2 + F\sigma/2)= \frac{1}{2} + \frac{1}{2} \tr F \sigma.$$
In order to bound $ \tr F \sigma$ note that
$$\tr F (X \otimes Y)=\sum_{i,j} \braket{ij | F (X \otimes Y) | ij} = \sum_{i,j} \braket{ji | X \otimes Y | ij}= \sum_{ij} X_{ji}Y_{ij}=\tr XY.$$
Hence, if $\sigma = \sum_i p_i \, \proj{\alpha_i} \otimes \proj{\beta_i}$ is a separable state then
\begin{align*}
\tr M \sigma  =  \frac{1}{2} + \frac{1}{2} \sum_i p_i \, \tr F \proj{\alpha_i} \otimes \proj{\beta_i}
 =\frac{1}{2} + \frac{1}{2} \sum_i p_i \, \lvert\braket{\alpha_i | \beta_i}\rvert^2\geq \frac{1}{2}.
\end{align*}

In order to see that 2.\ holds, note that
$$\rho_{\text{anti}}= \frac{2}{d(d+1)} \sum_{1 \leq i<j\leq d} \frac{\ket{ij}-\ket{ji}}{\sqrt{2}}\frac{\bra{ij}-\bra{ji}}{\sqrt{2}}.$$
Consider now the following state, known as a \emph{Slater determinant},
$$\ket{\psi}= \frac{1}{d!} \sum_{\pi \in S_d} \sgn(\pi) \, \ket{\pi(1)} \otimes \cdots \otimes \ket{\pi(n)},$$
where we introduced the sign of a permutation
$$\sgn(\pi)= (-1)^{L}= \det(\sum_i^n \ket{\pi(i)}\!\bra{i}),$$
with $L$ the number of transpositions in a decomposition of the permutation $\pi$.
A quick direct calculation shows that the Slater determinant extends $\rho_{\text{anti}}$, i.e.\ that
$$\tr_{3 \cdots d} \proj{\psi} = \sum_{i_3 \cdots i_n} (I \otimes \bra{i_3 \cdots i_n}) \proj{\psi} (I \otimes \ket{i_3 \cdots i_n}) = \rho_{\text{anti}}.$$
Note that the extension we constructed was actually antisymmetric! But if we take two copies of the antisymmetric state, the negative signs cancel out:
$$\ket{\psi} \otimes \ket{\psi} \in \Sym^n(\CC^d \otimes \CC^d)$$
is the desired symmetric $(d-1)$-extension of $\rho_{\text{anti}} \otimes \rho_{\text{anti}}$.

\begin{ex*}
Use 1.\ and 2.\ to show that the upper bound of \cref{thm:ext} is essentially tight.
\end{ex*}

%% file: 30may_ex.tex
\problemsession{3}{30 May, 2013}{Michael Walter}

\begin{ex}[Tsirelson's bound]
\label{ex:tsirel}

In Fernando's \cref{lec:complex sep} on Thursday you have seen that a quantum strategy for the CHSH game can reach a winning probability of $\frac{1}{2}(1+\frac{1}{\sqrt{2}}) \approx 0.85$. It is the goal of this exercise to prove this is optimal. That is, there does not exist a quantum strategy that reaches a value higher than $\frac{1}{2}(1+\frac{1}{\sqrt{2}})$. This result is known as Tsirelson's bound.

\smallskip\noindent\emph{Hint: Show first that the claim is equivalent to showing
\begin{equation}
\label{tsirel}
  \max_{A_0, B_0, A_1, B_1}\max_{\lVert\psi\rVert=1} \bra{\psi} A_0\otimes B_0+A_1\otimes B_0+A_0\otimes B_1-A_1\otimes B_1 \ket{\psi}\leq 2\sqrt{2}.
\end{equation}
where the maximization over $A_0, B_0, A_1, B_1$ is over square matrices with eigenvalues $\{-1, 1 \}$.
Note that the left-hand side is the operator norm of the ``Bell operator'' $A_0\otimes B_0+A_1\otimes B_0+A_0\otimes B_1-A_1\otimes B_1$ (optimized over choices of observables $A_0, B_0, A_1, B_1$). Use properties of the norm and the explicit form of the matrices appearing in the Bell operator in order to conclude the proof. The calculation involves a few steps, but I am sure you can do it :)}

\begin{sol}
  Let us consider a quantum strategy where Alice and Bob share a pure quantum state $\psi_{AB}$. On input $r$, Alice performs a projective measurement $\{A_r^a\}$, where $a$ labels her output bit. Similarly, on input $s$, Bob performs a projective measurement $\{B_s^b\}$, labeled by his output bit $b$.
  (Exercise: Why is it enough to restrict to pure states and projective measurements?)
  Let us define corresponding observables $A_r = A_r^0 - A_r^1$ and $B_s = B_s^0 - B_s^1$. Then,
  \begin{align*}
      &\bra{\psi} A_0 \otimes B_0 + A_1 \otimes B_0 + A_0 \otimes B_1 - A_1 \otimes B_1 \ket{\psi} \\
    = &(p(00|00) + p(11|00) - p(01|00) - p(10|00))
    + (p(00|10) + p(11|10) - p(01|10) - p(10|10)) \\
    + &(p(00|01) + p(11|01) - p(01|01) - p(10|01))
    - (p(00|11) + p(11|11) - p(01|11) - p(10|11)) \\
    = &\sum_{r,s} p_{\text{win}}(rs) - p_{\text{lose}}(rs)
    = 4 ( p_{\text{win}} - p_{\text{lose}} )
    = 4 ( 2 p_{\text{win}} - 1 )
  \end{align*}
  Thus, $p_{\text{win}} \leq \frac{1}{2}(1+\frac{1}{\sqrt{2}})$ is indeed equivalent to the inequality~\eqref{tsirel}
  To prove~\eqref{tsirel}, we use the Cauchy-Schwarz and triangle inequalities to obtain
  \begin{align*}
    &\quad \bra\psi A_0 \otimes B_0 + A_1 \otimes B_0 + A_0 \otimes B_1 - A_1 \otimes B_1 \ket\psi \\
    &\leq \lVert \left( A_0 \otimes B_0 + A_1 \otimes B_0 + A_0 \otimes B_1 - A_1 \otimes B_1 \right) \ket\psi \rVert \\
    &\leq \lVert \left( A_0 \otimes (B_0 + B_1) \right) \ket\psi \rVert + \lVert \left( A_1 \otimes (B_0 - B_1) \right) \ket\psi \rVert \\
    &= \lVert \left( I \otimes (B_0 + B_1) \right) \ket\psi \rVert + \lVert \left( I \otimes (B_0 - B_1) \right) \ket\psi \rVert \\
    &= \lVert \ket{\psi_0} + \ket{\psi_1} \rVert + \lVert \ket{\psi_0} - \ket{\psi_1} \rVert,
  \end{align*}
  where $\ket{\psi_s} := \left( I \otimes B_s \right) \ket\psi$ are vectors of norm $\leq 1$. Note that
  \begin{equation*}
    \lVert \ket{\psi_0} + \ket{\psi_1} \rVert + \lVert \ket{\psi_0} - \ket{\psi_1} \rVert
    \leq \sqrt{2 + 2 \Re \langle{\psi_0}\vert{\psi_1}\rangle} + \sqrt{2 - 2 \Re \langle{\psi_0}\vert{\psi_1}\rangle}
    = \sqrt{2 + 2x} + \sqrt{2 - 2x}
  \end{equation*}
  for some $x \in [-1,1]$. By optimizing over all $x$ we get the desired upper bound.
\end{sol}

\end{ex}

\bigskip

\begin{ex}[Entanglement classes]
\label{ex:entanglement classes}

Matthias mentioned in \cref{lec:exact} on Wednesday that every three-qubit state $\ket{\psi}$ belongs to the entanglement class of one of the following six states:

\begin{align*}
&\ket{000}_{ABC}, \ket{\Phi^+}_{AB} \otimes \proj 0_C, \quad
\ket{\Phi^+}_{AC} \otimes \proj 0_B, \quad
\ket{\Phi^+}_{BC} \otimes \proj 0_A, \\
&\ket{GHZ}_{ABC}=\frac{1}{\sqrt{2}}(\ket{000}+\ket{111}),
\ket{W}_{ABC}=\frac{1}{\sqrt{3}}(\ket{100}+\ket{010}+\ket{001})
\end{align*}

It is the goal of this exercise to prove this. This means, we want to show that for all  $\ket{\psi} \in \CC^2 \otimes \CC^2 \otimes \CC^2$ there exist invertible two-by-two matrices $a, b, c$ such that $a \otimes b\otimes c \ket{\psi}$ equals one of the six states above.

\smallskip\noindent\emph{Hint: Do a case by case analysis where the different cases correspond to the ranks of the reduced density matrices of $\ket{\psi}$ (which are either one or two). Start with the easy cases, where at least one of the single particle reduced density matrices has rank one. When all single-particle ranks equal to two, things are a little more tricky. Here, use the Schmidt decomposition between $A$ and $BC$ and the fact (which you may also easily prove) that the range of the density operator on system $BC$ contains either one or two product vectors.}

\begin{sol}[Sketch of Solution]
  Suppose that $\rho_C$ has rank 1 (i.e.\ it is a pure state). Then it follows from the Schmidt decomposition that $\ket{\psi}_{ABC} = \ket{\psi}_{AB} \otimes \ket{\psi}_C$. If $\ket{\psi}_{AB}$ is a product state then $\ket{\psi}_{ABC}$ belongs to the class of $\ket{000}_{ABC}$. Otherwise, if $\ket{\psi}_{AB}$ is entangled then it can be obtained from an EPR pair by SLOCC (consider the Schmidt decomposition), hence $\ket{\psi}_{ABC}$ belongs to the class of $\ket{\Phi^+}_{AB} \otimes \proj 0_C$. We can similarly analyze the case where $\rho_B$ or $\rho_A$ have rank 1.
  The four classes thus obtained are all different, since the local rank is invariant under invertible SLOCC operations (exercise).

  It remains to analyze the case where all single-particle ranks are equal to 2. For this, consider the Schmidt decomposition
  \begin{equation*}
    \ket{\psi_{ABC}} = \ket{\psi^1_A} \otimes \ket{\psi^1_{BC}} + \ket{\psi^2_A} \otimes \ket{\psi^2_{BC}}.
  \end{equation*}
  Let $V$ be the two-dimensional vector space spanned by the $\ket{\psi^k_{BC}}$.
  Our approach to distinguishing between the remaining classes is to consider the \emph{tensor rank} of $\psi_{ABC}$, i.e.\ the minimal number of product vectors into which the state can be decomposed. For example, the tensor rank of the GHZ state is 2. More generally, the tensor rank of the state $\ket{\psi_{ABC}}$ can be 2 only if there are at least two product vectors in $V$. Thus we are lead to study the number of product vectors in $V$.

  Note that there is always at least one product vector in $V$. To see this, observe that $\ket{\phi_{BC}} = \sum \phi_{i,j} \ket{ij_{BC}}$ is a tensor product if and only if the determinant of its coefficient matrix $(\phi_{i,j})$ is non-zero. Thus, we need to find zeros of determinant of the state
  \begin{equation*}
    X \ket{\psi^1_{BC}} + Y \ket{\psi^2_{BC}}.
  \end{equation*}
  This is a non-constant homogeneous polynomial in $X$ and $Y$ (or the zero polynomial), and therefore always has a non-trivial zero. For example, for the $W$ state, where $\psi^1_{BC} = (\ket{10} + \ket{01})/\sqrt{2}$ and $\psi^2_{BC} = \ket{00}$, product vectors correspond to zeros of the polynomial
  \begin{equation*}
    \det \left( \frac X {\sqrt 2} \bpm 0 & 1 \\ 1 & 0\epm + Y \bpm 1 & 0 \\ 0 & 0 \epm \right)
    = \det \bpm Y & \frac X {\sqrt 2} \\ \frac X {\sqrt 2} & 0\epm
    = - \frac {X^2} 2.
  \end{equation*}
  Thus there is only a single linearly independent product vector in $V$. By what we saw above, it follows that the tensor rank of the W state is at least three. In particular, the W and the GHZ class are inequivalent.

  \emph{Case 1:} Suppose that there are (at least) two product vectors in $V$, say $\ket{\phi^1_B} \otimes \ket{\phi^1_C}$ and $\ket{\phi^2_B} \otimes \ket{\phi^2_C}$. Denote by $\ket{\xi^1_{BC}}$ and $\ket{\xi^2_{BC}}$ the ``dual basis'' in $V$, i.e.\
  $\langle \xi^i_{BC} \vert \phi^j_B \otimes \phi^j_C \rangle = \delta_{i,j}$. Then
  \begin{equation*}
    \ket{\psi_{ABC}} =
    \langle \xi^1_{BC} \vert \psi_{ABC} \rangle \otimes \ket{\phi^1_B} \otimes \ket{\phi^1_C} +
    \langle \xi^2_{BC} \vert \psi_{ABC} \rangle \otimes \ket{\phi^2_B} \otimes \ket{\phi^2_C},
  \end{equation*}
  which is of GHZ type.

  \emph{Case 2:} Suppose that there is only a single product vector in $V$. If we write
  \begin{equation*}
    \ket{\psi_{ABC}} =
    \ket{\phi^1_A} \otimes \ket{\phi^1_B} \otimes \ket{\phi^1_C} +
    \ket{\phi^2_A} \otimes \ket{\phi^2_{BC}}
  \end{equation*}
  with $\ket{\phi^2_{BC}}$ an entangled state orthogonal to $\ket{\phi^1_B} \otimes \ket{\phi^1_C}$, then it can be shown that this vector is of $W$ type: Suppose for simplicity of notation that $\ket{\phi^1_B} = \ket{\phi^1_C} = \ket{0}$ (we can always achieve this by using a local unitary). Then $\ket{\phi^2_{BC}}$ is a linear combination of the other computational basis states, $\ket{01}_{BC}$, $\ket{10}_{BC}$ and $\ket{11}_{BC}$. Finally, the assumption that there is only a single product vector in $V$ implies that there is in fact no contribution of $\ket{11}_{BC}$ (consider the corresponding ``determinant polynomial'').
  Thus, $\ket{\psi_{ABC}}$ is of the form
  $$
    \ket{\phi^1_A} \otimes \ket{00_{BC}} +
    \ket{\phi^2_A} \otimes \left( \gamma \ket{10_{BC}} + \delta \ket{01_{BC}} \right)
  $$
  After another rotation that maps $\ket{\phi^2_A}$ to $\ket 0$, we arrive at a state of the form
  \begin{align*}
    &\left( \alpha \ket{0_A} + \beta \ket{1_A} \right) \otimes \ket{00_{BC}} +
    \ket{0_A} \otimes \left( \gamma \ket{10_{BC}} + \delta \ket{01_{BC}} \right) \\
    = &\alpha \ket{000}_{ABC} + \beta \ket{100}_{ABC} + \gamma \ket{010}_{ABC} + \delta \ket{001}_{ABC}.
  \end{align*}
  which is certainly in the W class.
\end{sol}

\end{ex}

\bigskip

\begin{ex}[Entanglement polytope of W class]
\label{ex:W polytope}
In \cref{lec:entanglement polytopes} on Thursday, Michael discussed the polytopes associated to the different three-qubit entanglement classes. In particular, he noted that states of the W class obey the following eigenvalue inequality:
$$\lambda_A^{\max}+\lambda_B^{\max}+\lambda_C^{\max}\geq 2$$
As a warmup, show that this inequality is violated for the $\ket{GHZ}$. Then show that the inequality holds for all states in the W class.

\begin{sol}
  The GHZ state has maximal local eigenvalues $\lambda_A^{\max} = \lambda_B^{\max} = \lambda_C^{\max} = 0.5$, and hence violates the inequality. Now consider an arbitary state in the W class, which we can always write in the form
  \begin{equation*}
    \ket{\psi_{ABC}} = \alpha \ket{000} + \beta \ket{100} + \gamma \ket{010} + \delta \ket{001}
  \end{equation*}
  for an orthogonal basis $\ket 0, \ket 1$.
  By the variational principle,
  \begin{equation*}
    \lambda_A^{\max}+\lambda_B^{\max}+\lambda_C^{\max}
    = \max_{\phi_A,\dots,\phi_C} \langle \psi_{ABC} | \underbrace{\proj{\phi_A} \otimes I_B \otimes I_C + \dots}_{=M} | \psi_{ABC} \rangle
  \end{equation*}
  The operator $M$ is positive semidefinite, with eigenvalues $0,\dots,3$. The unique eigenvector for eigenvalue $3$ is
  $\ket{\phi_A,\phi_B,\phi_C}$, and the eigenspace for eigenvalue $2$ is spanned by the vectors
  $\ket{\phi_A^\perp,\phi_B,\phi_C}$, $\ket{\phi_A,\phi_B^\perp,\phi_C}$, $\ket{\phi_A,\phi_B,\phi_C^\perp}$.
  Thus, by choosing $\ket{\phi_A} = \ket{\phi_B} = \ket{\phi_C} = \ket{0}$ we find that
  \begin{equation*}
    \lambda_A^{\max}+\lambda_B^{\max}+\lambda_C^{\max}
    \geq \lvert\alpha\rvert^2 3 + (1 - \lvert\alpha\rvert^2) 2 \geq 2. \qedhere
  \end{equation*}
\end{sol}
\end{ex}

\bigskip

\begin{ex}[Secret bit]

An important tool in Aram's \cref{lec:de finetti proof} on Wednesday on the security of quantum key distribution was the following observation:
If the reduced density matrix of Alice and Bob's system is in a pure state: $\tr \proj{\psi}_{ABE}= \proj{\phi}_{AB}$, then $\ket{\psi}_{ABE}= \ket{\phi}_{AB}\ket{\gamma}_E$ for some $\ket{\gamma}$. Hence, Eve is completely decoupled from Alice and Bob. Prove this statement.

\begin{sol}
  Use the Schmidt decomposition.
\end{sol}

Assume now that $\ket{\phi}_{AB}=\frac{1}{\sqrt{2}} (\ket{00}+\ket{11})$ is an EPR pair. Let Alice, Bob measure in the $\{\ket{0}, \ket{1}\}$ basis and Eve with an arbitrary POVM (and denote the outcomes by $x, y, z$, respectively). Show that the joint probability distribution of the outcomes is of the form
$$p(x, y, z)= \frac{1}{2} \delta_{xy} q(z).$$
Thus, Alice and Bob's outcomes are maximally correlated, but uncorrelated to Eve's.

\begin{sol}
  Let us denote the elements of Eve's POVM by $\{M_z\}$. Then,
  \begin{align*}
    p(x,y,z)
    = \bra{\phi_{AB} \otimes \gamma_E} \left( \proj x \otimes \proj y \otimes M_z \right) \ket{\phi_{AB} \otimes \gamma_E}
    = \underbrace{\lvert \langle \phi_{AB} | x y \rangle \rvert^2}_{=\frac 1 2 \delta_{xy}} \,
    \underbrace{\bra{\gamma_E} M_z \ket{\gamma_E}}_{=: q(z)}.
  \end{align*}
\end{sol}

\end{ex}

%% file: 31may_1.tex
\lecture{13}{31 May, 2013}{Fernando G.S.L. Brand\~ao}{LOCC distinguishability}

\section{Data hiding}

\paragraph{Review of bad news from yesterday.}
\benum
\item The weak membership problem (that is, determining whether
  $\rho_{AB}\in\Sep$ or $D(\rho_{AB}, \Sep)\geq \eps$ given the
  promise that one of these holds) is $\NP$-hard for $\eps =
  1/\poly(\dim)$.
\item $k$-extendability  does not give a good approximation in trace
  norm until $k \geq d$, which corresponds to an algorithm that takes
  time exponential in $d$.
\eenum

Let's look more closely at what went wrong with using $k$-extendable
states to approximate $\Sep$. We considered the anti-symmetric state on $\CC^d \otimes \CC^d$,
$$W^-_{AB} = \rho_{\text{anti}} = \frac{I-F}{d(d-1)}.$$
It is (anti-symmetrically) $k$-extendible for $k=d-1$, and satisfies
$\min_{\sigma\in\Sep}
\frac 1 2 \|W^-_{AB} - \sigma_{AB}\|_1 = \frac 1 2$.

Here the trace distance describes our ability to distinguish $W^-_{AB}$ and
$\sigma$ using arbitrary two-outcome measurements $\{M,I-M\}$
satisfying only $0 \leq M \leq I$.  However, since arbitrary
measurements can be hard to implement, it is often reasonable to
consider the smaller class of measurements that can be implemented
with LOCC.

\paragraph{Locality-restricted measurements}
\bit
\item Define the LOCC norm to be
$$\frac 1 2 \|\rho_{AB} - \sigma_{AB} \|_{\text{LOCC}} :=
\max_{\substack{0 \leq M \leq I\\ \{M,I-M\}\in \text{LOCC}}}
|\tr (M(\rho-\sigma))|.$$
\item Define the 1-LOCC norm to be analogous, but with LOCC replaced
  with 1-LOCC.  This stands for ``one-way LOCC.''  This means that one
  party (by convention, Bob) makes a measurement $(B_k)$, sends the outcome $k$ to
  Alice and she makes a measurement $A_k$ based on this message. The
  resulting operation has the form
  $$M = \sum_k A_k \ot B_k,$$
  where $B_k \geq 0$ for all $k$, $\sum_k B_k = I$, and $0 \leq A_k \leq I$ for all k.
\eit

Is $W^-_{AB}$ still far from $\Sep$ in the LOCC norm?  Observe that
$$
\min_{\sigma\in\Sep}\frac 1 2 \|W^-_{AB} - \sigma_{AB}\|_{\text{LOCC}} \leq
\frac 1 2 \|W^-_{AB} - W^+_{AB}\|_{\text{LOCC}},$$
since $W^+_{AB} := \frac{I+F}{d(d+1)} = \int d\ket\theta\, \proj\theta^{\ot
  2}$ is separable.

Now if $\{M,I-M\}\in\text{LOCC}$ then we can decompose $M = \sum_k A_k
\ot B_k$ as well as $I-M = \sum_k A_k' \ot B_k'$ with each $A_k, B_k, A_k',
B_k' \geq 0$.  In particular, $0 \leq M^{T_A} \leq I$.
We can then further relax
\bas &\quad \frac 1 2 \|W^-_{AB} - W^+_{AB}\|_{\text{LOCC}}
\leq
\max_{\substack{0 \leq M \leq I \\ 0 \leq M^{T_A} \leq I}}
\tr (M(W^+_{AB} - W^-_{AB}))
\\ & =
\max_{\substack{0 \leq M \leq I \\ 0 \leq M^{T_A} \leq I}}
\tr (M^{T_A}((W^+_{AB})^{T_A} - (W^-_{AB})^{T_A}))
\leq
\frac 1 2 \|(W^+_{AB})^{T_A} - (W^-_{AB})^{T_A}\|_1.\eas

To evaluate this last quantity, observe that $F^{T_A} = d \, \proj{\Phi^+}$ -- the partial transpose of the swap operator is proportional to a maximally entangled state. Thus
\bas
(W^-_{AB})^{T_A} &= \L(\frac{I-F}{d(d-1)}\R)^{T_A}
= \frac{I - F^{T_A}}{d(d-1)}
= \frac{I - d\Phi^+}{d(d-1)}
\\
(W^+_{AB})^{T_A} &= \L(\frac{I+F}{d(d+1)}\R)^{T_A}
= \frac{I + F^{T_A}}{d(d+1)}
= \frac{I + d\Phi^+}{d(d+1)}
\eas
Now we can calculate
\bas
&\quad \frac 1 2 \L\|(W^-_{AB})^{T_A} - (W^+_{AB})^{T_A}\R\|_1
= \frac 1 2
\L\| \frac{I-d\Phi^+}{d(d-1)} - \frac{I + d \Phi_+}{d(d+1)}\R\|_1 \\
&= \frac 1 2
\L\| \frac{I}{d(d-1)(d+1)} - \frac{\Phi_+}{(d-1)(d+1)}\R\|_1
\leq \frac{1}{d}.
\eas

This is an example of {\em data hiding}.  The states $W^+_{AB}, W^-_{AB}$ are
perfectly distinguishable with global measurements, but can only be
distinguished with bias $\leq 1/d$ using LOCC measurements.

\section{Better de Finetti theorems for 1-LOCC measurements}
This data hiding example raises the hope that a more useful version of
the de Finetti theorem might hold when we look at 1-LOCC
measurements.  Indeed, we will see that the following improved de
Finetti theorem does hold:
\begin{thm}\label{thm:BCY}
If $\rho_{AB}$ is a $k$-extendible state on $\CC^{d_A} \ot \CC^{d_B}$ then
\bas \min_{\sigma\in\Sep}
\|\rho_{AB} - \sigma_{AB}\|_{\mathrm{1-LOCC}} \leq
\sqrt{\frac{2\ln(2) \log(d_A)}{k}}.\eas
\end{thm}
This was first proved in [Brand\~ao, Christandl, Yard;
\href{http://arxiv.org/abs/1010.1750}{arXiv:1010.1750}], but in Aram's lecture you will see a simpler proof
from [Brand\~ao, Harrow; \href{http://arxiv.org/abs/1210.6367}{arXiv:1210.6367}].
It can be shown [Matthews, Wehner, Winter; \href{http://arxiv.org/abs/0810.2327}{arXiv:0810.2327}] that
$$
\|\rho_{AB} - \sigma_{AB}\|_{\text{1-LOCC}} \geq
\frac{1}{\sqrt{127}} \|\rho_{AB}-\sigma_{AB}\|_2 :=
\frac{1}{\sqrt{127}} \sqrt{\tr((\rho_{AB}-\sigma_{AB})^2)}.$$
Thus, \cref{thm:BCY} also gives a good approximation in the 2-norm.

\paragraph{Application to weak membership.}
Let's consider the weak membership problem for $\Sep$, but now with
the distance measure given by 1-LOCC norm:
$$D(\rho,\Sep) := \min_{\sigma \in \Sep}
\|\rho - \sigma\|_{\text{1-LOCC}}.$$

We will solve this problem using {\em semidefinite programming} (SDP),
which means optimizing a linear function over matrices subject to
semidefinite constraints (i.e., constraints that a given matrix is
positive semidefinite).  Algorithms are known that can solve SDPs in
time polynomial in the number of variables.
The SDP for checking whether $\rho_{AB}$ is
$k$-extendable is to search for a $\pi_{AB_1,\dots,B_k}$ satisfying
\be
\begin{aligned}
\label{eq:sdp}
 \pi_{AB_1,\dots,B_k} & \geq 0, \\
\pi_{AB_j} &= \rho_{AB} \quad(\forall j).
\end{aligned}
\ee
The algorithm is to run the SDP for $k = \frac{4\ln(2)
  \log(d_A)}{\eps^2}$.  If $\rho\in\Sep$ then the SDP will be feasible
because $\rho$ is also $k$-extendable.  The harder case is to show
that the SDP is infeasible when $D(\rho_{AB},\Sep) \geq \eps$.  But
this follows from \cref{thm:BCY}.
(To be precise, the SDP~\eqref{eq:sdp} amounts to checking for permutation-invariant extensions $\rho_{AB^k}$, for which \cref{thm:BCY} also holds, rather than symmetric extensions, which is the same distinction as between permutation-invariant and permutation-symmetric states that we discussed in \cref{subsec:perm inv}.)

The run time is polynomial in $d_A d_B^{k+1}$, which is dominated by
the $d_B^k$ term.  This is
$$\exp(c\log(d_A)\log(d_B) / \eps^2),$$
which is slightly more than polynomial-time.  It is called
``quasi-polynomial,'' meaning that it is  $\exp(\poly(\log(\text{input size})))$.

\paragraph{Idea behind proof of \cref{thm:BCY}.}
Suppose we had a ``magic'' entanglement measure: $E:\mathcal D(\CC^{d_A} \ot
\CC^{d_B}) \mapsto \RR_+$ with the following properties.
\benum
\item {\bf Normalization:} $E(\rho_{AB}) \leq \min(\log(d_A), \log(d_B))$.
\item {\bf Monogamy:} $E(\rho_{A:B_1B_2}) \geq E(\rho_{A:B_1}) +
  E(\rho_{A:B_2})$.
\item {\bf Faithfulness:} $E(\rho_{AB}) \leq \eps$ implies that
  $D(\rho,\Sep)\leq f(\eps)$ where $f\ra 0$ as $\eps\ra 0$,
  e.g. $f(\eps) = c\sqrt\eps$.
\eenum

If we had such a measure, the proof would be very easy:
\bas
& \quad \log(d_A) \\
& \geq E(\rho_{A:B_1\dots B_k})
& \text{by normalization} \\
& \geq E(\rho_{A:B_1}) + E(\rho_{A:B_2\dots B_k})
& \text{by monogamy} \\
& \geq \sum_{i=1}^k E(\rho_{A:B_i})
& \text{repeating the argument}
\\& = kE(\rho)
\eas
Rearranging, we have $E(\rho) \leq \log(d_A)/k$, and finally we use
faithfulness to argue that $D(\rho,\Sep) \leq f(\log(d_A) / k)$.

Such a measure does exist!  It is called {\em squashed entanglement}
and was introduced in 2003 by our very own Matthias Christandl and
Andreas Winter [\href{http://arxiv.org/abs/quant-ph/0308088}{quant-ph/0308088}].
Normalization and monogamy are straightforward to prove for it (and were proved in the original
paper), but faithfulness was not proved until 2010 [Brand\~ao, Christandl, Yard; \href{http://arxiv.org/abs/1010.1750}{arXiv:1010.1750}].

%% file: 31may_2.tex
\lecture{14}{31 May, 2013}{Matthias Christandl}{Representation theory and spectrum estimation}
\label{lec:kron}

We have seen that the spectrum of density matrices plays an important role in the understanding of the quantum marginal problem and the SLOCC classification of entanglement. In this section, we will introduce some tools from representation theory in order to find an elegant way to estimate the spectrum by measuring a number of copies of a given quantum state. An unexpected relation to the marginal problem and the entanglement invariants will arise.

\section{Representation theory}

Given a group $G$, a \emph{(finite-dimensional, unitary) representation} of $G$ is a mapping $g \mapsto U(g)$ from the group $G$ into the unitaries on a finite-dimensional Hilbert space $V$ such that
\begin{equation*}
  U(g) U(h) = U(gh).
\end{equation*}

We say a representation is \emph{irreducible} if any invariant subspace $W \subseteq V$ (i.e., $U(g) W \subseteq W$ for all $g \in G$) is either zero or all of $V$.

Finally, we say that two representations $g \mapsto U(g)$ and $g \mapsto \widetilde U(g)$ are \emph{equivalent} if there exists an isomorphism $A$ such that $A U(g) A^{-1} = \widetilde U(g)$ for all $g$ (also called an \emph{intertwiner}).

We have the following important theorem:

\begin{thm}
\label{thm:complete-reducibility}
For $G$ a finite or a compact Lie group, any representation $g \mapsto U(g)$ can be decomposed into (i.e., is equivalent to) a sum of irreducible representations. In other words, there exists an isomorphism $A$ from
\begin{equation*}
  V \cong \bigoplus_{i \in \hat G} V_i \otimes \CC^{m_i}
\end{equation*}
such that $A U(g) A^{-1} = \bigoplus_i U_i(g) \otimes I_{\CC^{m_i}}$. Here, $\hat G$ is the set of equivalent classes of irreducible representations, and $m_i$ is called the \emph{multiplicity} in $V$ of an irreducible representation $V_i$ with action $g \mapsto U_i(g)$.
\end{thm}

Thus irreducible representations are the building blocks of general representations.

\begin{exl*}
Let us consider $G = \SU(2)$. In this case $V_j$ is the representation of spin $j \in \{0,1/2,1,\dots\}$, of dimension $\text{dim}(V_j) = 2j+1$. A basis for $V_j$ is $\{ \ket{j, m} \}_{m = -j}^{j}$.

The action of the Lie algebra of $\SU(2)$ on $V_j$ is given by
\begin{align*}
  \sigma_{z} \cdot \ket{j, m} &= m \ket{j, m}, \\
  \sigma_{\pm} \cdot \ket{j, m} &\propto \ket{j, m \pm 1},
\end{align*}
where $\sigma_\pm = \sigma_x \pm i \sigma_y$ are the spin raising/lowering operators.
The action of the group $\SU(2)$ on $V_j$ is obtained by exponentiating the Lie algebra.

More concretely, we can also write $V_j = \Sym^{2j} (\mathbb{C}^2) \subseteq (\mathbb{C}^2)^{\otimes 2j}$.
Then the action of the group on $V_j$ is given by $g \mapsto \Psym^{2,2j} \, g^{\otimes n} \, \Psym^{2,2j}$.
\end{exl*}

\subsection{Schur-Weyl duality}

An important theorem in the representation theory of $\SU(d)$ and the symmetric group is the so-called Schur-Weyl duality.

Consider the representation $g \mapsto g^{\otimes n}$ of $\SU(2)$ on $V^{(n)} = (\mathbb{C}^2)^{\otimes n}$. By \cref{thm:complete-reducibility}, we can decompose this representation as follows into irreducible representations:
\begin{equation} \label{eq1}
V^{(n)} = \bigoplus_{j} V_j \otimes \mathbb{C}^{m_j^{(n)}}
\end{equation}

For $S_n$, the symmetric group of order $n$, there is also a natural representation in $V^{(n)}$, given by $\pi \ket{i_1, \dots, i_n} = \ket{\pi^{-1}(1), \dots, \pi^{-1}(n)}$ for $\pi \in S_n$. This representation clearly commutes with the one for $SU(d)$, so that the multiplicity spaces $\CC^{m_j^{(n)}}$ of the $\SU(d)$-action become representations of $S_n$.
\emph{Schur-Weyl duality} asserts that these representations are irreducible. Moreover, the $j$ that occur in the decomposition~\eqref{eq1} are $n/2$, $n/2-1$, \dots. In fact, one particular copy of $V_j$ is given by
\begin{equation}
\label{eq:onecopy}
  \left( \frac 1 {\sqrt 2} \left( \ket{01} - \ket{10} \right) \right)^{\otimes (n/2 - j)} \otimes \Sym^{2j}(\CC^2) \subseteq (\CC^2)^{\otimes n},
\end{equation}
and the summand $V_j \otimes \mathbb{C}^{m_j^{(n)}}$ in~\eqref{eq1} can be obtained by acting with the symmetric group.

We will not need any further details of Schur-Weyl duality in this lecture. Instead our goal is to compute the dimensions $m_j^{(n)}$ in~\eqref{eq1}.

\subsection{Computing \texorpdfstring{$m_j^{(n)}$}{m\_j}}

Let us consider $m_{n/2}^{(n)}$. We have
\begin{align*}
&\quad V^{(n+1)}
\cong V^{(n)} \otimes V^{(1)}
\cong \left( \bigoplus_j V_j \otimes \mathbb{C}^{m_j^{(n)}} \right) \otimes V_{1/2}
\cong \bigoplus_j (V_j \otimes V_{1/2}) \otimes \mathbb{C}^{m_j^{(n)}} \\
&\cong \bigoplus_j \left( V_{j+1/2} \oplus V_{j-1/2} \right) \otimes \mathbb{C}^{m_j^{(n)}}
\cong \bigoplus_{j'} V_{j'} \otimes \left( \mathbb{C}^{m_{j'-1/2}^{(n)}} \oplus \mathbb{C}^{m_{j'+1/2}^{(n)}} \right)
\end{align*}
Here we set $V_{-1/2} = 0$. Comparing with \cref{eq1}, we find the following recursion relation:
$$m_{j}^{(n+1)} = m_{j+1/2}^{(n)} + m_{j -1/2}^{(n)}.$$
Its solution can be checked to be
\begin{equation} \label{approxm}
m_{j}^{(n)} = \binom{n}{n/2 - j} - \binom{n}{n/2 - j - 1} \leq 2^{n \, h(1/2 \pm j/n)}
\end{equation}
with $h(p) := H(p,1-p) = -p \log p -(1-p) \log (1-p)$ is the binary entropy function.

\section{Spectrum estimation}

Let us for a moment forget about representation theory and consider the problem of spectrum estimation. In this problem we are given a source of quantum states which emits $n$ copies of an unknown density matrix: $\rho^{\otimes n}$. The goal is to perform a measurement which gives a estimate of the eigenvalues of $\rho$.

For a qubit state $\rho$, the eigenvalues of $\rho$ we can be written as $(1/2 + r, 1/2 - r)$, with $r \in [0, 1/2]$. The problem of estimating $r$ was considered by Keyl and Werner, who observed an interesting connection of the problem to the representation theory of $SU(2)$. They proposed measuring $j$ according to the decomposition~\eqref{eq1}. Then, with high probability $j/m \approx r$.

More precisely, the claim is that
\begin{equation*}
\Pr[j] = \tr P_j \rho^{\otimes n} \leq \text{const} \cdot 2^{-n \, \delta(1/2 + j/n \Vert 1/2 + r)},
\end{equation*}
where $P_j$ denotes the projector onto $V_j \otimes (\CC^2)^{\otimes n}$,
and where $\delta(x \Vert y)$ is the \emph{relative entropy} between two binary random variables, given by $\delta(x \Vert y) = x \log(x/y) + (1-x) \log((1-x)/(1-y))$.

Let us now sketch the proof. We can compute
\begin{equation}
\frac 1 2 (\bra{0, 1} - \bra{1, 0}) \rho^{\otimes 2} \rho (\ket{0, 1} - \ket{1, 0}) = \det \rho = (1/2 + r)(1/2 - r).
\end{equation}
Then, using \cref{eq:onecopy},
\begin{align*}
  \tr P_j \rho^{\otimes n}
  &= m_j^{(n)} \, (1/2 + r)^{n/2-j} (1/2 - r)^{n/2-j} \, \sum_{m=-j}^j (1/2 + r)^{j+m} (1/2 - r)^{j-m}.
\end{align*}
We can rewrite and upper-bound this as follows by using \cref{approxm}:
\begin{align*}
  &\quad m_j^{(n)} \, (1/2 + r)^{n/2-j} (1/2 - r)^{n/2-j} \, \sum_{m=-j}^j (1/2 + r)^{j+m} (1/2 - r)^{j-m} \\
  &= m_j^{(n)} \, (1/2 + r)^{n/2-j} (1/2 - r)^{n/2-j} \, \sum_{m=-j}^j (1/2 + r)^{j-m} (1/2 - r)^{j+m} \\
  &= m_j^{(n)} \, (1/2 + r)^{n/2-j} (1/2 - r)^{n/2-j} \, \sum_{m=-j}^j \frac {(1/2 - r)^{m+j}} {(1/2 + r)^{m-j}} \\
  &= m_j^{(n)} \, (1/2 + r)^{n/2+j} (1/2 - r)^{n/2-j} \, \sum_{m=0}^{2j} \frac {(1/2 - r)^m} {(1/2 + r)^m} \\
  &\leq 2^{n \, h(1/2 + j/n)} \, (1/2 + r)^{n/2+j} (1/2 - r)^{n/2-j} \, \sum_{m=0}^{\infty} \frac {(1/2 - r)^m} {(1/2 + r)^m} \\
  &= 2^{-n \, \delta(1/2+j/n \Vert 1/2+r)} \underbrace{\sum_{m=0}^{\infty} \frac {(1/2 - r)^m} {(1/2 + r)^m}}_{= \text{ const.}}.
\end{align*}

\subsection{Application of the Keyl-Werner relation}

Let us finish by mentioning one application. Suppose we have $\ket{\psi}_{ABC}^{\otimes n}$ and we measure $\{ P_{j_A} \}$, $\{ P_{j_B} \}$, and $\{ P_{j_C} \}$ on $A$, $B$ and $C$, respectively.
We just learned we will obtain with high probability outcomes $j_A, j_B, j_C$ such that $j_A / n \approx r_A$, $j_B / n \approx r_B$, and $j_C / n \approx r_C$. Therefore we must have
\begin{equation} \label{eq:cov}
  \tr \left( P_{j_A} \otimes P_{j_B} \otimes P_{j_C} \, \proj\psi_{ABC}^{\otimes n} \right) \neq 0.
\end{equation}
Therefore, there exists $g \in G$ such that
$$(\bra{\omega_{j_A}} \otimes \bra{\omega_{j_B}} \otimes \bra{\omega_{j_C}}) g \ket{\psi}_{ABC}^{\otimes n} \neq 0$$ for some highest weight vectors $\ket{\omega_{j_A}}$, etc.\ in the subspaces associated with $P_{j_A}$, etc.  The function given in \cref{eq:cov} is a polynomial in the entries of $\ket{\psi}$ and transforms covariantly under SLOCC with associated labels given by $j_A, j_B$ and $j_C$. We thus see quite concretely that solving the quantum marginal problem is related to the study of covariant polynomials and their asymptotics, as anticipated in \cref{lec:entanglement polytopes}. This connection can in turn be used to compute the entanglement polytopes.

%% file: 31may_3.tex
\lecture{15}{31 May, 2013}{Aram Harrow}{Proof of the 1-LOCC quantum de Finetti theorem}

\section{Introduction}

In this lecture, I will give a proof of the following theorem first mentioned by Fernando, on the way introducing useful properties of von Neumann and Shannon entropy.

\begin{thm}
\label{thm:BCY max min}
Let $\rho_{AB}$ be $k$-extendible and $M'$ a 1-LOCC measurement. Then there exists a separable state $\sigma$ such that
\be |\tr M' (\rho-\sigma)| \leq \text{const}\cdot \sqrt{\frac{\log
    d_A}{k}}.
\label{eq:BCY-bound}\ee
\end{thm}

It is possible to swap the quantifiers with help of von Neumann's
minimax theorem; that is, if $\rho$ is $k$-extendable then there
exists a separable $\sigma$ such that for {\em any} 1-LOCC measurement
$M'$ in \cref{eq:BCY-bound} holds.
This is the version of the theorem that Fernando described in \cref{thm:BCY}.
For the purposes of the proof, we will
work with the easier version stated in \cref{thm:BCY max min}.

Recall that $M'$ can be written in the form $M'= \sum_{x=1}^m A_x \otimes B_x$ for $0\leq A_x \leq I  $ and $0\leq B_x \leq I  $.  Define the measurement $M_B(\rho)= \sum_x \tr (B_x\rho) \proj{x}$.
This corresponds to Bob measuring his state and output the outcome $x$ as a classical register $\proj x$.
It is the goal to show that
\begin{equation}
\label{eq:goal one way}
	I_A \otimes M_B(\rho_{AB}) \approx I_A \otimes M_B(\sigma_{AB })
\end{equation}
for some separable state $\sigma_{AB}$. 

Now, the fact that $\rho_{AB}$ is $k$-extendible implies that there exists a state $\pi_{AB_1 \cdots B_k}$ such that $\rho_{AB}= \pi_{AB_l}$ for all $l=1,\dots,k$ (cf.\ the discussion below the SDP~\eqref{eq:sdp}).
We now consider the state
$$\omega_{AB_1 \cdots B_k} := (I_A \otimes M_B^{\otimes k})(\pi_{AB_1 \cdots B_k})$$



\bit
\item Case 1: $\omega_{AB_1}\approx \omega_A \otimes \omega_{B_1}$. Then we are done.
\item Case 2: $\omega_{AB_1}$ is far from $\omega_A \otimes \omega_{B_1}$. Then we condition on $B_1$ and are  looking at system $B_2$, having reduced the uncertainty about that system. This way we get a little closer to case 1. When continuing to $B_3$ etc, this will prove the theorem.
\eit

This was the high-level view. We will now make this precise by using a
measure of correlation based on entropy. Since the maximum of an
entropy is $\log d$, this will give the bound of theorem, as opposed
to the linear scaling in $d$ that we encountered in the trace-norm
quantum de Finetti theorem.

\section{Conditional entropy and mutual information}
Recall that the Shannon entropy of a probability distribution $p$ of a random variable $X$ is given by
$$H(p)= -\sum_x p_x \log p_x = H(X)_p.$$
When we have joint distributions $p(x,y)$, we can look at the marginal distributions $p(x)= \sum_y p(x,y)$ and $p(y)= \sum_x p(x,y)$ and their entropies. The conditional entropy of $X$ given $Y$ is defined as
$$H(X|Y)_p = \sum_y p(y) H(X)_{p(x|y)}$$
where $p(x|y)=\frac{p(x,y)}{p(y)}$.
Writing the conditional entropy out explicitly we find the formula
$$H(X|Y)_p= H(XY)_p-H(Y)_p.$$
This gives us a beautiful interpretation of the conditional entropy: it is just the entropy of the joint distribution of $XY$ minus the entropy of $Y$.

We can measure the correlation between two random variables $X$ and $Y$ by looking at the difference between the entropy $H(X)$ and $H(X|Y)$
$$I(X:Y)= H(X)-H(X|Y)=H(X)+H(Y)-H(XY)$$
Note that this quantity is symmetric with respect to interchange of $X$ and $Y$.
It is known as the \emph{mutual information} between $X$ and $Y$ and quantifies by how much our uncertainty about $X$ reduces when we are given the random variable $Y$ (and vice versa, of course). It is also the amount of bits that you save by compressing $XY$ together as opposed to compressing $X$ and $Y$ separately.

The mutual information has a few nice properties:
\begin{itemize}
\item $I(X:Y) \geq 0$.
\item $I(X:Y) \leq \min \{ \log |X|, \log |Y| \}$, where $|X|$ denotes the number of symbols in $X$.
\item \textbf{Pinsker's inequality:} $I(X:Y) \geq \frac{1}{2\ln 2} \left( \sum_{x,y}|p(xy)-p(x)p(y)|\right)^2$.
\end{itemize}

Let us now look at the quantum version of all this. I will use the notation that $S(A)_\rho=S(\rho_A)$. When we have a joint state $\rho_{AB}$, then $S(A)_\rho = S(\tr_B \rho_{AB})$. Note that it is not immediately clear how to define the conditional entropy in the quantum case, since we cannot directly condition on the quantum system $B$. Luckily, we had a second way of writing the conditional entropy, and we are just going to define the \emph{quantum conditional entropy} as the difference
$$S(A|B):= S(AB)-S(B)$$
and the \emph{quantum mutual information} as
$$I(A:B)=S(A)+S(B)-S(AB).$$
Similarly to its classical counterpart, it has the following properties:
\begin{itemize}
\item $I(A:B) \geq 0$.
\item $I(A:B) \leq 2 \min \{ \log d_A, d_B \}$ -- note the factor of two.
\item \textbf{Pinsker's inequality:} $I(A:B) \geq \frac{1}{2\ln 2} ||\rho_{AB}-\rho_A \otimes \rho_B ||_1^2$
\end{itemize}

This last property looks like it could be useful in proving the theorem. But it all cannot be that easy, because we know that we should use the 1-LOCC norm rather than the trace norm! In order to proceed, we need the \emph{conditional mutual information}
$$I(A:B|C)= S(A|C)+S(B|C)-S(AB|C)= S(AC)+ S(BC)-S(ABC)-S(C).$$
This formula is a little difficult to grasp and it is difficult to develop an intuition for it.
The conditional mutual information has a nice property, though; it satisfies the following \emph{chain rule},
$$I(A:BC)=I(A:C)+I(A:B|C),$$
which you can easily check. It is called the chain rule, in part, because we can iterate it: Let's assume we have a $k$-extendible state and we measure all of the $B$ systems. How much does $A$ know about the $B$'s? The chain rule gives us
\begin{align*}
I(A: B_1 \cdots B_k)&=   I(A: B_1)+ I(A: B_2 \cdots B_k|B_1)\\
  &=   I(A: B_1)+ I(A: B_2|B_1)+ I(A: B_3 \cdots B_k|B_1B_2)\\
    &=   I(A: B_1)+ I(A: B_2|B_1)+\cdots + I(A: B_k|B_1B_2 \cdots B_{k-1})
\end{align*}
There are $k$ terms and the sum is smaller than $\log d_A$.
Hence there is some $l$ such that
$$I(A:B_l|B_1B_2 \cdots B_{l-1})_\omega \leq \frac{\log d_A}{k}.$$
Since $B_1\dots B_{l-1}$ are classical, we can write this conditional mutual information as an average.
To see this explicitly, we apply the first $l-1$ measurements to $\pi_{AB_1\dots B_l}$ to obtain a state of the form
\begin{equation}
\label{eq:nice trick}
	I_{AB_l} \otimes M_B^{\otimes l-1} (\pi_{AB_1\dots B_l}) = \sum_{\vec x = (x_1, \dots, x_{l-1})} p_{\vec x} \, \pi_{AB_l}^{\vec x} \otimes \proj{\vec x}_{B_1\dots B_{l-1}}.
\end{equation}
Then, $\omega_{AB_1\dots B_l} = \sum_{\vec x} p_{\vec x} \, \omega_{AB_l}^{\vec x} \otimes \proj{\vec x}_{B_1\dots B_{l-1}}$, where $\omega_{AB_l^{\vec x}} = I_A \otimes M_{B_l}(\pi_{AB_l}^{\vec x})$, which implies that
\[ I(A:B_l|B_1B_2 \cdots B_{l-1})_\omega = \sum_{\vec x} p_{\vec x} \, I(A:B_l)_{\omega_{AB_l^{\vec x}}}. \]
On the other hand, the measurements in \cref{eq:nice trick} leave $\pi_{AB_l}$ unchanged and so $\rho_{AB} = \pi_{AB_l} = \sum_{\vec x} p_{\vec x} \, \pi_{AB_l}^{\vec x}$.
Thus we obtain that
\begin{align*}
  &\quad \frac 1 {2 \ln 2} \lVert (I_A \otimes M_B)(\rho_{AB} - \sum_{\vec x} p_{\vec x} \, \pi_A^{\vec x} \otimes \pi_B^{\vec x} ) \rVert_1^2 \\
  &\leq \frac 1 {2 \ln 2} \sum_{\vec x} p_{\vec x} \left( \lVert (I_A \otimes M_B)(\pi_{AB_l}^{\vec x} - \pi_A^{\vec x} \otimes \pi_B^{\vec x} ) \rVert_1 \right)^2 \\
  &\leq \sum_{\vec x} p_{\vec x} \, I(A:B_l)_{\omega_{AB_l^{\vec x}}} = I(A:B_l|B_1B_2 \cdots B_{l-1})_\omega \leq \frac {\log d_A} k,
\end{align*}
where we used the triangle inequality and Jensen's inequality, Pinsker's inequality, and, lastly, the bound on the mutual information computed above.
This establishes \cref{eq:goal one way}, which in turn immediately implies our theorem (see [Brand\~ao, Harrow; \href{http://arxiv.org/abs/1210.6367}{arXiv:1210.6367}] for more details).